\documentclass[preprint,12pt]{elsarticle}

\usepackage[utf8]{inputenc}
\usepackage[T1]{fontenc}
\usepackage{lmodern}
\usepackage{microtype}
\usepackage[scaled]{helvet}
\usepackage[english]{babel}

\usepackage{amsmath,amsfonts,amssymb}
\usepackage{graphicx}
\usepackage{placeins}
\usepackage{float}       % [H] float placement (supplementary tables)
\usepackage{rotating}    % sidewaysfigure (supplementary)
\usepackage[bookmarks=false]{hyperref}
\usepackage{subfig}
\biboptions{sort&compress}
\journal{Acta Materialia}

\graphicspath{{figures/}}

\newcommand{\startsupplement}{%
  \clearpage
  \setcounter{section}{0}%
  \setcounter{figure}{0}%
  \setcounter{table}{0}%
  \setcounter{equation}{0}%
  \renewcommand{\thesection}{S\arabic{section}}%
  \renewcommand{\thesubsection}{S\arabic{section}.\arabic{subsection}}%
  \renewcommand{\thesubsubsection}{S\arabic{section}.\arabic{subsection}.\arabic{subsubsection}}%
  \renewcommand{\thefigure}{S\arabic{figure}}%
  \renewcommand{\thetable}{S\arabic{table}}%
  \renewcommand{\theequation}{S\arabic{equation}}%
  \renewcommand{\theHsection}{S.\arabic{section}}%
  \renewcommand{\theHsubsection}{S.\arabic{section}.\arabic{subsection}}%
  \renewcommand{\theHsubsubsection}{S.\arabic{section}.\arabic{subsection}.\arabic{subsubsection}}%
  \renewcommand{\theHfigure}{S.\arabic{figure}}%
  \renewcommand{\theHtable}{S.\arabic{table}}%
  \renewcommand{\theHequation}{S.\arabic{equation}}%
}

\begin{document}

\begin{frontmatter}

\title{Faulted loop nucleation and dopant activation in Al-implanted 4H-SiC}

\author[label1]{Sabine Leroch\corref{cor1}}
\ead{leroch@iue.tuwien.ac.at}
\author[label1]{Robert Stella}
\ead{stella@iue.tuwien.ac.at}
\author[label2]{Andreas Hössinger}
\ead{andreas.hoessinger@silvaco.com}
\author[label1]{Lado Filipovic\corref{cor1}}
\ead{filipovic@iue.tuwien.ac.at}
\cortext[cor1]{Corresponding authors.}

\affiliation[label1]{organization={Christian Doppler Laboratory for Multi-Scale Process Modeling of Semiconductor Devices and Sensors at the
 Institute for Microelectronics, TU Wien},
  postcode={1040}, city={Vienna}, country={Austria}}
\affiliation[label2]{organization={Silvaco Europe Ltd.},
  addressline={St Ives}, postcode={PE27 5JL}, state={Cambridgeshire}, country={United Kingdom}}

\begin{abstract}
%% ----------------------
%% Write your abstract here. Do not enclose it in an "abstract"
%% environment.
%% ----------------------
We present a molecular dynamics (MD) study of Al implantation in 4H-SiC to resolve how implantation temperature and dose control defect evolution and dopant activation during annealing. Simulations use the Gao-Weber potential combined with a Morse Al-SiC interaction reparameterized to density functional theory (DFT) diffusion and kick-in/kick-out barriers. 
%We identify two regimes separated by the Al solubility concentration scale ($\sim2\times10^{20}$\,cm$^{-3}$): a low-concentration regime dominated by isolated point defects and small complexes, and a high-concentration regime characterized by defect clustering and extended planar defects. 
At aluminum (Al) concentrations above the saturation limit of $\sim2\times10^{20}$\,cm$^{-3}$, implantation at an elevated temperature of $900$\,K promotes interstitial-rich planar clustering already during implantation. During annealing, these clusters act as sinks for Al and evolve into faulted interstitial loops, thereby reducing substitutional Al incorporation. In contrast, lower implantation temperature preserves stronger local disorder that is subsequently consumed by epitaxial regrowth, leading to higher chemical activation within MD-accessible annealing times. Atomistic trajectories reveal a thermally activated transformation into faulted loops once a critical planar cluster size of about $60$ interstitials is reached. The stacking fault does not nucleate as a single coherent disk, as assumed in the textbook picture. Instead, several locally transformed regions within one connected or separated irregular clusters appear first, and as they grow their partials meet
and the individual faults merge into one. The extracted activation energy for dislocation nucleation is about $1.1$\,eV, while stacking-fault growth follows an Arrhenius behavior with an activation energy of about $2.1$\,eV. The dominant loops are Frank-type at high temperature, whereas transient Shockley partials are mainly observed at early stages and at temperatures below $2000$\,K. 
As the large, compact, and planar defect structures form/dissolve, additional thermodynamically and kinetically stable, compensating Al-C complexes form. These could confirm the experimental hypothesis that, in the case of Al supersaturation, secondary defects are responsible for the decrease in Al activation with progressing annealing time.

\end{abstract}


\end{frontmatter}

%% ---------------------------
%% If you wish to include additional packages, define new environments or
%% new commands, put them in the file includes.tex
%%
%% Write your abstract in the file abstract.tex.
%% ---------------------------

%% ---------------------------
%% Introduction
%% ---------------------------

%% Introduction
%% ---------------------------

\section{Introduction}
%\label{}
Due to its wide band gap of $3.26$\,eV, high breakdown voltage, and excellent thermal stability, 4H-SiC is particularly well suited for various semiconductor applications in microelectronics~\cite{KimotoCooper2014,Wellmann2022}. 
To enable electrical conductivity under normal operating conditions, SiC is doped with elements from either group III (p-type) or group V (n-type) of the periodic table~\cite{KimotoCooper2014}.
This study focuses on p-type doping of 4H-SiC using aluminum. Activated Al introduces acceptor states with thermal ionization energies of $198$\,meV and $210$\,meV above the valence band maximum (VBM) in the hexagonal and cubic lattice sites, respectively~\cite{Ivanov2005,KimotoCooper2014}.
Implantation occurs under non-equilibrium conditions allowing the thermal solubility limit of Al in 4H-SiC to be exceeded, in contrast to epitaxial incorporation~\cite{Linnarsson2001}.
Ion implantation is thus the preferred method for introducing Al dopants into precisely defined regions of the crystal~\cite{hallen2016,mueting2020}.  As a result, a wide range of dopant concentrations can be achieved, varying from $10^{17}$\,cm$^{-3}$ to $10^{21}$\,cm$^{-3}$~\cite{Negoro2004,Nipoti2018}.
To prevent the amorphization of SiC at high Al doses, implantation is typically performed at wafer temperatures between $500$ and $900$\,K.
%%AH: We should add some comments here mentioning why high temperature implantation is used in SiC technology 
%%AH: Ideally later in the text we would refer to those motivations
%%DONE

Ideally, dopant atoms should occupy substitutional lattice sites in an otherwise perfect crystal to become electrically active and contribute free carriers. However, ion bombardment displaces host atoms from their lattice positions, generating interstitial-vacancy pairs known as Frenkel pairs (FPs). These primary defects can agglomerate into larger interstitial clusters or lead to the formation of amorphous pockets embedded inside the crystalline matrix. Moreover, implanted Al ions may occupy interstitial sites or get trapped in larger intrinsic clusters instead of being at substitutional sites.
Even after high-temperature annealing at $1600$ to $2500$\,K, numerous thermally stable defects remain in the crystal such as intrinsic point defects and clusters as well as small Al-related complexes and extended defect structures ~\cite{Wendler1998,PhysRevB.71.165210,Kumar2024}.

Evidence of an increase in the number of compensating defects during annealing at Al supersaturation had already been noted in the work by Negoro \emph{et al.}~\cite{Negoro2004} and later by Nipoti \emph{et al.}~\cite{Nipoti2019} who observed that, after reaching a maximum hole concentration, the concentration tends to decrease with increasing annealing time, although Al rapidly occupied lattice sites. Experiments \cite{Negoro2004,Persson2003} have shown that, in the case of Al supersaturation, stacking faults occur during annealing. However, Negoro \emph{et al.} \cite{Negoro2004} could not detect any significant difference in the density of stacking faults after $1$ minute and $180$ minutes of annealing. They therefore attributed the compensation to secondary defects that were not resolvable in the transmission electron microscopy (TEM) measurements. 
Since then, there have been several attempts to study the origin of the observed compensation. TEM studies by Nipoti \emph{et al.}~\cite{Nipoti2018} revealed that nano-sized defect clusters and orthogonally arranged loops
appear in the implanted volume of annealed 4H-SiC for all Al concentrations,
while at concentrations exceeding the Al solubility of $2\times10^{20}$\,cm$^{-3}$, additional defect structures appeared, including Al precipitates and stacking faults free of Al. These stacking faults formed near the center of the implantation peak by insertion of an additional Si-C bilayer in the $(0001)$ plane and increased in size with increasing Al concentrations, extending over large areas in the crystal. 
The density and size of such planar defects increase not only with the implantation dose but also with the implantation temperature. 
An implantation window in which the formation of faulted loops is avoided is thus bounded from below by the critical amorphization temperature of SiC of about $500$\,K, and from above by the onset of dislocation loop formation near $870$\,K, below which the induced defects remain mainly isolated compact clusters~\cite{ZANG2025}. In addition, several experimental studies~\cite{Wendler1998,KONDO2008160,BARBOT2011,Michaud2013} suggest that moderate implantation temperatures around $500$\,K can be more favorable at Al supersaturation to suppress the formation of extended defects during annealing, which has recently been confirmed by MD simulations \cite{leroch2026_JMCC} using the GW-Morse potential applied in the current study. Those MD simulations have shown that the formation of stacking faults can be prevented or significantly reduced for Al concentrations above $2\times10^{20}$\,cm$^{-3}$ if the implantation temperature is controlled accordingly to achieve a degree of disorder in the SiC crystal in the range of $40$–$60\%$ during implantation, which also increases Al activation via epitaxial incorporation at the subsequent annealing stage.

Extended defects such as large planar interstitial clusters and faulted loops primarily degrade electrical properties by trapping dopants or recombining carriers. Although these extended defects can also introduce electronic states within the band gap, they are generally not treated as conventional dopant compensation centers. In contrast, small point defects and defect complexes in 4H-SiC typically compensate dopants through localized compensating levels in the band gap. 
To identify smaller defects recent spectroscopic studies of Kumar \emph{et al.}~\cite{Kumar2024}, combining low-energy muon spin rotation (LE-$\mu$SR) and deep-level transient spectroscopy (DLTS) have been done. The experiments  have shown increased intensities of carbon- and, with rising Al dose, Al-related defect signals indicating that Al- and C-related complexes contribute to the observed acceptor compensation.
In parallel with these experiments, intensive efforts were made to characterize point defects including their recombination and migration barriers as well as the structure of smaller defect complexes using quantum-mechanical methods~\cite{bockstedte2003,bockstedte2004,mattausch2004,liao2009,Zheng2013,Guillaume2007,gali2007,huang2022}. Numerous density functional theory (DFT) calculations addressed the most common intrinsic point defects in 4H-SiC, such as the carbon vacancy and its complexes and stable carbon clusters~\cite{mattausch2004,jiang2012,Li2023}, including the dicarbon antisite complex, which has been associated with the D$_\mathrm{II}$ center. 
Pure carbon interstitial clusters 
%containing up to four atoms 
exhibit exceptionally high thermal stability in the whole band gap, with dissociation energies reaching up to $5$\,eV per atom for trimers, while larger neutral interstitial clusters (size $>10$) made up of Si and C interstitials in stoichiometric distribution have dissociation energies of about $2$\,eV per atom~\cite{Ko2017}.
Moreover,  Mattausch \emph{et al.} \cite{Mattausch2005_Al}, Gali \emph{et al.}~\cite{gali2007} and Hornos~\cite{hornos2008} investigated small Al-related complexes. They demonstrated that substitutional and interstitial Al can form stable complexes with one or two carbon interstitials (with binding energies ranging from $1.7$ -$6$\,eV) and with one vacancy, all of which introduce deep compensating levels in the band gap.
While the neutral Al$_\mathrm{Si}$V$_\mathrm{C}$ complex %causes deep levels in the band gap and 
has been associated with the experimentally observed trap level A$_3$, 
none of the studied (Al-C) interstitial complexes could be clearly assigned to any of the shallow DLTS levels investigated by Hornos~\cite{hornos2008}.
In our recent publication~\cite{leroch2026_JMCC}, MD simulations identified additional stable trimers, one class composed of two substitutional Al atoms and one carbon or Si interstitial and a second class consisting of two carbon antisites binding one Al interstitial.  Complementary DFT calculations showed that most of the identified complexes are thermodynamically stable in the neutral state. While the Al$_\mathrm{Si}$-related complexes are mostly stable only in the positive and neutral state up to mid-gap, the Al$_\mathrm{I}$ related complexes are very stable across the entire bandgap. Their stability rises with the Fermi level, up to $7$\,eV for certain Al$_\mathrm{I}$2C$_\mathrm{Si}$ complexes in the negative state close to the conduction band minimum (CBM) making them even more stable than the pure C trimer. Moreover, several of these trimers have shallow compensating levels in the range of $0.2-0.5$\,eV above the VBM, as found in DLTS measurements and discussed by Hornos~\cite{hornos2008}.

Despite extensive experimental and theoretical work, substantial uncertainty remains regarding how implantation and annealing conditions affect the formation, stability, and interaction of defects in Al-implanted 4H-SiC.
In particular, the atomistic origin of extended defect nucleation, as well as 
the systematic identification and characterization of the structure and stability of small Al- and C-related complexes, is incomplete, and their role in acceptor compensation therefore remains poorly understood.
\textit{Ab initio} methods are the most accurate approach to systematically investigate the structure of  small defect complexes of up to size $10$, but they become computationally prohibitive for extended defect structures containing thousands of defects. Such large-scale evolution can be treated effectively with MD, which also enables simulation times of several hundred nanoseconds. The agglomeration of larger, interstitial clusters and extended defect structures is less sensitive to the choice of the potential, as it is predominantly driven by entropy. 
Experimentally, dislocation loops are typically detected only after they have reached sizes of several tens of nanometers, whereas in MD simulations their formation can be followed from the nucleation stage onward.
To the best of our knowledge, no MD study has yet reported spontaneous dislocation-loop nucleation during annealing of ion-implanted 4H-SiC. Earlier MD work on 4H-SiC mainly addressed epitaxial recrystallization from amorphous layers or implantation-induced point-defect evolution in cascades and short-time annealing over a few nanoseconds~\cite{Gao_zhang_2006,Gao2007_0001,Wang2024multiscale,Fang2024,wu2021}. Accordingly, the atomistic mechanisms of ion-implantation-driven faulted loop formation remain unresolved. 

The aim of this study is to track the temporal evolution of defects during high-temperature annealing, from initial compact interstitial clusters to planar, extended defect structures, at the atomic level, where we seek to determine the influence of implantation and annealing conditions on the defect structure. Moreover, we aim to reveal the connection between the stacking faults and the experimentally assumed secondary compensating defects. In addition, rates for dislocation-loop nucleation and growth are derived from the temporal defect evolution. Furthermore, diffusivities of point defects at different times/stages of annealing corresponding to varying underlying defect structures (amorphous, small compact clusters, extended planar defects) are calculated to determine time- and defect-dependent activation energies and cluster binding energies. 

This manuscript is structured as follows.
Section~\ref{sec:Methods} describes the simulation methods and setup.
In Section~\ref{sec:Results}, we follow the temporal evolution of the defect structures during annealing for varying implantation conditions and discuss the thermal stability of point defects and clusters and their influence on the Al activation. 
Finally, we describe the formation of faulted loops as a function of annealing temperature and time.
%\newpage

\section{Methods}
\label{sec:Methods}
We selected molecular dynamics which assumes neutral charge states for all defects to study the influence of the implantation conditions on  defect evolution and formation. As has been shown by our previous publication \cite{leroch2026_JMCC} this can be justified to a good extend since, 
at the first stage of annealing considered here, the large number of intrinsic defects can even compensate high dopant concentrations such that the Fermi level is pinned around or above mid-gap ($>1.6$\,eV). The carbon split interstitial ${\langle100\rangle}$\,(C-C) in the positive and neutral charge state, as well as the Si split interstitial ${\langle110\rangle}$\,(Si-Si) in the neutral state, are the most stable intrinsic point defects~\cite{bockstedte2004} contributing to diffusion/recombination around mid-gap. The Al interstitial exists in both positively and neutrally charged states, with migration barriers of $0.6$--$1$\,eV in the intrinsic to n-type region differing only slightly as a function of charge state~\cite{leroch2026_JMCC}. Moreover, most of the Al- and C- related defect complexes are stable in the neutral state above a Fermi level of $1.6$\,eV.
\subsection{Interatomic potentials}
For the simulation of the Al implantation into 4H-SiC, the Gao-Weber (GW)~\cite{gao2002} in combination with the Ziegler-Biersack-Littmark (ZBL) potential was used to describe the Si-C interactions, while a reparameterized Morse potential~\cite{leroch2023}, originally proposed by Dandekar and Shin~\cite{dandekar2011molecular}, was applied for the Al-Si and Al-C interactions 
\begin{equation}
        V(r) = D_0 \cdot \left( e^{-2\alpha(r-r_0)} - 2e^{-\alpha(r-r_0)}\right),
\end{equation}
with $D_0$ the strength of the potential, $\alpha$ the inverse decay length and $r_0$ the equilibrium distance. 
The parameters of the Morse potential, which were deduced by fitting Al diffusion and kick-in/kick-out barriers to DFT energies~\cite{leroch2026_JMCC} are listed in Table~\ref{tab:morse}. Pure Al-Al interactions were modeled using an embedded atom method (EAM) potential from the LAMMPS repository~\cite{thompson2022}.

\begin{table}[ht!]
    \centering
    \setlength{\tabcolsep}{9pt}   % adjust once; applies to all columns equally
    \begin{tabular}{lcc}
    \textbf{System} & \textbf{Parameters} & \textbf{Value} \\  
    \hline
    Al-Si & $D_0$ &  0.4\,eV \\
    & $\alpha$ & 1.6\,\AA{}$^{-1}$\\
    & $r_0$ & 2.8\,\AA{} \\ 
    Al-C & $D_0$ & 0.4\,eV \\
    & $\alpha$ & 1.4\,\AA{}$^{-1}$\\
    & $r_0$ & 2.05\,\AA{}\\ 
    \hline
    \end{tabular}
\caption{DFT-optimized Morse potential parameters used for the Al-Si and Al-C interactions. The interaction cutoff distance was set to 5\,\AA{}.}
\label{tab:morse}
\end{table}
%%AH: I fell this paragraph does not fit here - this has nothing to do with the methods
%DONE
\subsection{Molecular dynamics setup}
\begin{figure}[htbp!]
\centering
  \includegraphics[width=0.49\textwidth]{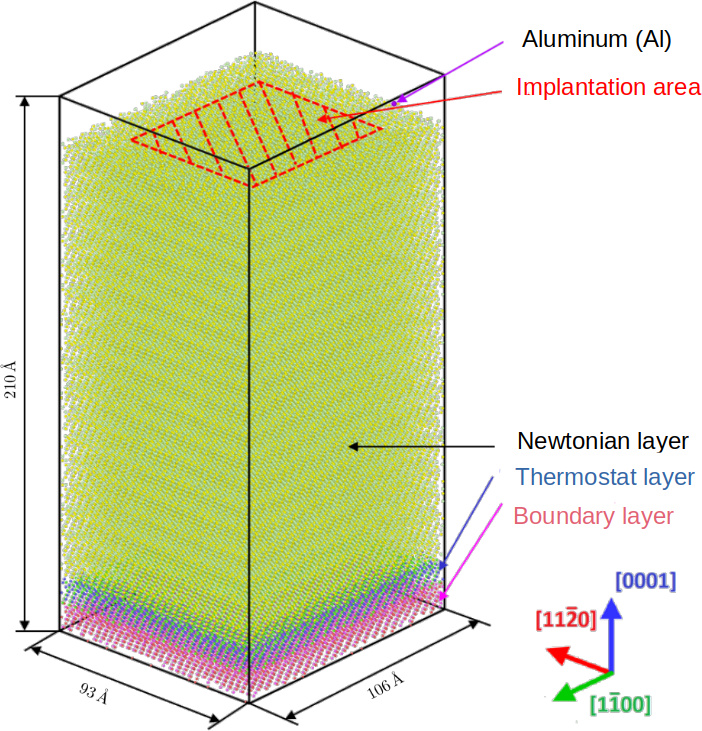} \hfill
  \caption{Molecular-dynamics (MD) implantation model setup.
 Simulation domain partitioned into a Newtonian (NVE) region, a thermostat region (heat sink), and a fixed boundary layer used during implantation.}
\label{fig:setup}
 \end{figure}
All MD simulations were performed using the LAMMPS software package~\cite{thompson2022}, while OVITO was employed for the subsequent visualization~\cite{stukowski2009visualization}.
A cuboidal 4H-SiC simulation cell with dimensions \mbox{$9$\,nm$\times10$\,nm$\times21$\,nm} with crystal orientation along the $\langle0001\rangle$ direction (z-direction) as shown in Figure~\ref{fig:setup} was constructed with periodic boundary conditions applied in the $x$ and $y$ directions (the (0001) plane), while the surface-normal ($z$) direction remained non-periodic.
For implantation, the simulation domain was partitioned into a Newtonian region, a thermostated heat-sink region, and a fixed bottom boundary layer to suppress rigid-body motion.
The system was first equilibrated at the selected implantation temperatures of $500$\,K and $900$\,K using the constant-temperature, constant-pressure (NPT)
ensemble with a Nos\'e-Hoover thermostat and barostat to relax residual stresses and to study the influence of the substrate temperature on defect formation. 
%%AH: In my opinion we should have an image here showing the below mentioned set-up.
%%     .) Also the implantation area should be drawn.
%%           Where does it sit in z-direction?
%DONE
%%     .) Does the 8x8 window mean that there was no implantation at the boundaries?
%% Implantation was done only in the 8x8 nm window following the set up in a MD paper from 2021 \cite{wu2021}.
 Al ions were implanted sequentially with a kinetic energy of $2$\,keV at an incidence angle of $7^\circ$ with respect to the surface normal of the $(0001)$ plane, using the microcanonical (NVE) ensemble in the Newtonian region. The implantation area was \mbox{$8$\,nm$\times8$\,nm}, and ion impact points were uniformly distributed within this surface plane. During implantation, an electron-stopping algorithm with adaptive time stepping was employed.
Following each implantation event, a relaxation period of $10$\,ps was applied to allow for recombination of closely spaced Frenkel pairs and dissipation of excess kinetic energy origin from the heat-spike that built up in the impact volume.
At the end of implantation, the systems were equilibrated at the implantation temperature in the NPT ensemble with a fixed time-step of $1$\,fs for $100$\,ps and then heated to the target annealing temperature at a rate of $80$\,K/ps.  Annealing temperatures between $1000$\,K and $2500$\,K were investigated, where the annealing was carried out in the NPT ensemble for $100$\,ns.
%%AH: Any idea how relevant this cool-down step is? In fact it is quite artificial, because in reality the annealing would just go on for longer at higher temperatures.
%% You are right it is of course very short. While in experiments recombination and diffusion reactions would continue, the cooling down more or less freezes the final annealing state. What one however could observe in MD, despite this very short cooling, is that close Frenkel pairs created at the high annealing temperature recombine, such that the number of intrinsic point defects decreases after relaxation to room temperature.
Finally, the annealed samples were cooled back to room temperature at a rate of $80$\,K/ps and equilibrated for an additional $100$\,ps before defect analysis and visualization.

As discussed in our recent work~\cite{leroch2026_JMCC}, implantation temperature plays a decisive role in extended-defect formation above the Al solubility limit (around \mbox{$2\times10^{20}$}\,cm$^{-3}$). For the shallow-implantation conditions used in this work (damage depth of approximately $9$\,nm), this corresponds to an implantation dose of about \mbox{$2\times10^{14}$}\,cm$^{-2}$. Thus, we focus primarily, on the investigation of the defect structures, at three Al ion implantation doses, namely
%%AH: It would be interesting (in brackets) to say how many atoms we have in total - I guess ~188000 - and how many Al atoms are added by implantation in each case.
% A table S3 has been added in the SI
\mbox{$5\times10^{13}$}\,cm$^{-2}$, which is far below this threshold and serves as a reference system, 
and doses \mbox{$5\times10^{14}$}\,cm$^{-2}$ and 
\mbox{$7.5\times10^{14}$}\,cm$^{-2}$, which correspond to strong supersaturation in the implanted region.

%\subsection{Data evaluation}

\subsection{Defect identification and analysis}

Defect structures were analyzed applying the Identify Diamond Structure (IDS) algorithm implemented in OVITO~\cite{stukowski2009visualization}.
To avoid bias from the highly defective top surface, defect evaluation was restricted to the bulk region (excluding the upper \mbox{$6$\,\AA{}}).
Antisites were identified from the local nearest-neighbor environment, i.e., lattice sites occupied by an atom whose four nearest neighbors are all of the same species as the occupying atom. Vacancies were identified by dangling-bond cage criteria using coordination analysis; for distorted clustered environments, a relaxed criterion (at least three dangling-bond atoms of identical species) was applied.
To distinguish isolated point defects from extended defect clusters, a connectivity-based clustering algorithm was implemented using the first minimum in the partial radial pair distribution function as cut-off. 
An Al atom was classified as chemically activated when it occupied a substitutional Si lattice site (Al$_\mathrm{Si}$). In the following, we distinguish between perfect activation (isolated Al$_\mathrm{Si}$) and general activation (all substitutional Al$_\mathrm{Si}$, including Al$_\mathrm{Si}$ bound in complexes or clusters).

For the dislocation analysis, OVITO’s Dislocation Extraction Algorithm (DXA) was applied. The algorithm locates the dislocations in the crystal lattice and determines the lengths of the partials together with their corresponding Burgers vectors. To determine the size of the stacking faults, a script was developed combining DXA and IDS analyses. The insertion of an additional crystal layer results in a cubic layer sandwiched between two other cubic layers (a 3C-SiC domain inside the 4H-SiC crystal). The script searches the region enclosed by the dislocation lines for the largest connected layer that satisfies this criterion and determines its convex hull to specify the stacking fault area.

\section{Results and discussion}
\label{sec:Results}
\subsection{Point defects and small defect complexes}
During implantation, Frenkel pairs (FPs) are initially generated; these point defects can either recombine or subsequently coalesce with neighboring defects into small complexes or compact clusters. During high temperature annealing, mobile interstitials diffuse through the crystal, whereas the less mobile vacancies tend to remain closer to their formation site.
Diffusing interstitials can also recombine with vacancies of other atom types forming antisites. Because carbon interstitials are typically more abundant than silicon interstitials, carbon antisites are formed as an additional recombination product during annealing. Another C-related point defect that is present in higher concentrations after annealing is the carbon vacancy, which has states in the band gap and can serve as a luminescence center~\cite{Ayedh2014,Wang2013}. Intrinsic interstitials and the silicon vacancy are less abundant. 

While stable C dimers and trimers, as well as complexes formed from carbon antisites predominate in pure SiC \cite{Li2023,mattausch2004,mattausch2005}, with rising Al dose, small Al-C complexes \cite{gali2007} become the most common species in Al-implanted systems \cite{Kumar2024}. The structures of some of these stable (Al-C) dimers and trimers were identified in a previous MD study \cite{leroch2026_JMCC}, in which we also determined their binding energies in different charge states using DFT. 
A detailed investigation of Al-C complexes is not the focus of this paper; however, to better interpret the following results, we briefly summarize here the most important findings regarding Al-related defect complexes.

\subsubsection{Al-C complexes}
The carbon antisite does not introduce levels in the band gap~\cite{kobayashi2019}, but plays an important role in the formation of Al-C complexes by acting as trap for diffusing Al interstitials during annealing.
As a first metastable configuration, an (Al-C)$_\mathrm{Si}$ split interstitial at the silicon site is formed after the capture of a diffusing Al$_\mathrm{I}$ by a carbon antisite, which initiates the kick-in of the Al to the Si site, as shown in Figure~S1, together with the activation barriers in Table~S3 of the supplementary material. 

In a subsequent transition, the (Al-C)$_\mathrm{Si}$ split interstitial transforms via a barrier of $0.3$\,eV into the thermodynamically stable Al$_\mathrm{Si}$C$_\mathrm{I}$ complex, which acts as a trapping state. This complex consists of an Al acceptor bound to a carbon interstitial forming a tilted split configuration corresponding to the isolated C split interstitial, and was reported by Gali \emph{et al.}~\cite{gali2007}. DFT nudged elastic band (NEB) calculations have shown that the neutral complex either returns with a barrier of $1.2$\,eV  to the (Al-C)$_\mathrm{Si}$ split interstitial or decays with a barrier of $1.6$\,eV toward isolated defects Al$_\mathrm{Si}$ and C$_\mathrm{I}$ increasing chemical activation~\cite{leroch2026_JMCC}. 
The binding energy of the neutral Al$_\mathrm{Si}$C$_\mathrm{I}$ complex is low, at 0.5\,eV. The complex is nevertheless persistent, not only because of its kinetic barriers, but also because it is repeatedly reformed during annealing after dissolution, as diffusing C interstitials recombine with another substitutional Al by overcoming a barrier of $1.2$\,eV. The thermodynamically stable neutral and the unstable negative Al$_\mathrm{Si}$C$_\mathrm{I}$ complex contribute to Al activation via the kick-in mechanism just described for Fermi levels above mid-gap, whereas the positive complex hardly contributes to activation because of the high migration barrier of Al$_\mathrm{I}^{3+}$  \cite{leroch2026_JMCC}. 
If the Al$_\mathrm{Si}$C$_\mathrm{I}$ complex captures another diffusing carbon interstitial, the ladder configuration Al$_\mathrm{Si}$(C$_\mathrm{I}$)$_2$, previously investigated by Gali \emph{et al.}~\cite{gali2007} is formed. While the neutral complex is metastable, the negatively and positively charged complexes have binding energies of $5$ and $6$\,eV, respectively.
Additional trimers observed in our MD simulations~\cite{leroch2026_JMCC} consist either of two Al acceptors and one carbon or silicon interstitial or of a single Al interstitial sandwiched by two carbon antisites in the basal plane or along the $c$-axis. DFT calculations have shown that these complexes are thermodynamically stable in the neutral and charged states~\cite{leroch2026_JMCC}, where the stability varies strongly with the Fermi level and the orientation of the defect in the crystal. In an earlier study we showed that the GW-Morse potential accurately describes the binding energy of neutral dimers, whereas for trimers, MD results differ from stability predictions based on DFT. 

\subsubsection{Al-Si complex}
The Al$_\mathrm{Si}$Si$_\mathrm{I}$ complex, which is formed in the course of the kick-out diffusion of Al$_\mathrm{I}$ has shallow and deep compensating levels in the band gap and preferentially decays with a low kick-out barrier of around $1$\,eV towards the free Al interstitial. The neutral Al$_\mathrm{Si}$Si$_{I,c}$ complex in the cubic plane is meta-stable near mid-gap, where it exhibits negative $U$-behavior. The neutral Al$_\mathrm{Si}$Si$_{I,h}$ complex in the hexagonal plane is a trapping state that decays into isolated defects and interstitial Al with nearly the same kinetic barriers of around $1.5$\,eV~\cite{leroch2026_JMCC}. Only the negative Al$_\mathrm{Si}$Si$_{I,h}$ complex at a Fermi level close to the CBM is unstable and contributes to Al activation.
The trimer formed by one Si interstitial sandwiched by two acceptor Al is thermodynamically stable across the whole band gap, where the stability decreases with rising Fermi level. The trimer in the hexagonal plane has a deep compensating level showing negative U-behavior around mid-gap, where the neutral trimer is meta-stable.

\subsubsection{Al-V$_C$ complex}
Besides (Al-C) complexes, stable Al$_\mathrm{Si}$V$_\mathrm{C}$ complexes with one or two carbon vacancies also form; their persistence is governed less by strong binding and more by the high migration barrier of the carbon vacancy. The complex has deep levels in the band gap, is meta-stable in the neutral and unstable in the negative state. 
Most V-complexes disappear during annealing through recombination with diffusing interstitials.
As a result of such a recombination reaction (the decay of the Al$_\mathrm{I}$V$_C$ complex), Al can occupy C lattice sites, leading to severe local lattice distortions and displacement of neighboring Si atoms. According to DFT, Al$_\mathrm{C}$ is stable~\cite{matsushima2019}, but exhibits significantly higher formation energies and introduces deep levels in the band gap; consequently, it does not function as an acceptor dopant.

\subsubsection{Al-Al complex}
As with the B$_\mathrm{Si}$B$_\mathrm{Si}$ complex in 3C-SiC, which contributes to hole conduction~\cite{Rurali2003}, the Al$_\mathrm{Si}$Al$_\mathrm{Si}$  complex is a shallow acceptor with a low binding energy of $0.2$\,eV under p-type conditions. The negative complex is unstable. Whereas the Al$_\mathrm{I}$Al$_\mathrm{I}$ complex has deep compensating levels in the band gap.
%%AH: The annealing time used in this study is technologically not relevant
%%     Therefore we should try to postulate what is going to happen for longer anneling times.
%%     In the following you sometimes say - at the end of annealing - but this is not realy relevant.
%%     You should rather say - for longer annealing times .... - 
%% I postulated what could go on after longer annealing in section 3.3 and in the conclusion. Stating there that the dissolution of exdented defects created at 900 K cause the formation of a higher concentration of thermally stable compensating (Al-C) complexes. Moreover, thermodnamically stable stacking faults act as recombinaton centers for carriers and life time killers. All together this leads to the necessity of longer annealing times, higher temperatures or the irradiation with hellium to suppress fault expansion.
 \subsection{Time evolution of defects during annealing}
\label{sec:time}
While the concentrations of point defects and Al- and carbon-related complexes are relatively insensitive to the implantation temperature, cluster formation is highly temperature-dependent at Al supersaturation. 
Moreover, we have previously demonstrated that increased clustering at elevated implantation temperatures of $900$\,K reduces Al activation, which is significantly higher at moderate temperatures of $500$\,K~\cite{leroch2026_JMCC}.
\subsubsection{Below the effective saturation dose}
\begin{figure}[htbp!]
\centering
  \subfloat[0 ns]{\includegraphics[width=0.45\textwidth]{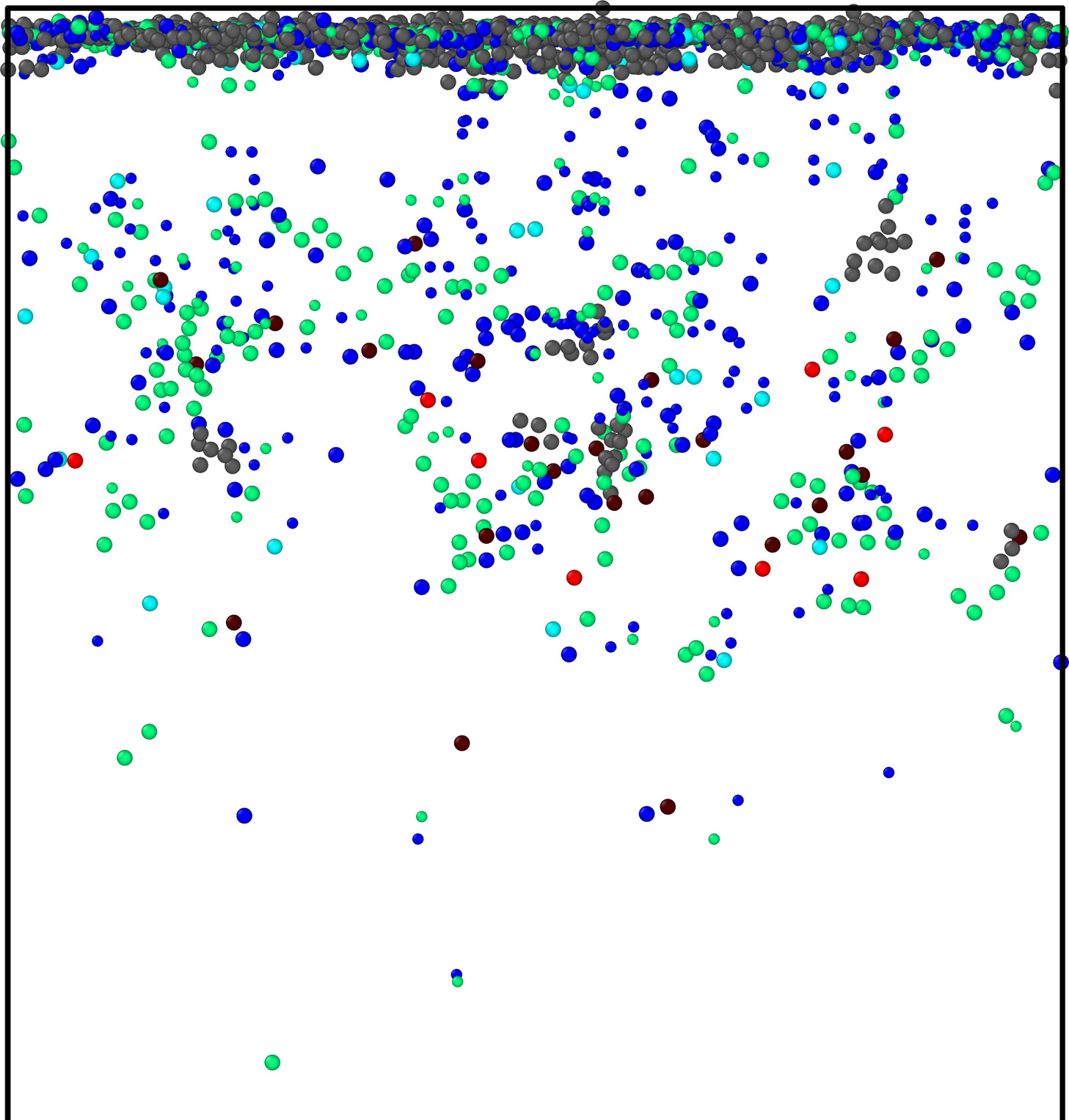}} \hfill
  \subfloat[1 ns]{\includegraphics[width=0.45\textwidth]{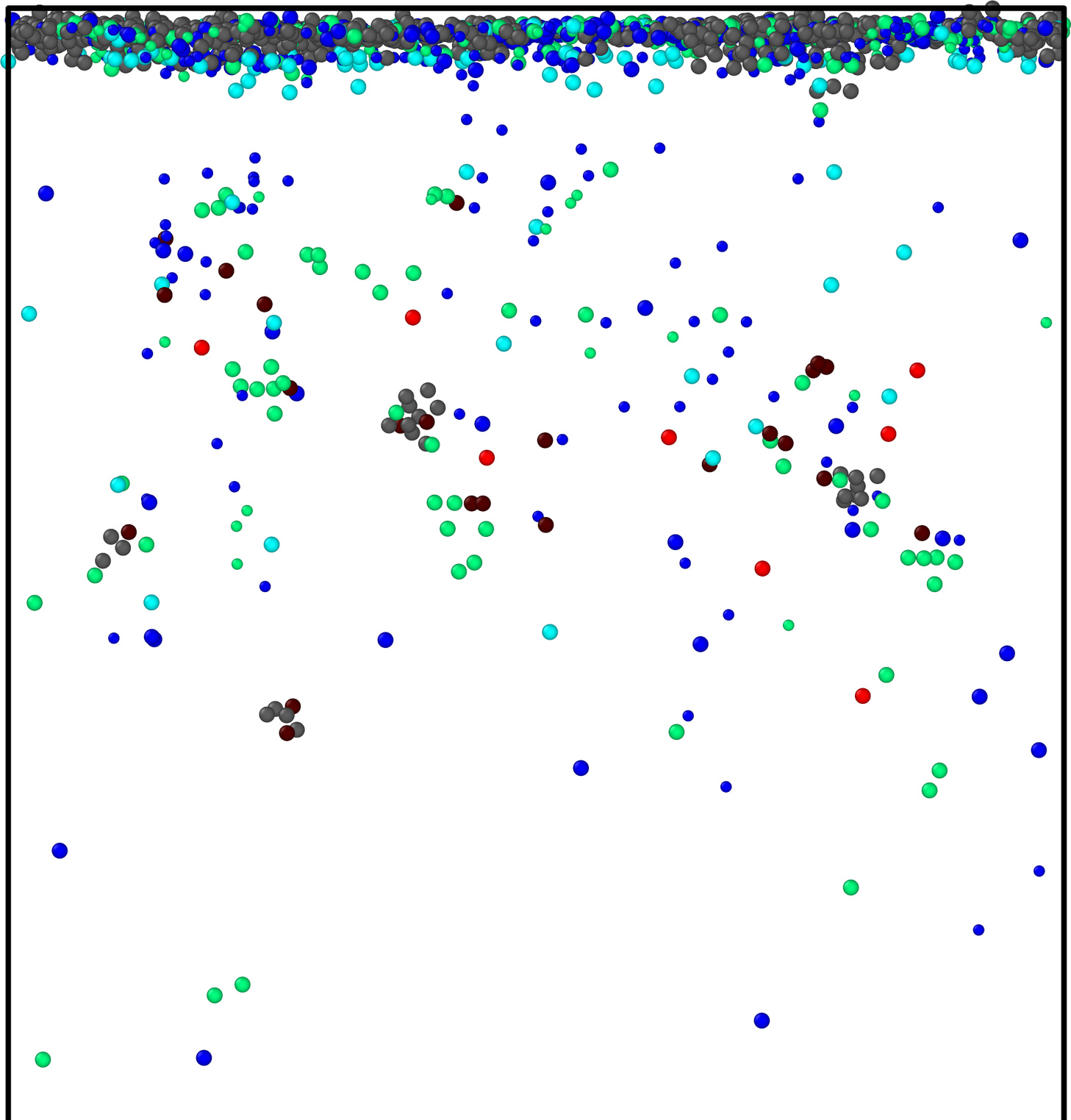}} \\
  \subfloat[30 ns]{\includegraphics[width=0.45\textwidth]{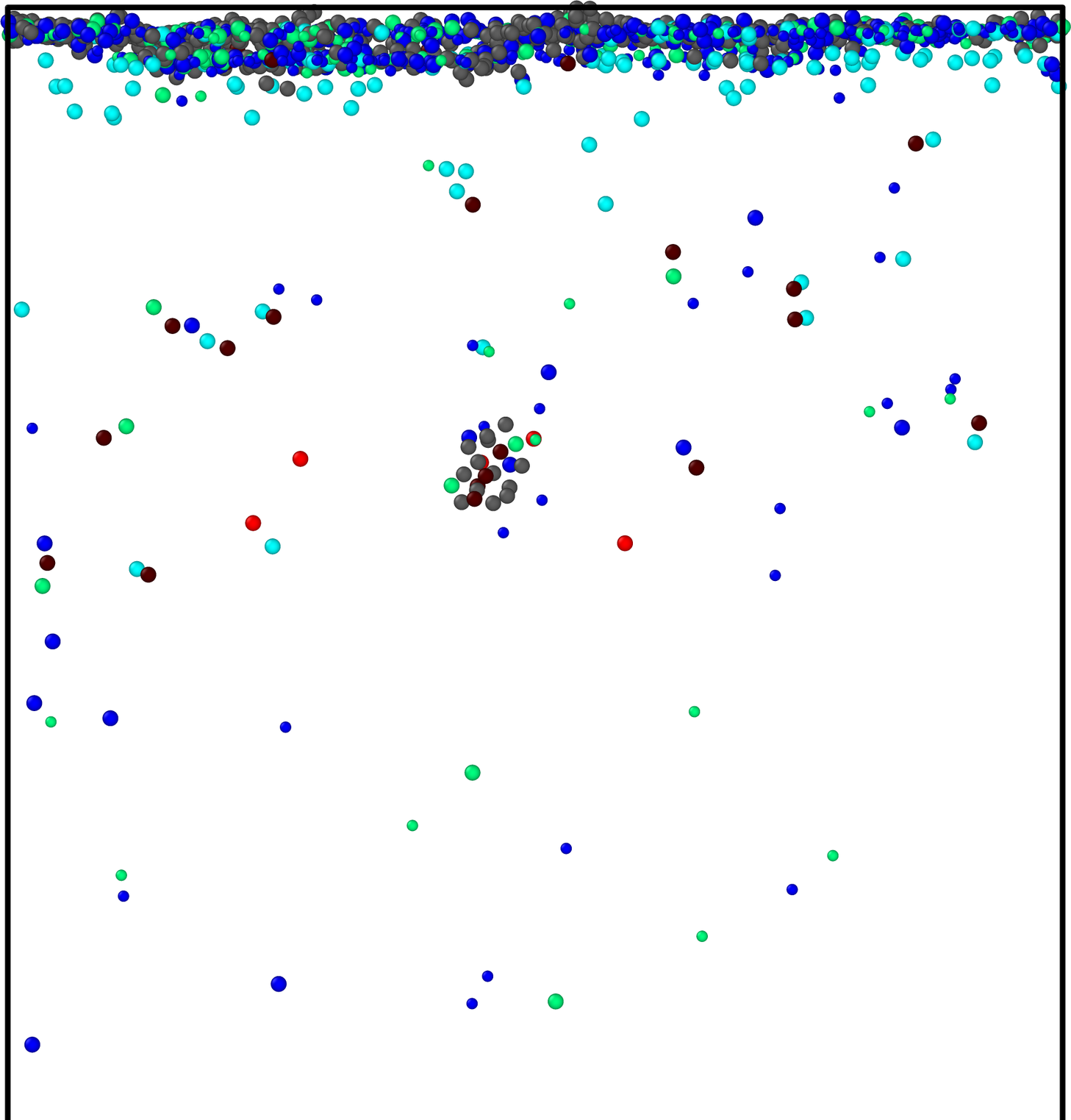}} \hfill
  \subfloat[80 ns]{\includegraphics[width=0.45\textwidth]{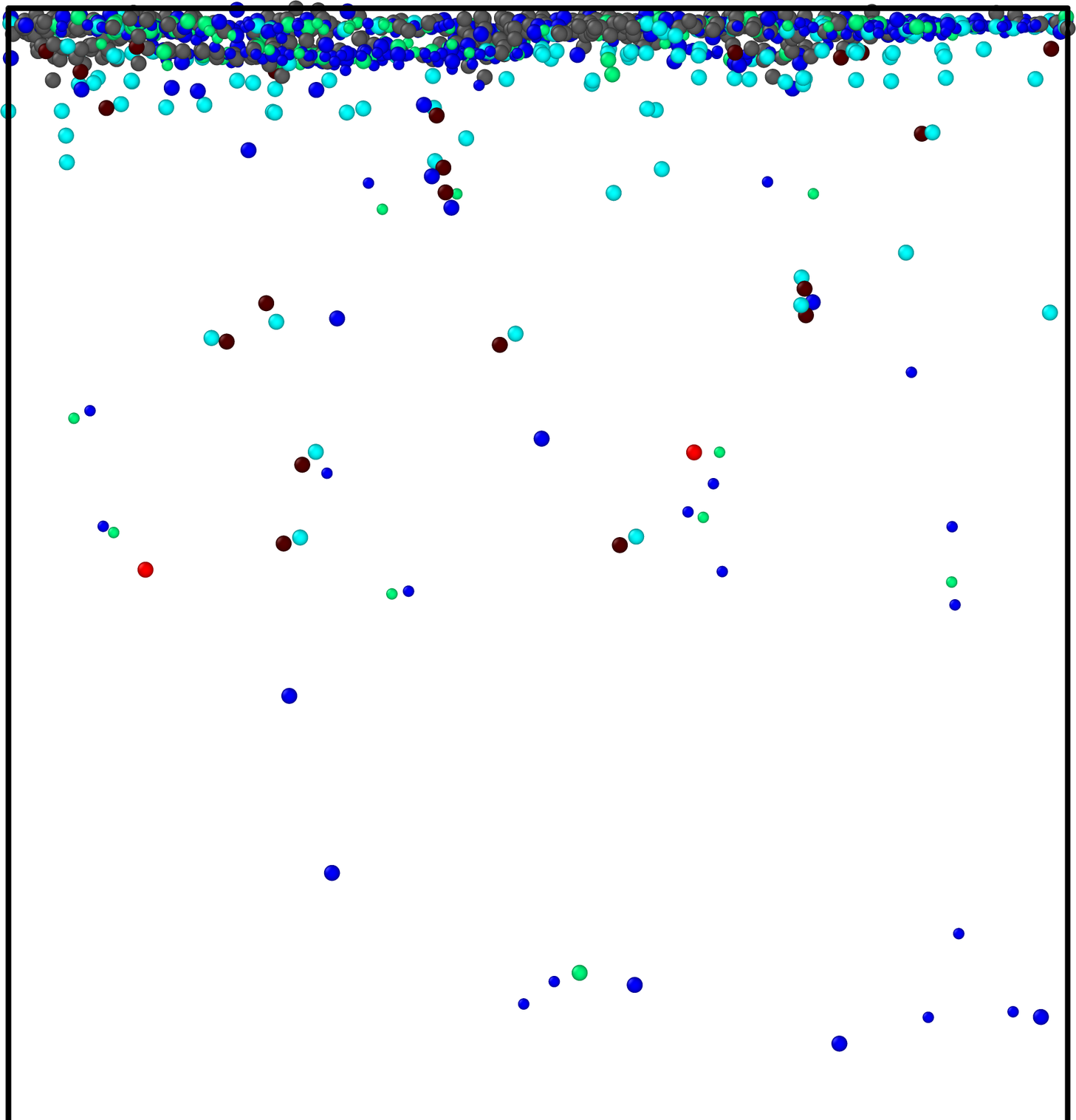}}
  \caption{Snapshots of the defect structure during Ostwald ripening at an Al dose of $5\times10^{13}$\,cm$^{-2}$ and anneal temperature of $2350$\,K (a) directly after the heat-up, (b) after 1\,ns, (c) after 30\,ns, and (d) after 80\,ns of annealing. Colors denote C defects (blue/cyan), Si defects (green), activated Al (red), other Al-related defects (brown), amorphous pockets (gray), and vacancies (small spheres).}
  \label{fig:TED_snapshots}
\end{figure}

\begin{figure}[htbp!]
\centering
  \subfloat[]{\includegraphics[width=0.49\textwidth]{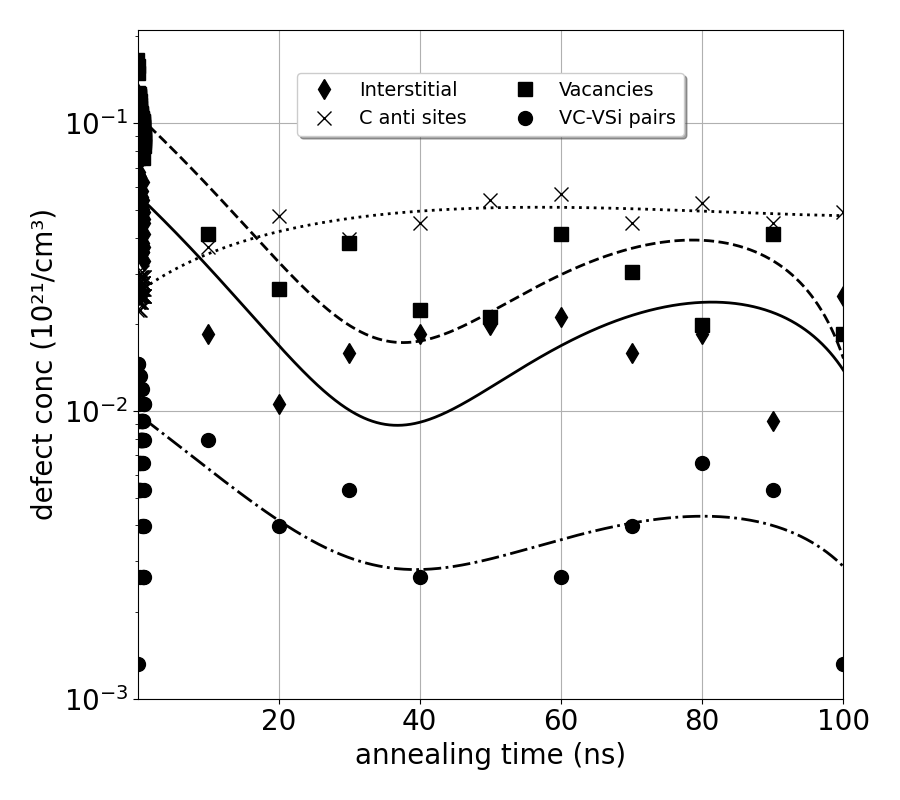}} \hfill
  \subfloat[]{\includegraphics[width=0.49\textwidth]{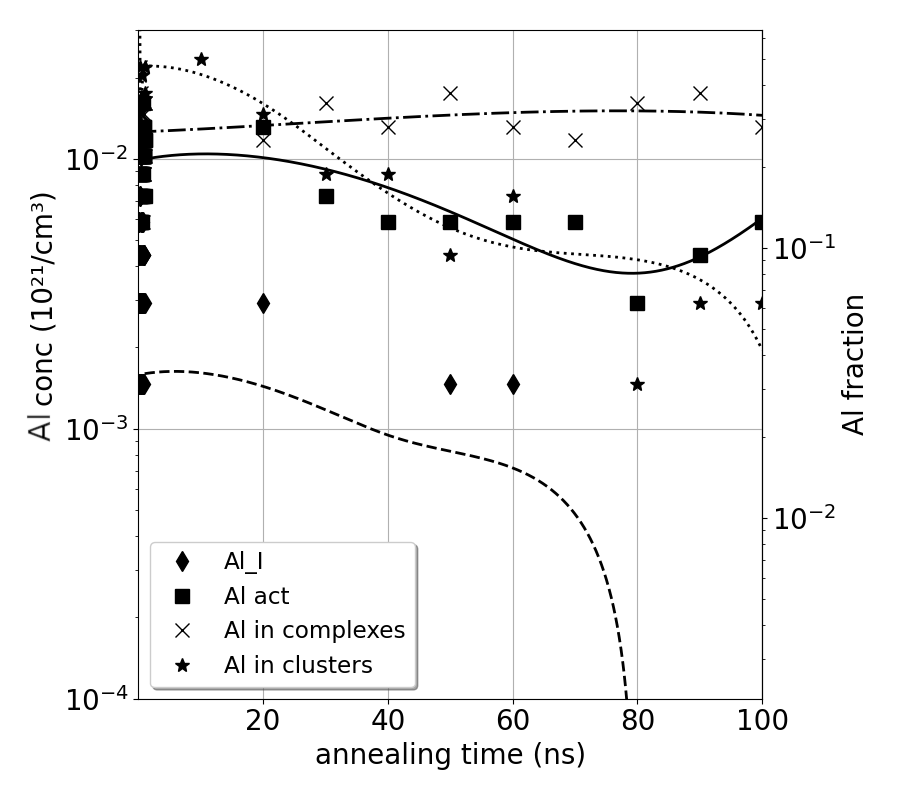}} \\
  \subfloat[]{\includegraphics[width=0.48\textwidth]{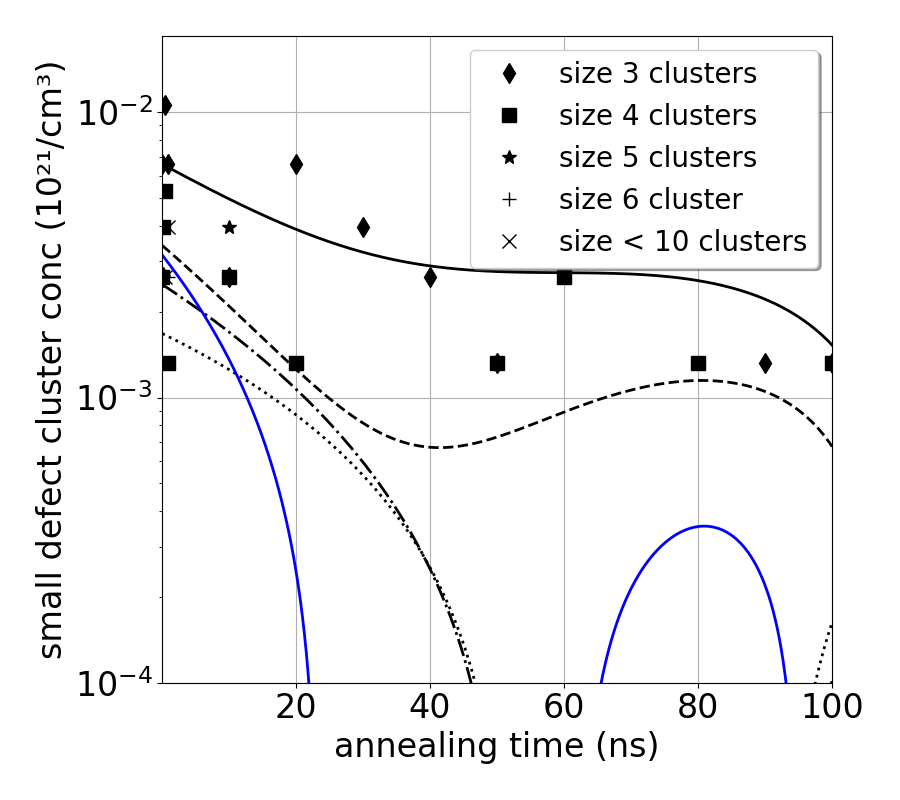}} \hfill
  \subfloat[]{\includegraphics[width=0.48\textwidth]{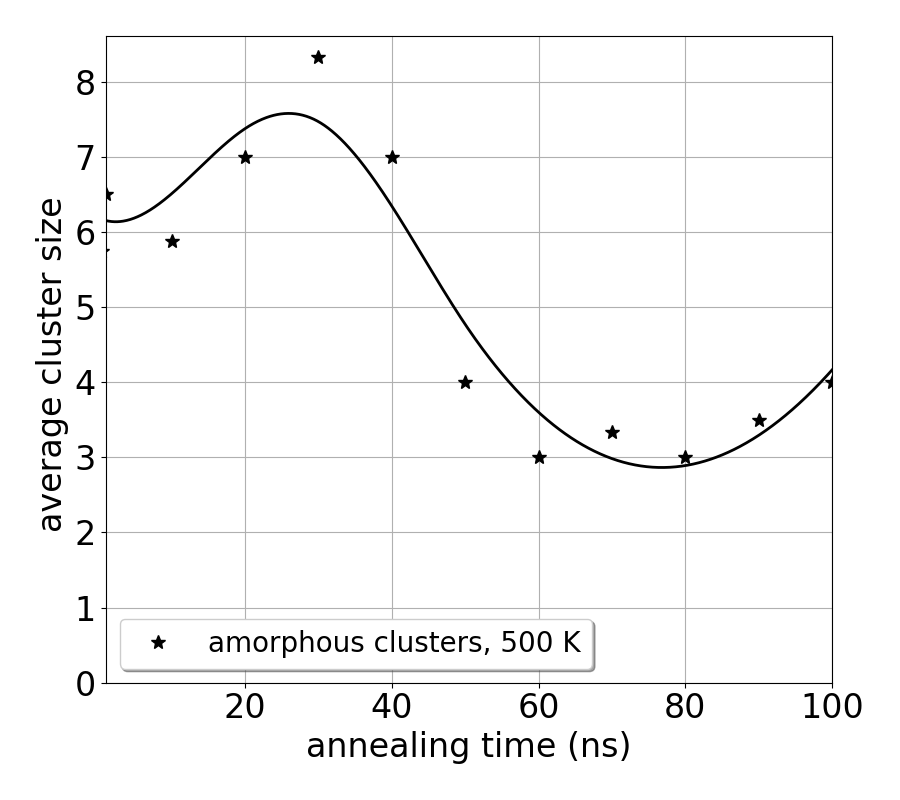}}
  \caption{Time evolution of point defects and clusters during annealing at $2350$\,K for a dose of $5\times10^{13}$\,cm$^{-2}$. (a) Evolution of intrinsic point defects (interstitials solid, vacancies dashed, carbon anti-sites dotted) including the V$_\mathrm{C}$V$_\mathrm{Si}$ di-vacancy (dash-dotted), (b) free (Al interstitials dashed, activated Al solid) and bound Al concentrations (complexes dash-dotted, clusters dotted), (c) stability of clusters as a function of cluster size (3 solid, 4 dashed, 5 dotted, 6 dash-dotted, 7-9 solid blue), and (d) average cluster size. Lines are included as visual guides.}
  \label{fig:TED_plots}
\end{figure}

Before turning our attention to concentrations above the solubility-related threshold, we study the process of Ostwald ripening at low Al concentrations. We use the low-dose system as a reference to identify similarities in recovery stages and to interpret the qualitatively different cluster evolution observed at high doses, and its dependence on implantation temperature. 
For systems far below the solubility limit (corresponding in the present setup to Al doses up to about \mbox{$1\times10^{14}$\,cm$^{-2}$}) complete dissolution of intrinsic defect clusters of size $>5$ can be observed for annealing temperatures above $2150$\,K irrespective of implantation temperature. 
%AH: You should also mention the implantation temperature here 
%DONE
As an example, we follow the full Ostwald ripening process for a dose of \mbox{$5\times10^{13}$\,cm$^{-2}$} in Figure~\ref{fig:TED_snapshots} and Figure~\ref{fig:TED_plots} to identify typical recovery regimes driven by recombination reactions and diffusion during cluster growth and subsequent dissolution and its impact on Al activation.

%%AH: The y-label in Fig2b is probably not good - I assume it should rather be 
%%      Al concentration
%%     I do not consider A_act as a defect
%% DONE
%    In Fig3c the cluster size 6 is problematic because it increases dramatically after 90ns
%%     Unfortunately we do not know what happens to those later.
%%    Fig2d indicates that some cluster size > 4 seems to be most stable in the end.
%%      We don't know which.
%% For this special set-up clusters of size 3 and a few of size 5 survive the 100 ns of annealing. So the average size becomes 4 at the end. This is however not a statistical valid statement/trend because at slightly higher or lower temperature there where no size 5 clusters found, but size 4 clusters instead for example. What can be seen in all the systems however, is the stability of size 3 clusters. Moreover, all the complexes remaining at the end of cluster dissolvation are Al-C complexes. We investigated some of them in the last paper. However, for a complete and clean structure analysis one has to make DFT calculations, since the Morse potential is not capable of predciting all the stable configurations especially for clusters bigger than size 3. We stated this also in the current publication in section 3.1 and in the introduction. That is why a deeper investigation of these Al-C clusters is avoided at this stage and the focus is on dislocation loop formation.

\paragraph{\textbf{Ostwald ripening}}

The snapshots in Figure~\ref{fig:TED_snapshots} illustrate the temporal course of Ostwald ripening during annealing immediately after the heating phase, where close Frenkel-pair recombination dominates; after $1$\,ns, when small clusters begin to form, driven by interstitial diffusion; after $30$\,ns, close to the point at which the average cluster size reaches its maximum; and after $80$\,ns, when the clusters have dissolved and only point defects and small (Al-C) related defect complexes remain.
In Figure~\ref{fig:TED_plots} the temporal evolution of intrinsic point defects including the V$_\mathrm{C}$V$_\mathrm{Si}$ di-vacancy in (a), the evolution of the free and clustered Al concentrations in (b), the stability of clusters with respect to cluster size in (c), as well as the resulting average cluster size in (d) is summarized. The concentrations of intrinsic interstitials (here, the sum of C and Si interstitials) and vacancies (the sum of C and Si vacancies) are strongly correlated and exhibit two recovery stages. 
A first quasi-steady minimum is reached after $\sim35$\,ns when the clusters have ripened reaching their maximum size. A second minimum is reached toward the end of the annealing, when most of the interstitials have been released from clusters and recombined with vacancies or diffused to the surface. Between these two minima the number of point defects is rising driven by the dissolution of interstitial clusters. 

\paragraph{\textbf{Residual defects}}
Aside from the V$_\mathrm{C}$V$_\mathrm{Si}$ divacancy, no other purely intrinsic dimers remain beyond $\sim20$\,ns of annealing.
Carbon antisites are generated primarily during implantation and the first $20$\,ns of annealing. 
In agreement with experimental and DFT studies, carbon-related point defects such as V$_\mathrm{C}$ and C$_\mathrm{Si}$ are more stable than silicon point defects. 
Similarly as found for intrinsic Si-C defect clusters studied by Ko \emph{et al.}~\cite{Ko2017}, the stability of small (Al-Si-C) clusters depends on size as shown in Figure~\ref{fig:TED_plots}(c). Clusters with more than ten defects dissolve faster than much smaller complexes consisting of
$2$-$5$ defects. The thermodynamically stable structures consist of pure C trimers and small (Al-C) complexes up to sizes of $5$ defects 
%are especially stable and do not completely dissolve within $100$\,ns, while larger clusters have
%already disappeared 
as shown in Figure~\ref{fig:TED_plots}(d).

\paragraph{\textbf{Al activation}}
Immediately after implantation, Al activation (perfect activation, i.e., all substitutional Al$_\mathrm{Si}$ as defined in Section~\ref{sec:Methods}) reaches approximately $15\,\%$, driven by vacancy recombination, and decreases slightly during Ostwald ripening over the first $\sim80$\,ns before it rises again toward the end of annealing, as shown in Figure~\ref{fig:TED_plots}(b) and inferred from (d). The detachment of Al interstitials bound to the surface of intrinsic clusters begins already after about $10$\,ns of annealing, long before the maximum cluster size is reached. The freed Al interstitials do not increase activation but instead contribute to the formation of Al$_\mathrm{Si}$C$_\mathrm{I}$ dimers by trapping Al$_\mathrm{I}$ at carbon antisites thereby initiating the kick-in process described above and shown in Figure~S1 of the supplementary material. The reduction in Al activation after around $30$\,ns of annealing is likewise caused by the formation of the same complex type, in which now diffusing C interstitials are captured by substitutional Al$_\mathrm{Si}$ involving a recombination barrier of $1.2$\,eV \cite{leroch2026_JMCC}. This process is further enhanced between $40$\,ns and $80$\,ns, when dissolving clusters release interstitials, reflected by the rise of point-defect concentrations in Figure~\ref{fig:TED_plots}(a) and the decrease of the average cluster size in Figure~\ref{fig:TED_plots}(d). Moreover, Al deactivation is driven by kick-out processes involving diffusing Si interstitials as shown in \cite{leroch2026_JMCC}. 
At the end of the Ostwald ripening cluster dissolution can nevertheless leave stable complexes behind as well as activated/free Al. 
After the clusters have been dissolved the concentration of Al$_\mathrm{Si}$C$_\mathrm{I}$ complexes effectively  decrease, thereby increasing Al activation. This is confirmed by a renewed increase in Al activation toward the end of the annealing cycle seen in Figure~\ref{fig:TED_plots}(b). 
A close inspection of the atomic trajectories reveals that some of the thermodynamically stable Al$_\mathrm{Si}$C$_\mathrm{I}$ complexes begin to dissolve or exchange C atoms among themselves. Consistently, the C diffusivity in Figure~S3(b) %\ref{fig:Diff_dose_4}(b) 
increases toward the end of annealing, indicating an elevated concentration of mobile C interstitials in the system. In contrast, the Al diffusivity in Figure~S3(a) 
%\ref{fig:Diff_dose_4}(a) 
decreases sharply after dissolution of the clusters because predominantly immobile Al remains. 
An Arrhenius analysis of the late-stage C diffusivity yields an apparent activation energy of approximately $3.2$\,eV. Subtracting the desorption barrier of $1.6$\,eV given in Table~S3 of the supplementary material results in an (Al-C) binding energy of $\sim1.6$\,eV for the neutral Al$_\mathrm{Si}$C$_\mathrm{I}$ complex. 
The value agrees well with the neutral DFT binding energy calculated for the Al$_\mathrm{Si}$C$_\mathrm{I}$ complex in \cite{leroch2026_JMCC}.
However, as discussed there \cite{leroch2026_JMCC}, the neutral binding energy is overestimated by up to $1$\,eV since it does not account for the destabilizing capture of electrons which takes place during dissolution.
Prolonged annealing is expected to further increase Al activation while reducing (Al-C) complex and carbon vacancy concentrations.
\subsubsection{Al supersaturation}
\label{subsec:Alsuper}
\begin{figure}[!htbp]
\centering
  % Column Headers
  \hfill
  \makebox[0\textwidth]{\textbf{500\,K}} \hfill
  \makebox[0.5\textwidth]{\textbf{900\,K}} \hfill \\
  % \vspace{-18pt}

  % Row 1: All size summary
%  \rotatebox{90}{\makebox[0.45\textwidth]{\textbf{All Sizes Summary}}} \hspace{0pt}%
%\begin{figure}[htbp!]
%\centering
  \subfloat[]{\includegraphics[width=0.49\textwidth]{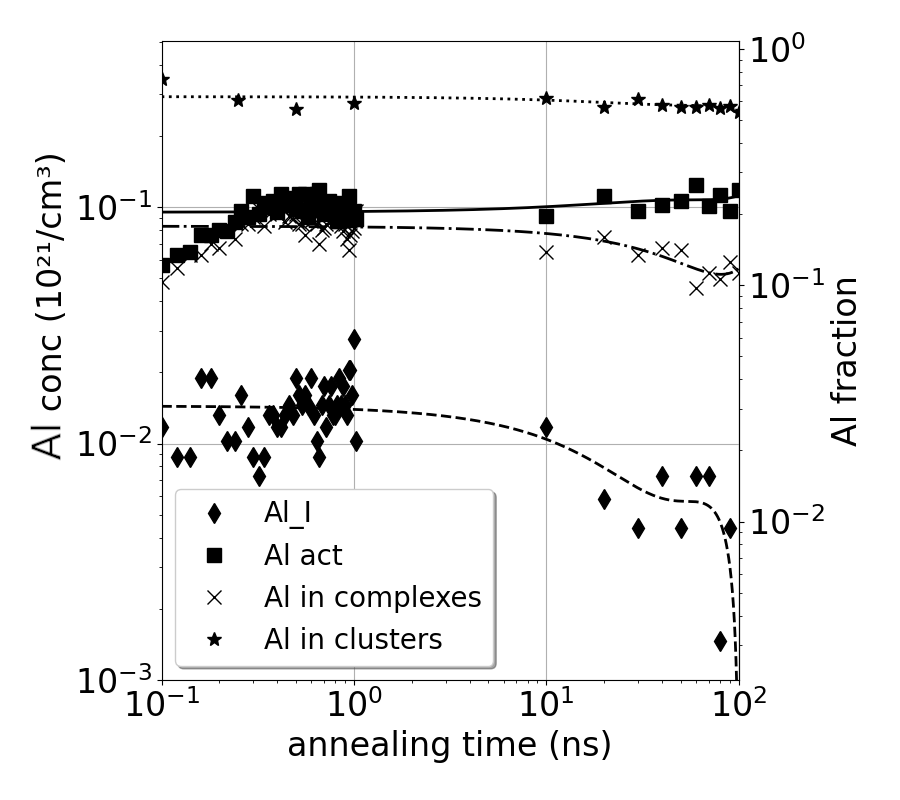}} \hfill
  \subfloat[]{\includegraphics[width=0.49\textwidth]{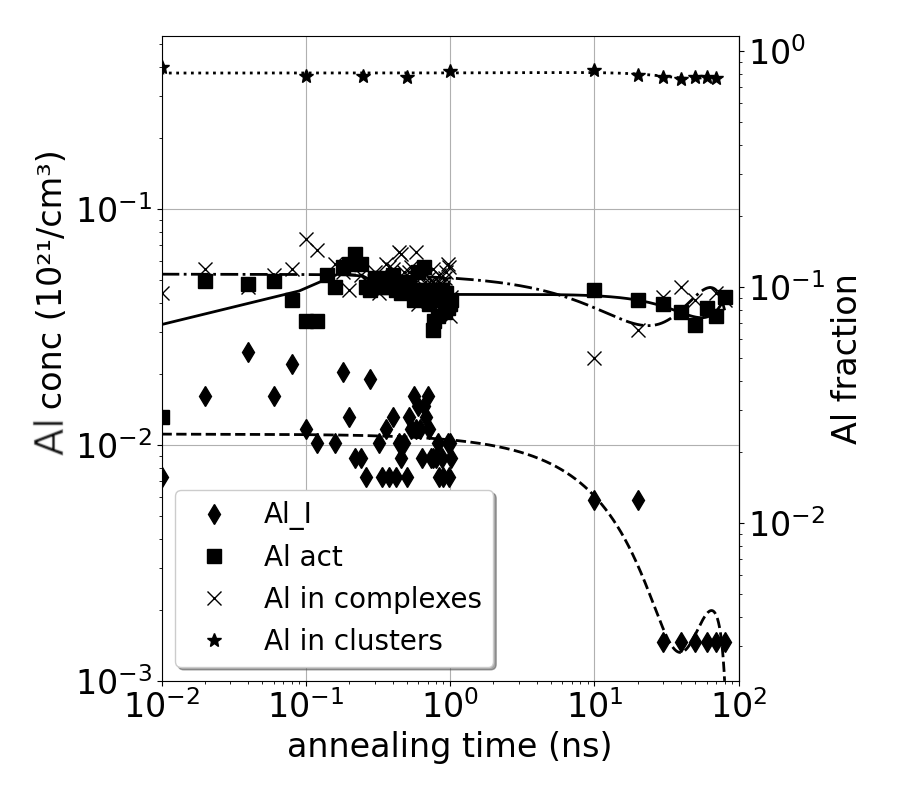}} \\
  \subfloat[]{\includegraphics[width=0.48\textwidth]{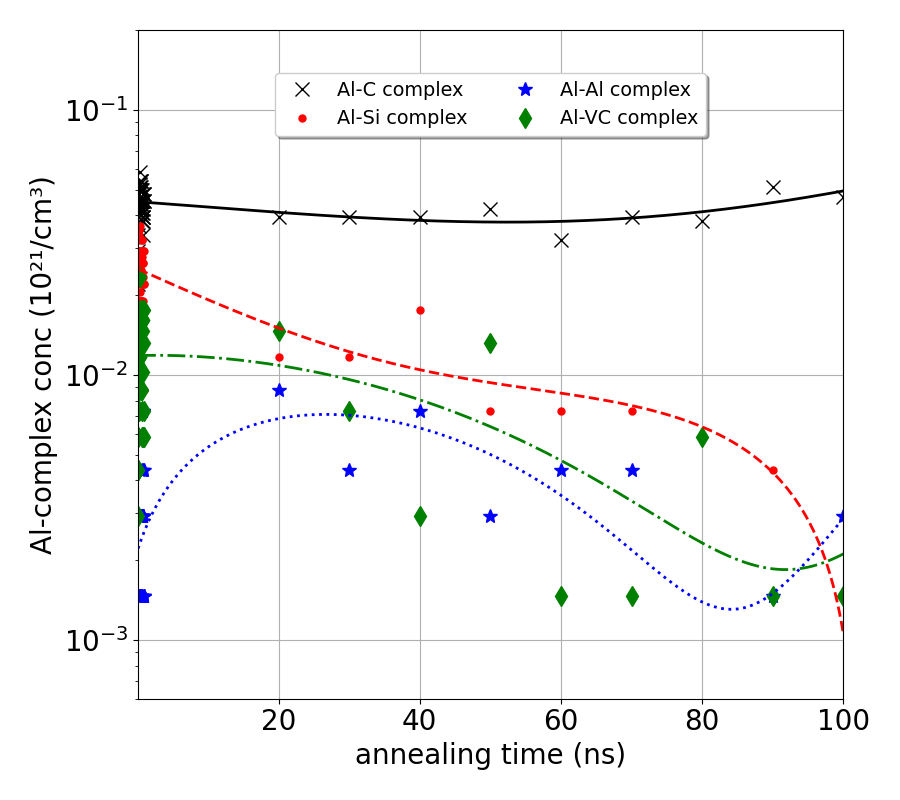}} \hfill
  \subfloat[]{\includegraphics[width=0.48\textwidth]{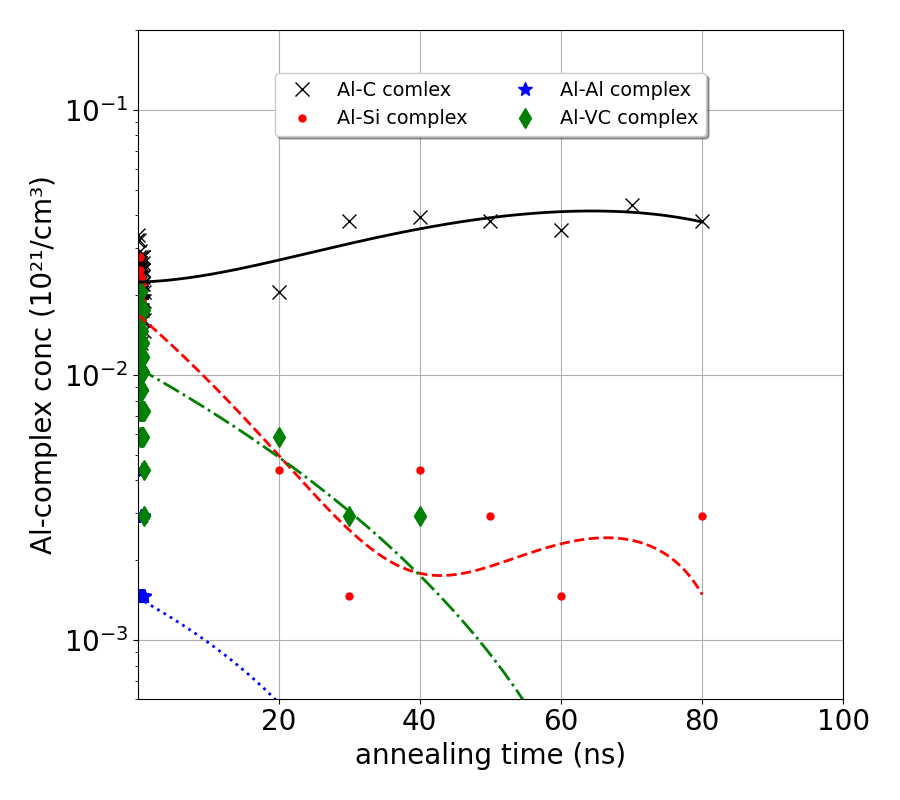}}
 \caption{Time evolution of Al-related defects during annealing at $2350$\,K for a dose of $5\times 10^{14}$\,cm$^{-2}$. (a, b) Concentration of Al point defects (dashed), activated Al (solid), and Al bound in clusters (dotted) or complexes (dashed-dotted). (c, d) Evolution of Al complexes: (Al-C: solid), (Al-V$_\mathrm{C}$: dashed-dotted), (Al-Si: dashed), and (Al-Al: dotted). The left column (a, c) corresponds to implantation at $500$\,K and the right column (b, d) to implantation at $900$\,K. Lines serve as visual guides only.}
  \label{fig:Al_time}
\end{figure}

\paragraph{\textbf{Al activation as a function of implantation temperature}}
Figure~\ref{fig:Al_time} shows the time evolution of Al defect concentrations for a dose of $5\times10^{14}$\,cm$^{-2}$ 
at implantation temperatures of $500$\,K (a,c) and $900$\,K (b,d),  
resolved into point defects (free Al), complexes (size $=2$) , and clusters (size $>2$). In both cases, the activation is low immediately after implantation and during the heating phase. The Al fraction that is bound in clusters — and is therefore not available for activation — accounts for the largest proportion and remains at $50$\% and $80$\%, respectively, at the end of the short annealing cycle for $500$ and $900$\,K. 

In the partially amorphized sample ($\sim40\,\%$ of disorder at lower implantation temperature), activation increases to about $20\,\%$ within the first nanosecond, while for the higher implantation temperature, it reaches only $\sim10\,\%$ over the same interval.
Thus, at high doses, Al activation is primarily driven by epitaxial recrystallization of 4H-SiC. 
In this regime, dopant incorporation is favored if the recrystallization front (activation energy $\sim1.4$\,eV~\cite{leroch2024}) advances at a similar rate as diffusion of free Al interstitials (migration barrier of $1.1$\,eV). Notably, the concentration of chemically activated Al correlates negatively with the amount of Al bound in complexes, where Al complexes comprise Al bound to intrinsic interstitials, to the carbon vacancy, or to Al itself.
%%AH: Same comment as above Fig2, I do not consider Al_I an Al defect
%%AH: Can you somehow put some sort of title 
%%      above (a)(c)  500K (Implant tempterature)
%%      above (b)(d)  900K (Implant tempterature)
%%  Like in Fig5
%DONE
The further increase in activation within $\sim60$\,ns after recrystallization at low implantation temperatures of $500$\,K is largely attributable to the decay of Al$_\mathrm{Si}$C$_\mathrm{I}$ and  Al$_\mathrm{Si}$V$_\mathrm{C}$ complexes and to the dissolution of compact clusters releasing Al$_\mathrm{I}$. 
For high implantation temperatures of $900$\,K, 
the decrease in activation is a direct consequence of  Al$_\mathrm{Si}$C$_\mathrm{I}$ complex formation in the first $\sim60$\,ns of annealing, resembling the behavior observed for the low-dose system discussed previously.
In addition, Al released from the smaller compact clusters is rapidly incorporated into larger clusters, so that essentially no free Al remains for further activation, as shown in Figure~\ref{fig:Al_time}(b,d) between $30$\,ns and $60$\,ns.
%The behavior depending on the implantation temperature could also be observed for the system with a dose of 10.
\paragraph{\textbf{Evolution of complex and cluster concentrations}}
Owing to the high displacement energy and migration barrier of the Si interstitial, Al$_\mathrm{Si}$Si$_\mathrm{I}$ complexes are less frequent than Al$_\mathrm{Si}$C$_\mathrm{I}$ complexes directly after implantation in Figure~\ref{fig:Al_time}(c,d).
Moreover, the neutral Al$_\mathrm{Si}$Si$_\mathrm{I}$ complex is kinetically less stable than the neutral Al$_\mathrm{Si}$C$_\mathrm{I}$ complex. In the GW-Morse potential description the complex is a trapping state that preferentially decays into the free Al interstitial~\cite{leroch2026_JMCC}. The potential thus correctly reflects the behavior of the system above mid-gap. 
The reformation of the complex also proceeds more slowly, owing to the higher migration barrier and the lower concentrations of Si interstitials in comparison to C.

\begin{figure}[htbp!]
\centering
  \subfloat[]{\includegraphics[width=0.32\textwidth]{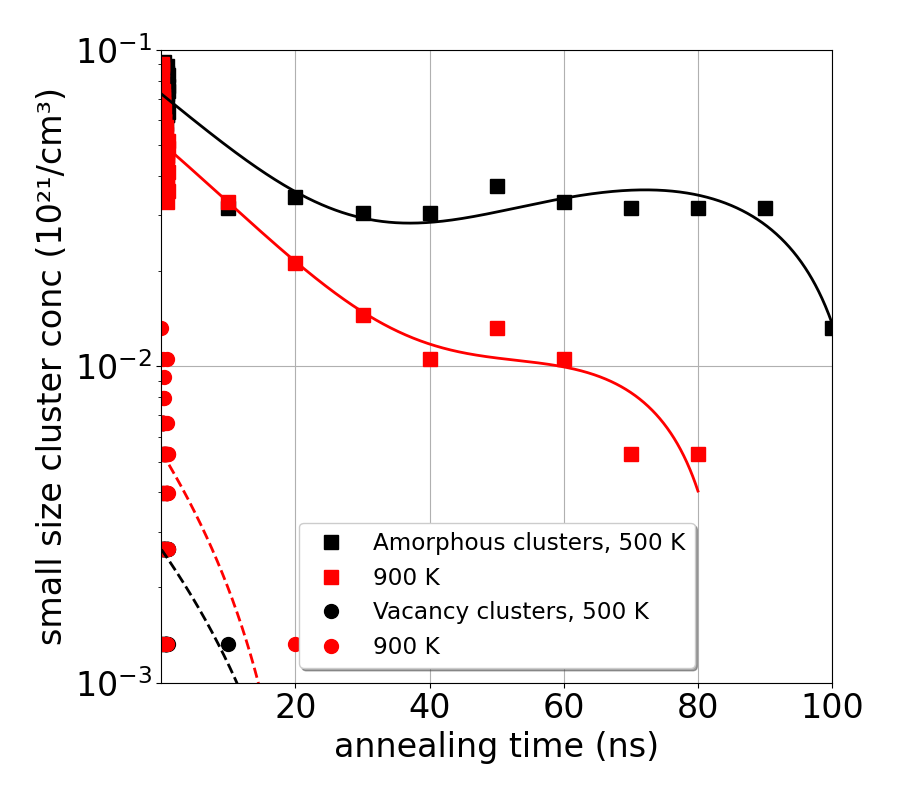}} \hfill
  \subfloat[]{\includegraphics[width=0.32\textwidth]{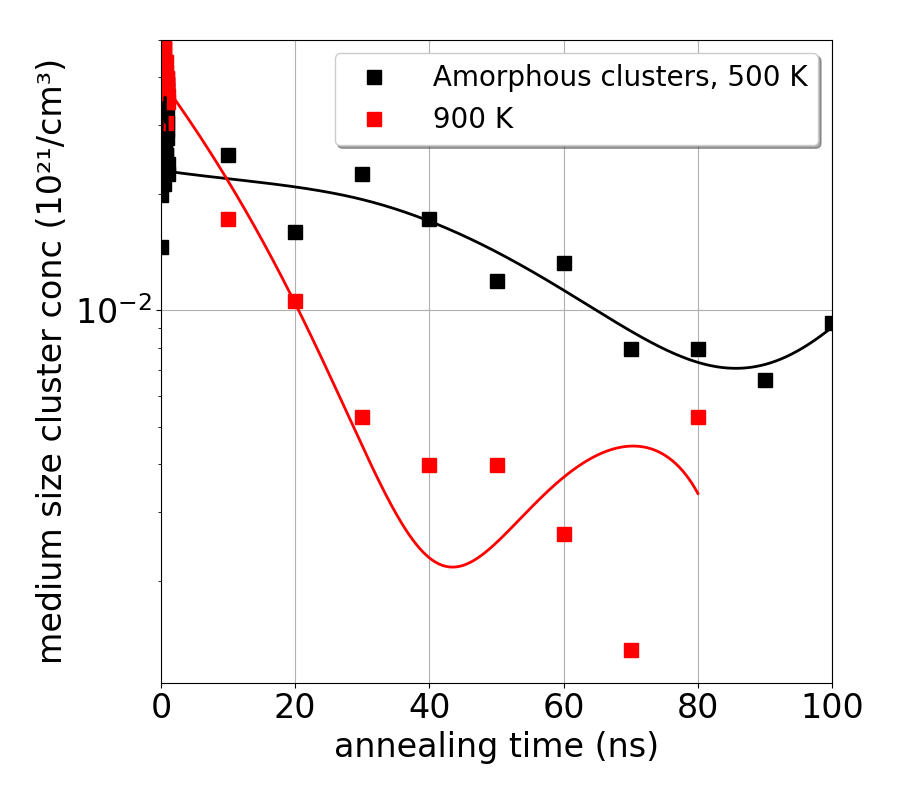}} \hfill
  \subfloat[]{\includegraphics[width=0.32\textwidth]{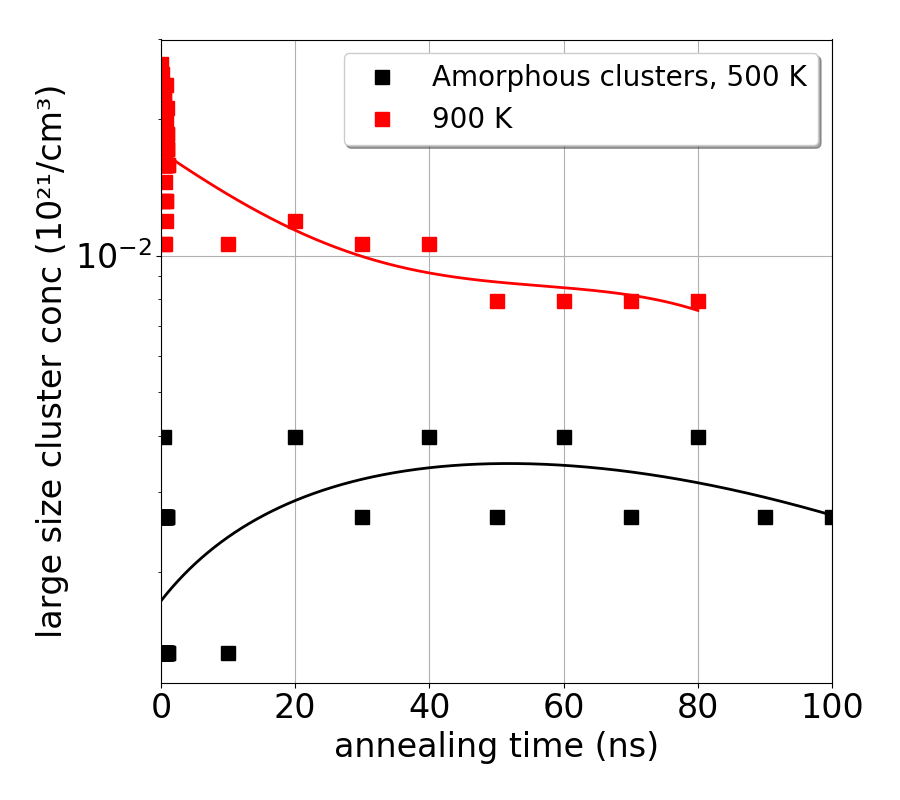}}
  \caption{Time evolution of cluster concentrations for a dose of $5\times 10^{14}$\,cm$^{-2}$ during annealing at $2350$\,K. Panels show (a) small clusters ($3\leq s<10$), (b) medium-sized clusters ($10\leq s<100$), and (c) large clusters ($s\geq100$). Black curves represent $500$\,K implantation and red curves represent $900$\,K. Lines are included as visual guides.}
  \label{fig:defect_cluster_conc}
\end{figure}

With the exception of the Al$_\mathrm{Si}$C$_\mathrm{I}$ complex, the concentrations of the Si-, V$_\mathrm{C}$-, and Al-related complexes are temperature dependent.
Already at the beginning of the annealing, the concentration of Al complexes at $500$\,K is higher than at $900$\,K.  Moreover, the (Al-Al) and the Al$_\mathrm{Si}$V$_C$ complexes are observed only for low implantation temperatures of $500$\,K till the end of the annealing.
In addition, the concentration of Si-, V$_C$-, and Al-related complexes drops much more slowly at $500$\,K than at $900$\,K.
The same statement holds also for the concentration of small and medium-sized compact clusters, as shown in Figure~\ref{fig:defect_cluster_conc}(a,b), respectively, where based on the distributions in Figure~\ref{fig:defect_cluster_sizes}(e,f), clusters are categorized into three size regimes ($s$): small clusters ($3 \leq s < 10$), medium-sized clusters ($10 \leq s < 100$), and large clusters ($s \geq 100$).
At higher annealing temperatures above $2200$\,K  at Al supersaturation, medium-sized clusters become less common and can even disappear entirely, as can also be seen in Figures~S4 and S5 of the supplementary material.

%%AH: This should rather be Fig6 (a)(b)
% Your are right but, in Fig 6 I have two different doses, while in Fig 5 I have only 5x10¹⁴/cm², so it is easier to understand like that.
\begin{figure}[!htbp]
\centering
  % Column Headers
  \hfill
  \makebox[0.35\textwidth]{\textbf{500\,K}} \hfill
  \makebox[0.4\textwidth]{\textbf{900\,K}} \hfill \\
  % \vspace{-18pt}

  % Row 1: All size summary
  \rotatebox{90}{\makebox[0.45\textwidth]{\textbf{All Sizes Summary}}} \hspace{0pt}%
  \subfloat[]{\includegraphics[width=0.48\textwidth]{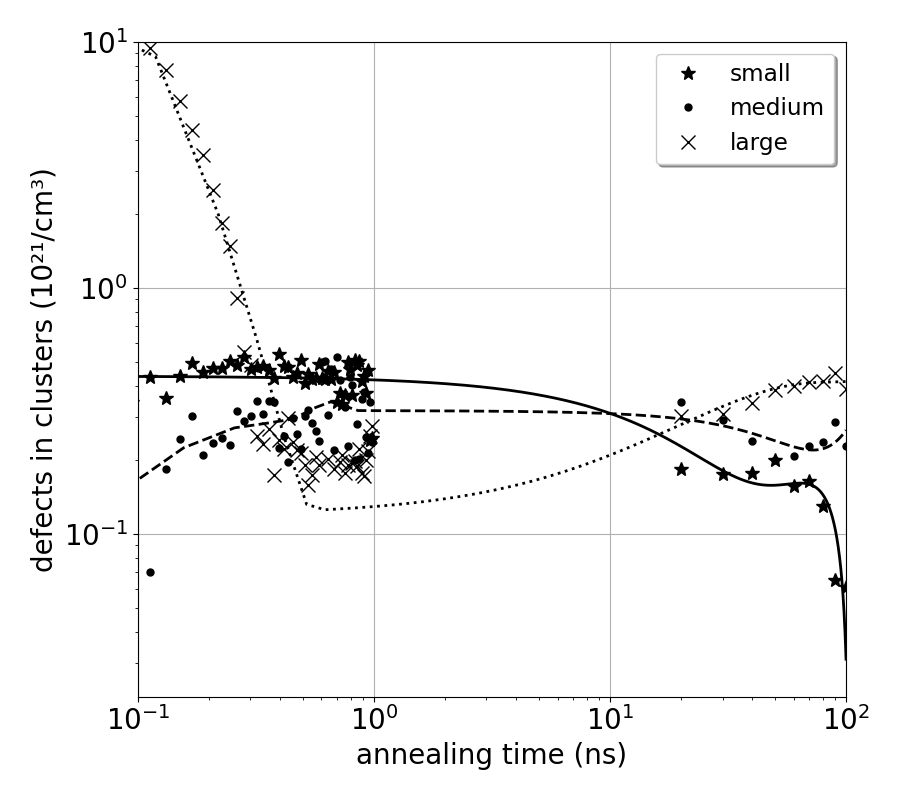}} \hfill
  \subfloat[]{\includegraphics[width=0.48\textwidth]{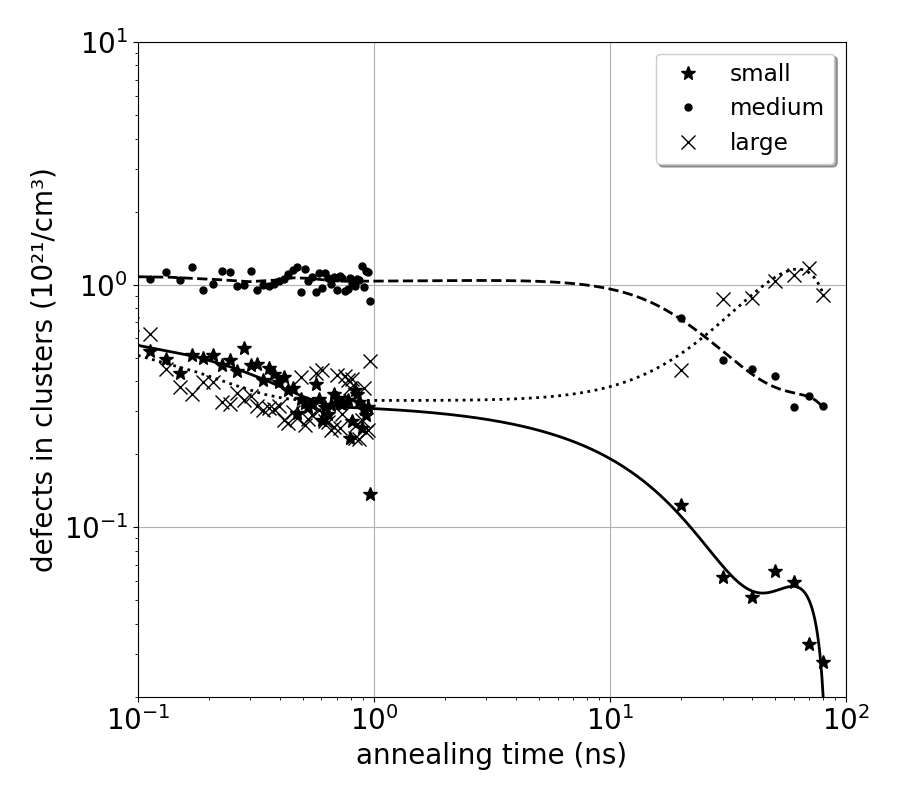}} \\
% \vspace{-18pt}
% Row 2: specific small sizes
  \rotatebox{90}{\makebox[0.45\textwidth]{\textbf{Selected Small Sizes}}} \hspace{0pt}%
  \subfloat[]{\includegraphics[width=0.48\textwidth]{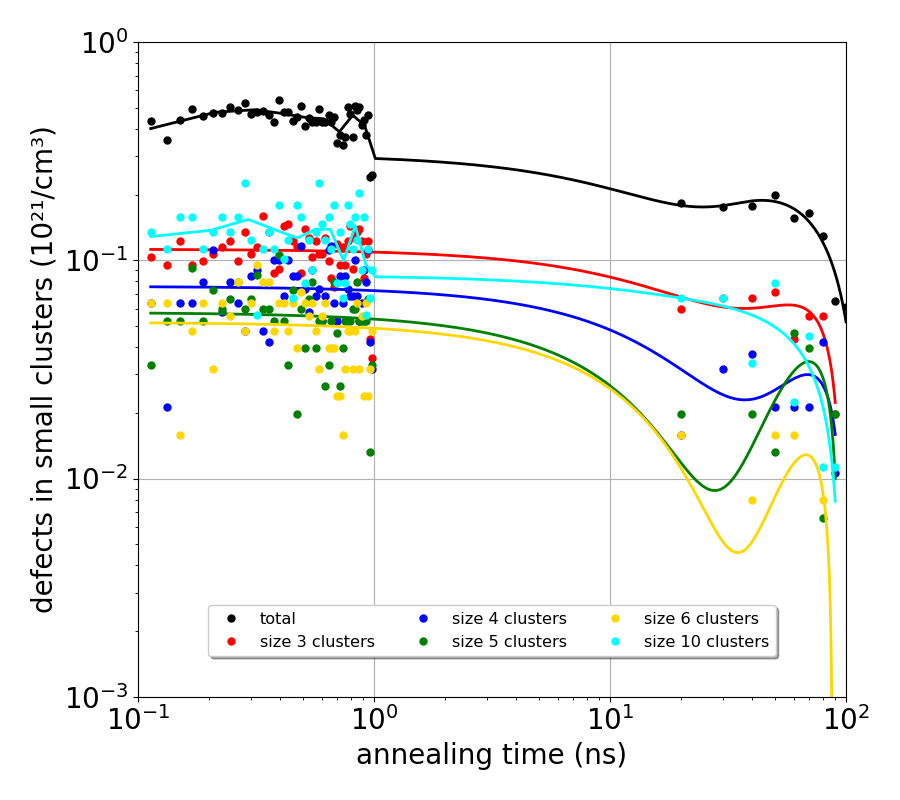}} \hfill
  \subfloat[]{\includegraphics[width=0.48\textwidth]{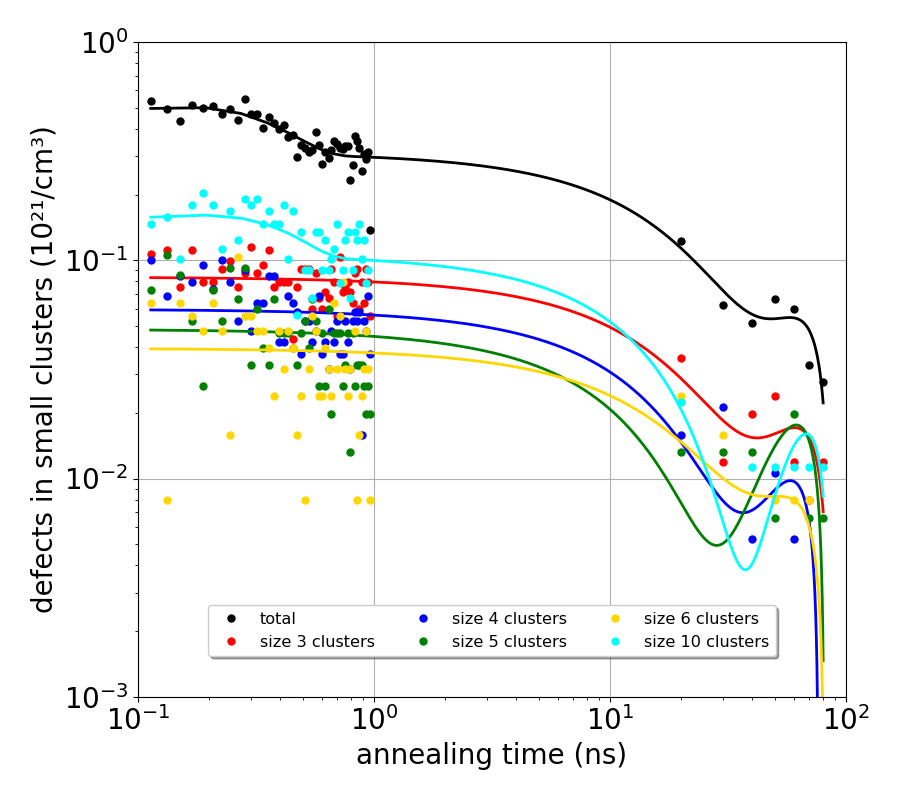}} \\
  \caption{Cluster-size evolution and distributions for $500$\,K (left column) and $900$\,K (right column) implantation at a dose of $5\times 10^{14}$\,cm$^{-2}$. (a, b) Cumulative concentration of small (solid), medium (dashed), and large clusters (dotted); (c, d) detailed evolution of selected stable sizes ($s=3,4,5,6$ and $s<10$). Lines are visual guides.}
  \label{fig:defect_cluster_sizes}
\end{figure}

\begin{figure}[!htbp]
\ContinuedFloat
\centering
  % Column Headers
  \hfill
  \makebox[0.35\textwidth]{\textbf{500\,K}} \hfill
  \makebox[0.4\textwidth]{\textbf{900\,K}} \hfill \\

  % Row 3: size Histograms
  \rotatebox{90}{\makebox[0.45\textwidth]{\textbf{Size Histograms}}} \hspace{0pt}%
  \subfloat[]{\includegraphics[width=0.48\textwidth]{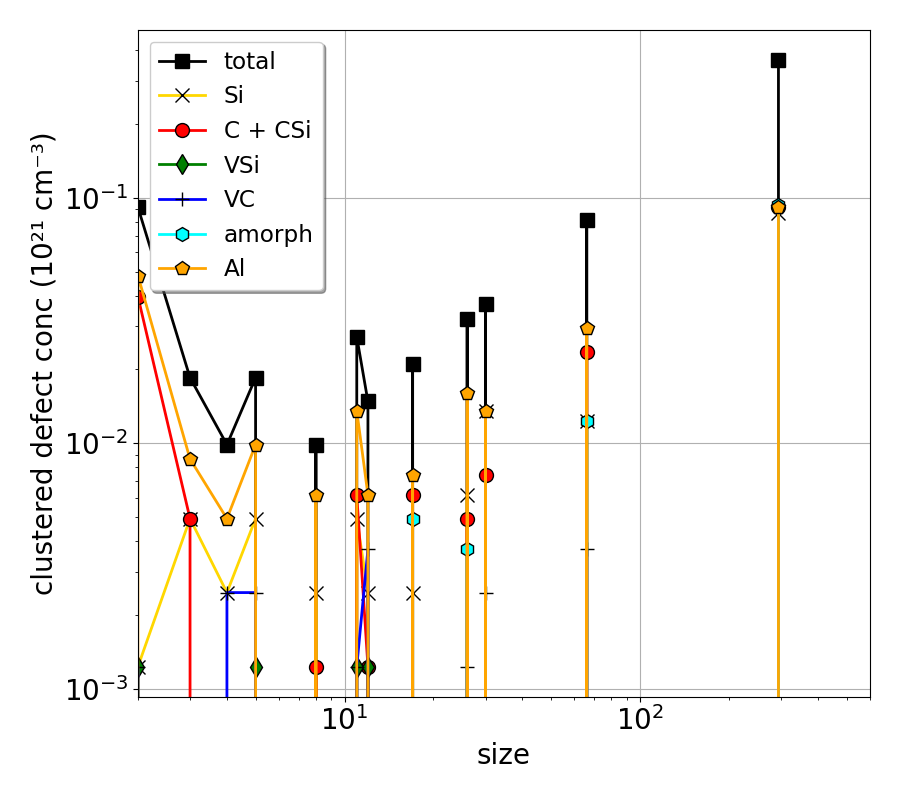}} \hfill
  \subfloat[]{\includegraphics[width=0.48\textwidth]{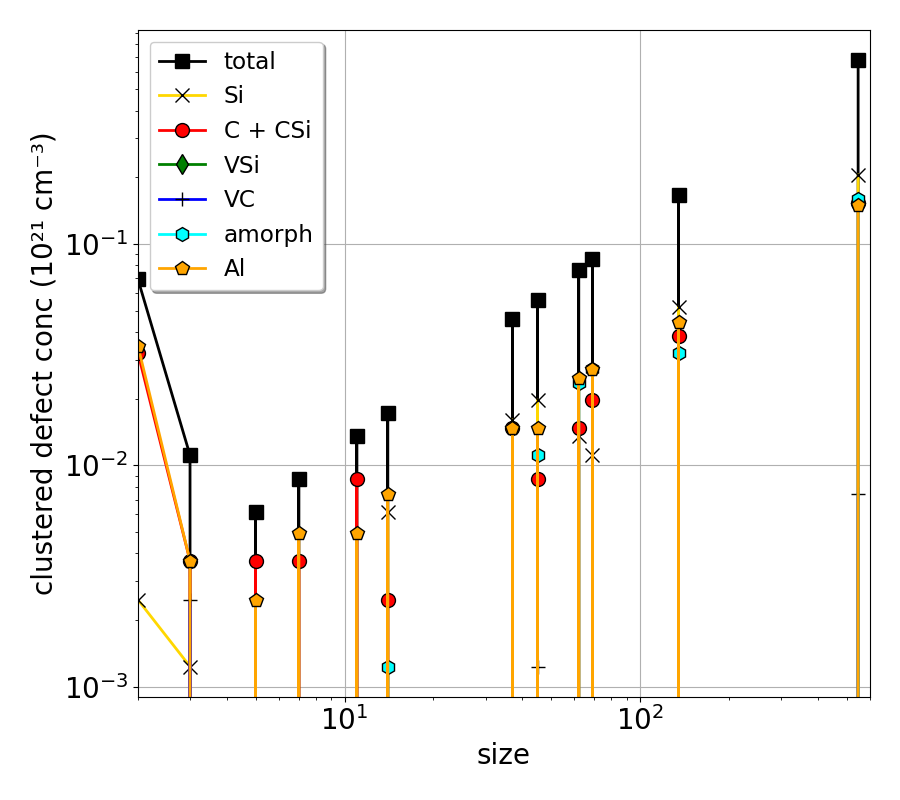}}

  \caption[]{Cluster-size evolution and distributions (continued).  (e, f) Defect distributions per cluster size and their chemical compositions, highlighting stable ripening peaks after $100$\,ns of annealing at $2350$\,K, for implantation at $500$\,K (e) and $900$\,K (f).}
\end{figure}
\paragraph{\textbf{Evolution of cluster size}}
At the elevated implantation temperature of $900$\,K, not only the total number of clusters but also the size of the compact small and medium-sized clusters decreases more strongly than for $500$\,K. This trend is evident in Figure~\ref{fig:defect_cluster_sizes}(a,b), which shows the defect concentration bound in clusters. 
In general, the ripening process, which at $900$\,K initiates from a system dominated by medium-sized clusters, proceeds much faster than at $500$\,K. At the lower temperature, the process begins from a partially amorphous phase, from which primarily compact small clusters emerge after $1$\,ns of annealing.
At $900$\,K, mobile interstitials separate more efficiently from their corresponding vacancies, producing widely separated Frenkel pairs and an enhanced accumulation of interstitials in clusters. 
Because diffusion in 4H-SiC is anisotropic, interstitials migrate preferentially within the basal plane rather than along the $c$-axis, favoring planar clustering. During annealing, these clusters act as seeds for extended planar defects.
Medium-sized clusters show no preferred cluster sizes and are kinetically less stable than small clusters ($s < 10$), which cluster around certain stable sizes, as seen in the size distribution histograms of Figure~\ref{fig:defect_cluster_sizes}(e,f) and in Figure~S8 of the supplementary information.
Among the small clusters, those containing $7-9$ defects dominate initially, eventually shrinking to preferred sizes of $3-5$ defects by the end of annealing. The high stability of trimers and $\sim$10-mer clusters is primarily responsible for the concentration plateau observed for small clusters in Figure~\ref{fig:defect_cluster_sizes}(c,d) between $30$\,ns and $70$\,ns, before their concentrations decrease further. 
A similar behavior in the cluster size distribution has been reported by Cowern \emph{et al.}~\cite{Cowern1999} for interstitial clusters grown during Ostwald ripening in boron-implanted silicon, where the binding energies of small silicon clusters show pronounced maxima at specific sizes, indicating high thermodynamic stability. 
Toward the end of the annealing process, even the large clusters start to shrink slightly. This indicates that the clusters are already fully ripened and will not grow any further.

\begin{figure}[htbp!]
\centering
%    \subfloat[]{\includegraphics[width=0.32\textwidth]{figures/average_cluster_sizes_dose_4_2350K_annealing_time_500.png}}
   \subfloat[]{\includegraphics[width=0.49\textwidth]{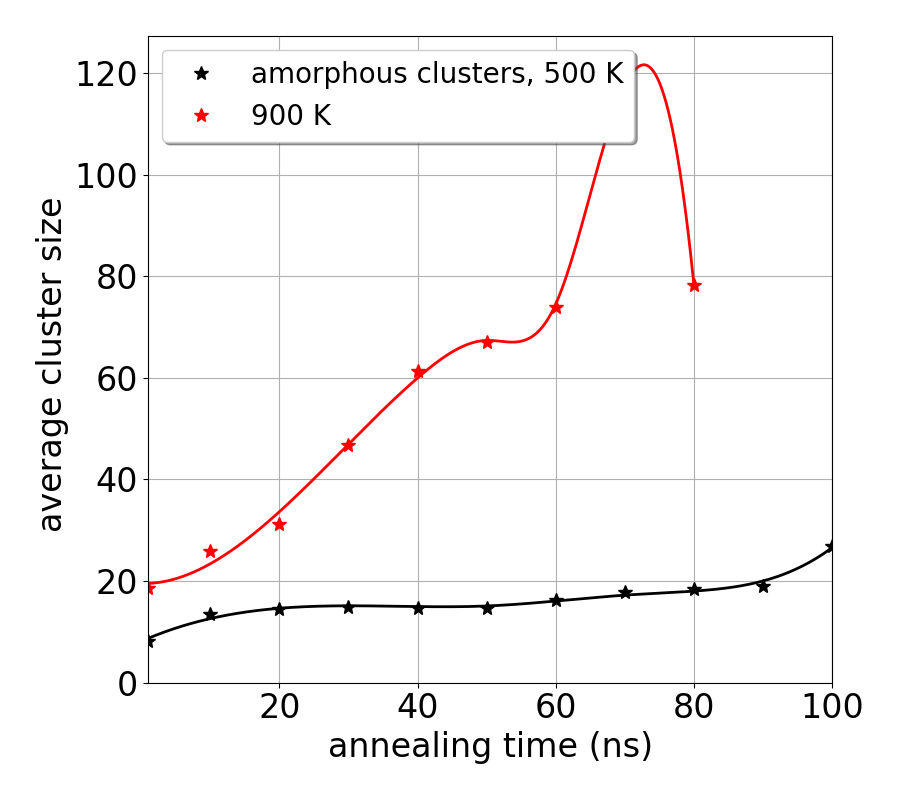}}
%  {\includegraphics[width=0.325\textwidth]{figures/cluster_size_300K_dose_4.png}
  \subfloat[]{\includegraphics[width=0.49\textwidth]{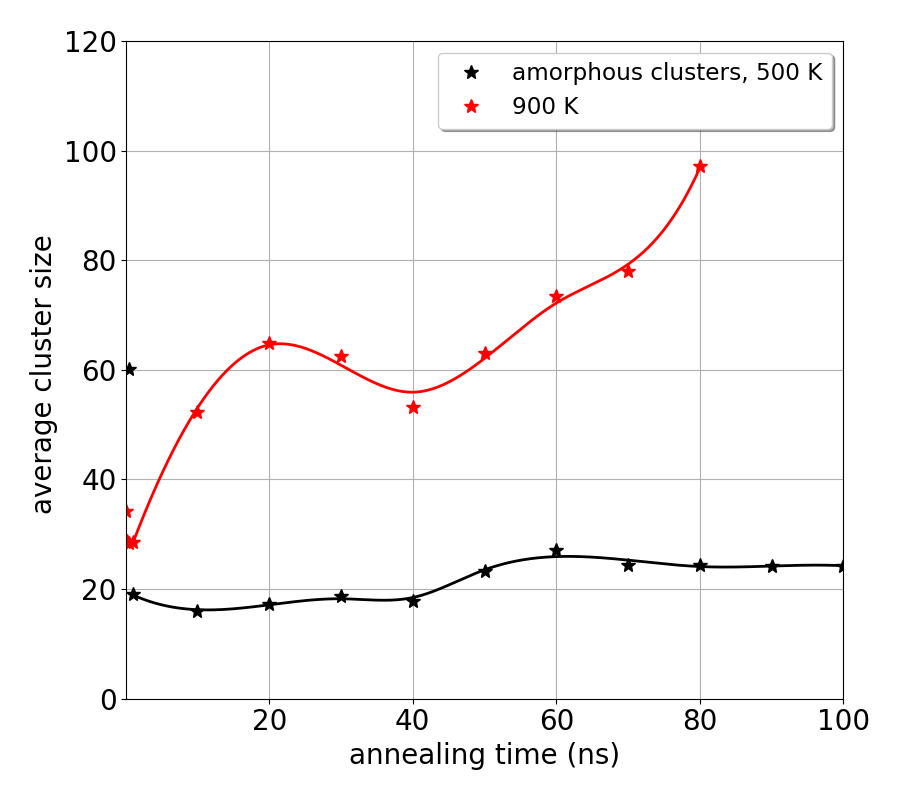}}
   \caption{Average cluster size (of all clusters $>2$) for annealing temperature of $2350$\,K over time for doses a) \mbox{$5\times 10^{14}$\,cm$^{-2}$} and b) \mbox{$7.5\times 10^{14}$\,cm$^{-2}$}. Lines provide visual guides only.}
  \label{fig:time_cluster_size}
 \end{figure}

In addition to trimers, which show the highest stability across all systems, defining other inherently stable sizes for Al-Si-C clusters is challenging. Furthermore, the accuracy of the Morse potential is insufficient for a systematic description of small clusters. We therefore limit ourselves to identifying general trends. Unlike pure Si clusters, the stoichiometry of these clusters is highly diverse and depends on both temperature and Al dose, as illustrated in Figure~\ref{fig:defect_cluster_sizes}(e,f). At low implantation temperatures of $500$\,K, for instance small clusters in the $4-10$ defect range generally show a larger proportion of Si and V$_\mathrm{C}$ defects. In contrast, at higher implantation temperatures of $900$\,K, pure (Al-C) clusters dominate by the end of the annealing process. Aluminum appears to contribute significantly to the stabilization of small clusters, where the Al content reaches at least $30$\,\%. In comparison, the Al proportion in medium and large clusters remains around $20$\,\%, as shown in Figure~S8 of the supplementary material.
The average cluster sizes for Al doses of $5\times$10$^{14}$\,cm$^{-2}$ and $7.5\times$10$^{14}$\,cm$^{-2}$ at different implantation temperatures are provided in Figure~\ref{fig:time_cluster_size}(a) and Figure~\ref{fig:time_cluster_size}(b), respectively. 
The evolution of the average cluster size at the highest dose behaves similarly to the system with a dose of \mbox{$5\times10^{14}$\,cm$^{-2}$} as a function of implantation temperature. However, at $900$\,K, a distinct plateau is observed between $10$\,ns and $40$\,ns. This feature coincides with the formation of an extrinsic stacking fault, as will be discussed in further detail in the next section. During the emergence of the additional layer, the precipitate surrounding the stacking fault maintains a constant size before further growth resumes after $40$\,ns.
In Figure~S4 and Figure~S5 of the supplementary material, we show the preferred cluster shape at certain cluster sizes.
Notably, kinetically stable clusters of size $3$ and $\sim 10$ are predominantly of planar shape (either in the basal plane or along the $c$-axis). Moreover, extended clusters which can exhibit several hundred defects are entirely located in the basal plane. In addition to large planar clusters, which are becoming increasingly energetically unstable with rising temperature and prone to transition into stacking faults, medium-sized clusters also play a role as seeds for dislocation formation as shown in Figure~S6 and discussed in detail in the supplementary information. The transition is thermally activated and can occur even in planar clusters of a size of ~$60$ interstitials . However, since this is a stochastic process, planar clusters can grow far beyond this size and remain stable even with hundreds of defects. 

 To summarize the findings of this section: At $500$\,K, Al activation is promoted by epitaxial recrystallization at the beginning of annealing. The concentration of chemically activated Al is inversely proportional to the concentration of small (Al-C) complexes. (Al-Si) and (Al-VC) complexes are kinetically less stable than (Al-C) complexes, where Al complexes with sizes of $2$–$10$ defects are thermodynamically much more stable than medium-sized compact clusters. Extended defects in the basal plane form preferentially at high implantation temperatures of $900$\,K, with medium-sized clusters playing an important role as seeds in the formation of extrinsic stacking faults.  

% \newpage
\subsection{General trends as a function of dose and implantation temperature}
To confirm that the behavior of the studied systems annealed at $2350$\,K follow a general trend as a function of implantation temperature, we show in Figure~\ref{fig:Al_conc_temp}  the evolution of free and bound Al concentrations at doses $5\times 10^{13}$\,cm$^{-2}$ and $7.5\times 10^{14}$\,cm$^{-2}$ at the end of annealing ($100$\,ns) for annealing temperatures ranging from $1000$ to $2500$\,K.
\subsubsection{Ostwald ripening and Al activation}
\begin{figure}[htbp!]
\centering
  % Column Headers
  \hfill
  \makebox[0.33\textwidth]{\textbf{$5\times 10^{13}$\,cm$^{-2}$, 500\,K}} \hfill
  \makebox[0.32\textwidth]{\textbf{$7.5\times 10^{14}$\,cm$^{-2}$, 500\,K}} \hfill
  \makebox[0.32\textwidth]{\textbf{$7.5\times 10^{14}$\,cm$^{-2}$, 900\,K}} \hfill \\
  \vspace{2pt}
  
  % Row 1: 1500K
  \rotatebox{90}{\makebox[0.25\textwidth]{\textbf{1500\,K}}} \hspace{-2pt}
  {\includegraphics[width=0.31\textwidth]{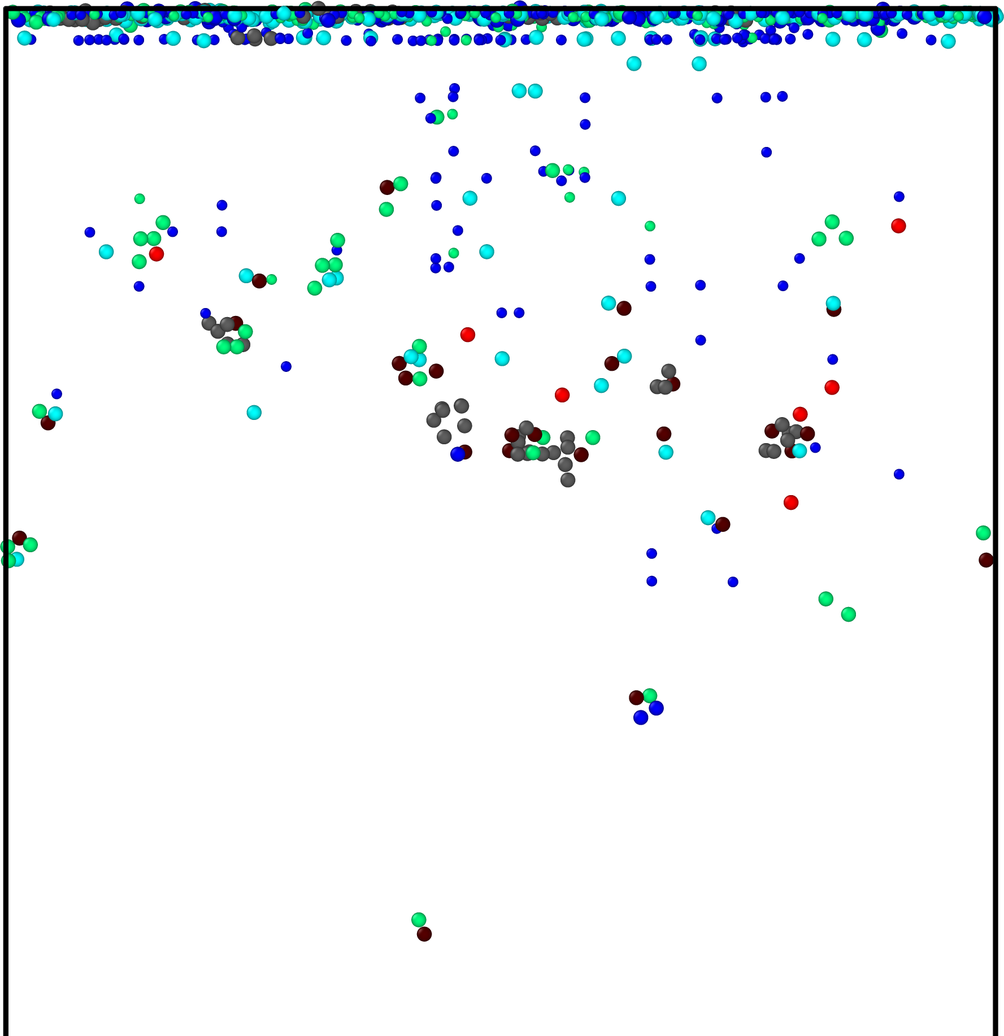}} \hfill
  {\includegraphics[width=0.31\textwidth]{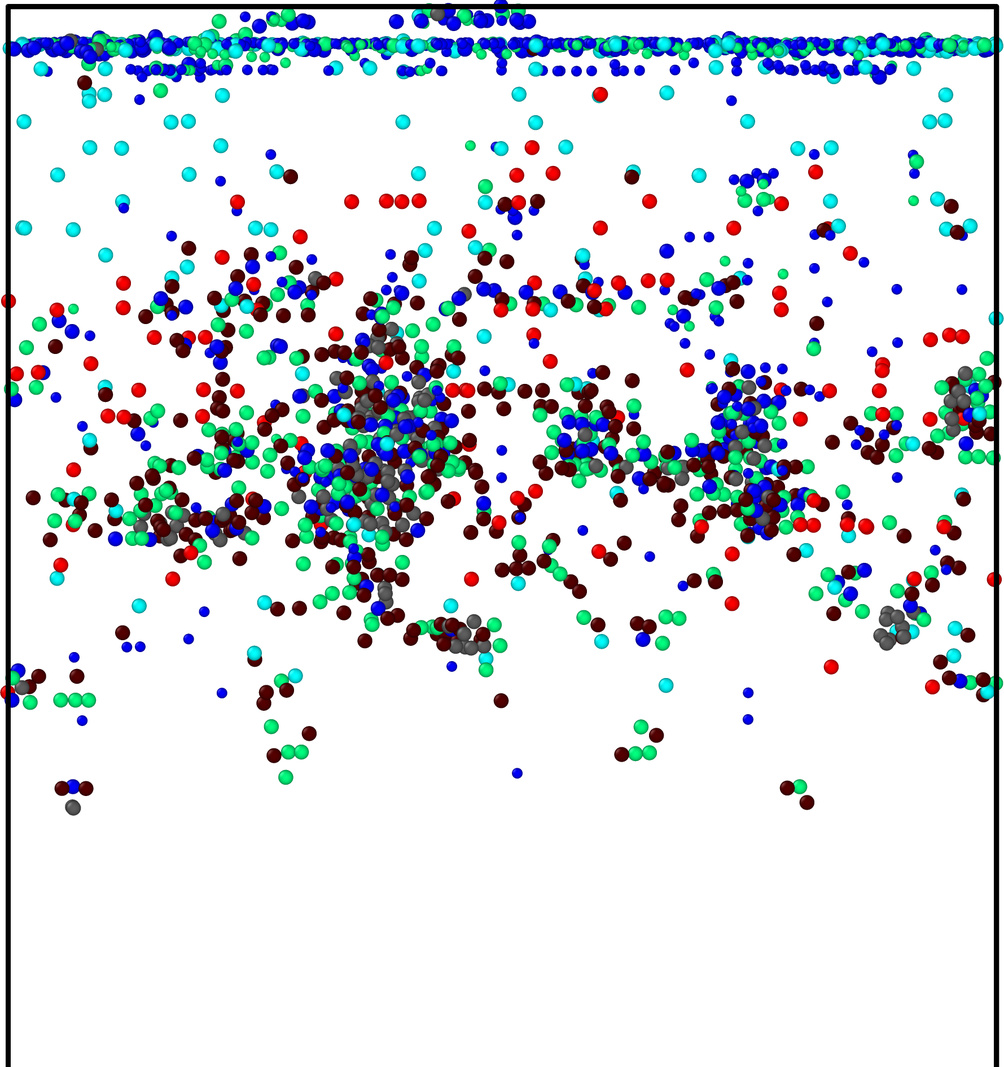}} \hfill
  {\includegraphics[width=0.31\textwidth]{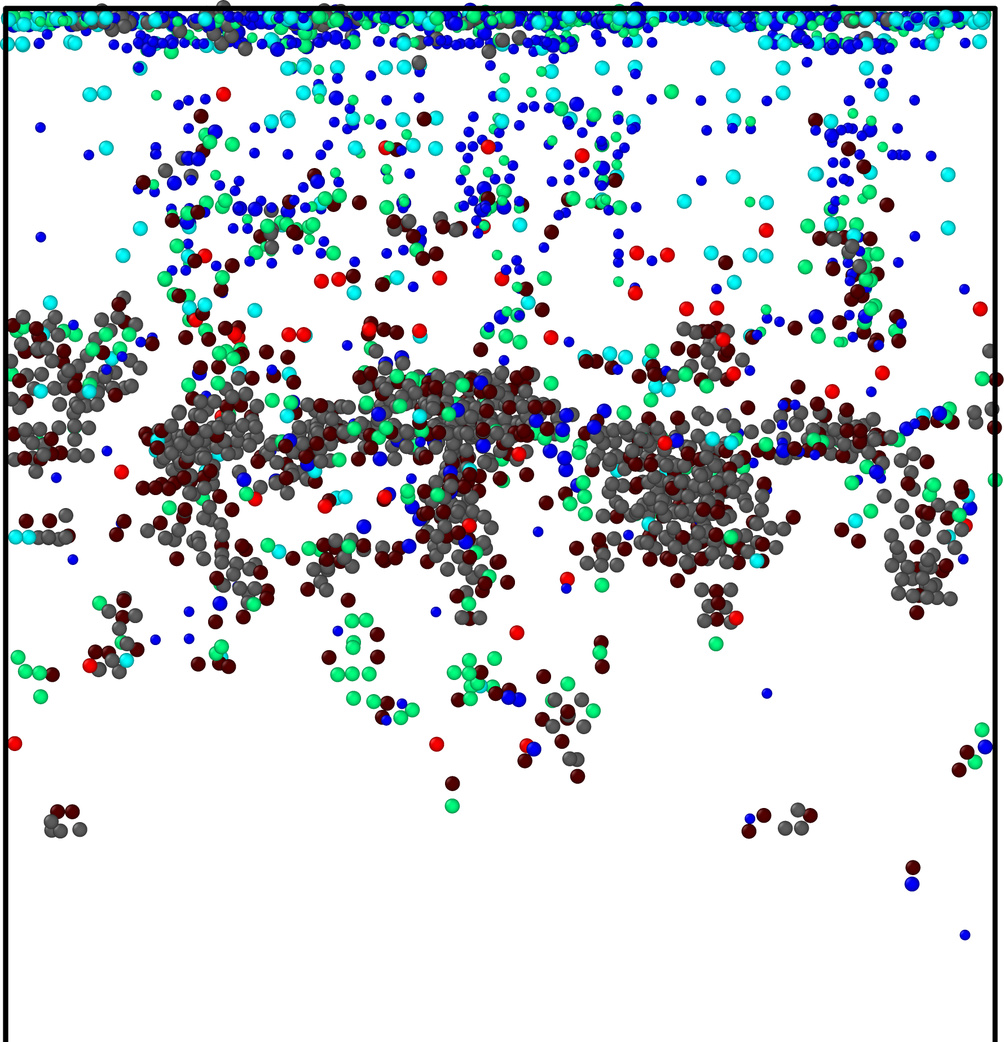}}\hfill \\
%  \vspace{-5pt}
  % Row 2: 2000K
  \rotatebox{90}{\makebox[0.25\textwidth]{\textbf{2000\,K}}} \hspace{-2pt}
  {\includegraphics[width=0.31\textwidth]{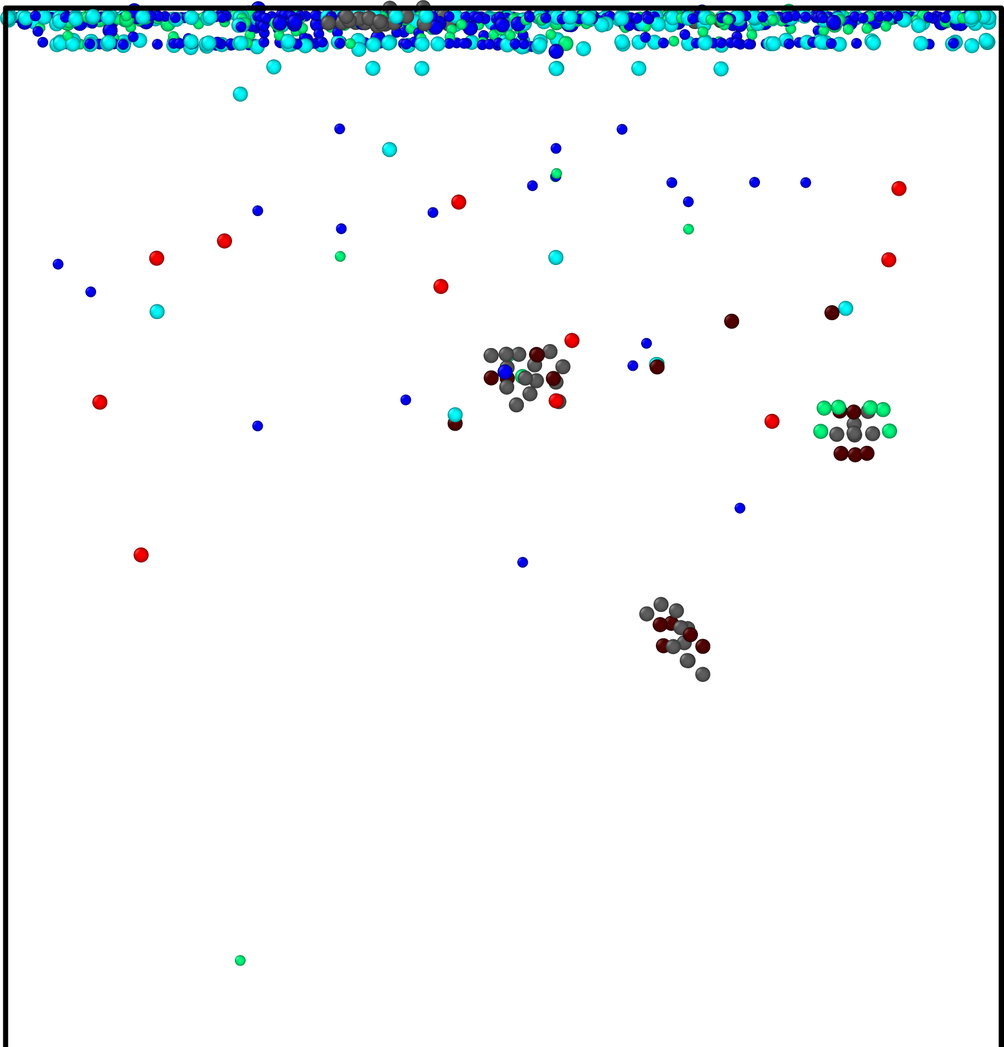}} \hfill
  {\includegraphics[width=0.31\textwidth]{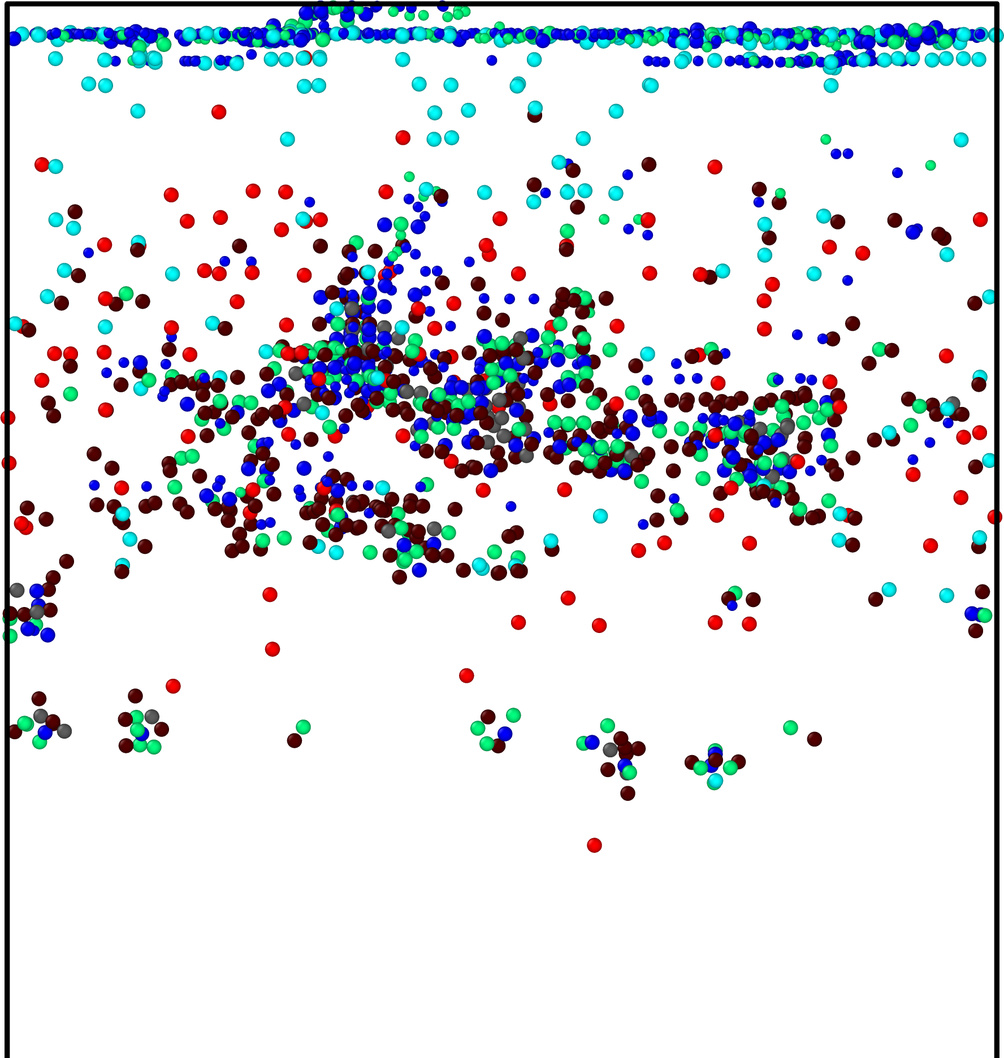}} \hfill
  {\includegraphics[width=0.31\textwidth]{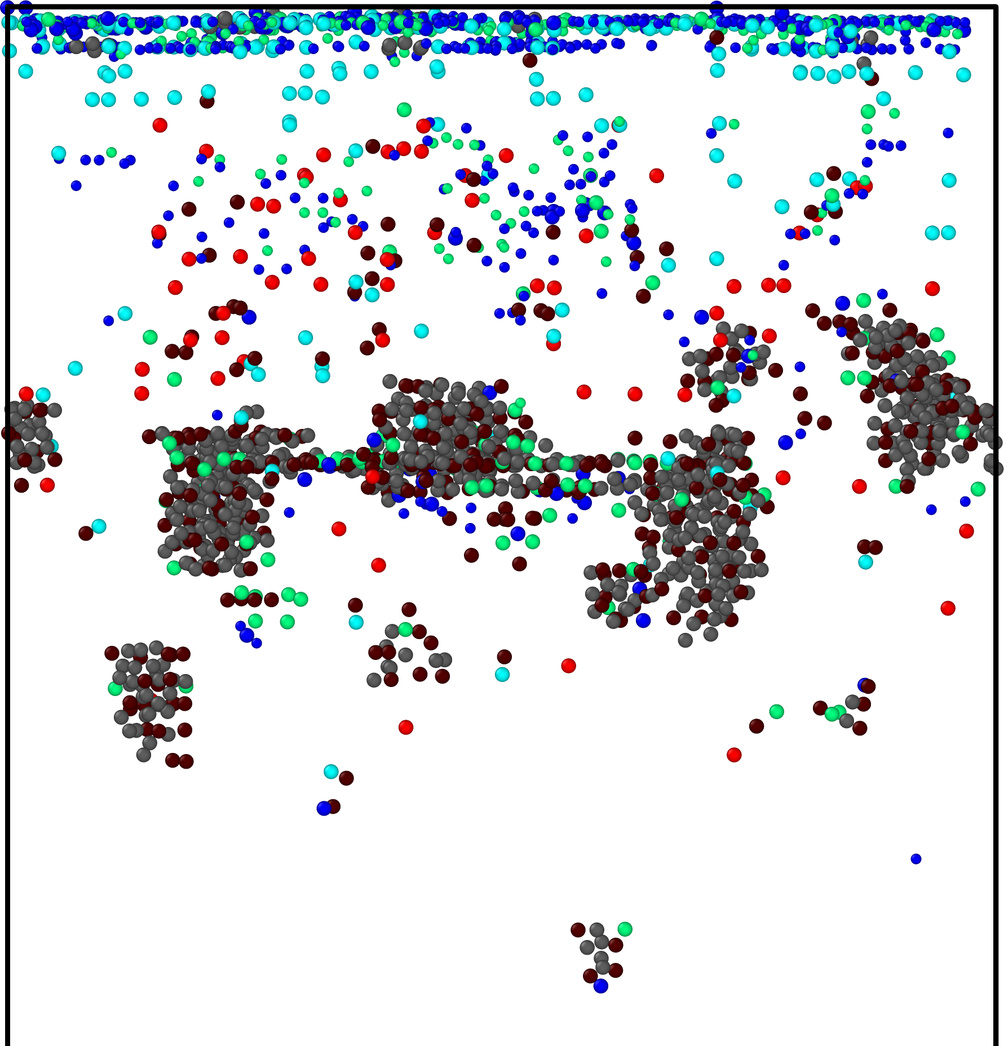}} \hfill\\
  \vspace{-5pt}
    % Row 3: 2500K
  \rotatebox{90}{\makebox[0.25\textwidth]{\textbf{2500\,K}}} \hspace{-2pt}
  \subfloat[]{\includegraphics[width=0.31\textwidth]{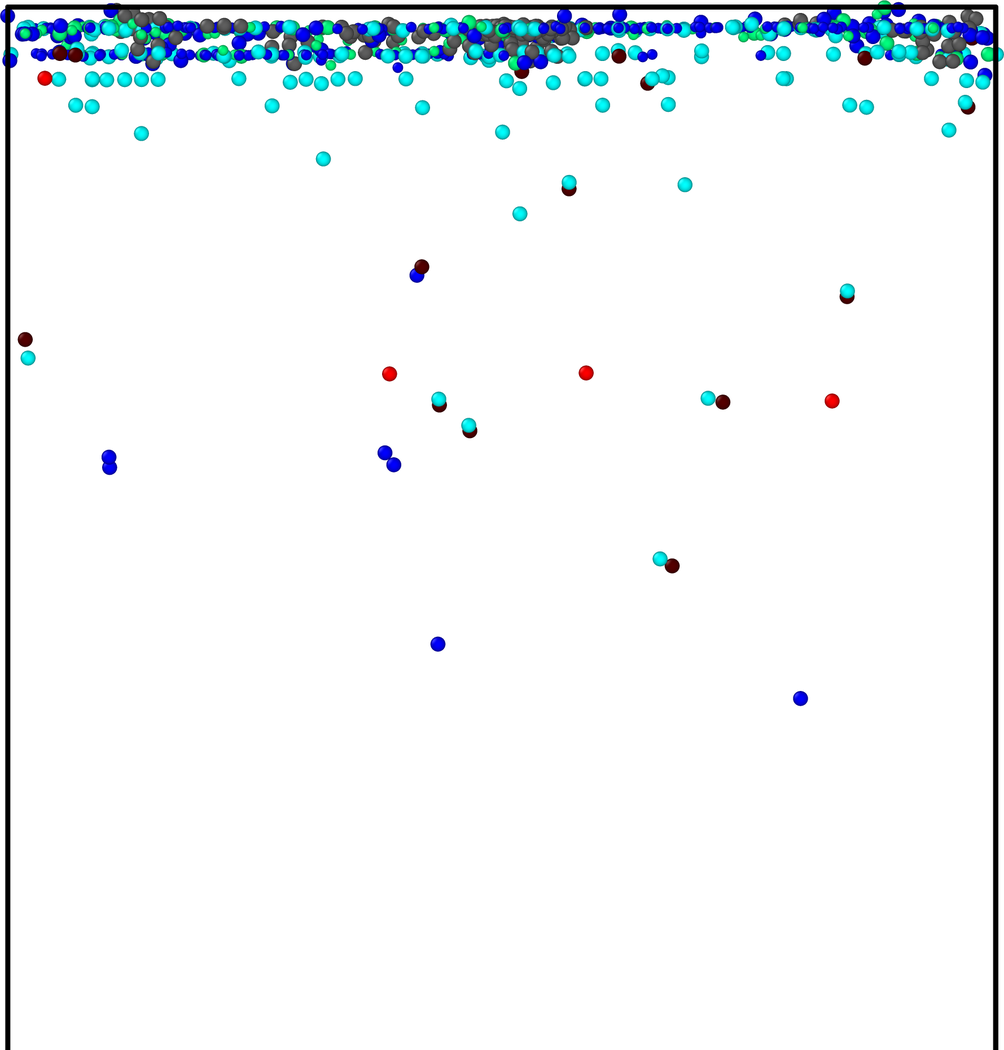}} \hfill
  \subfloat[]{\includegraphics[width=0.31\textwidth]{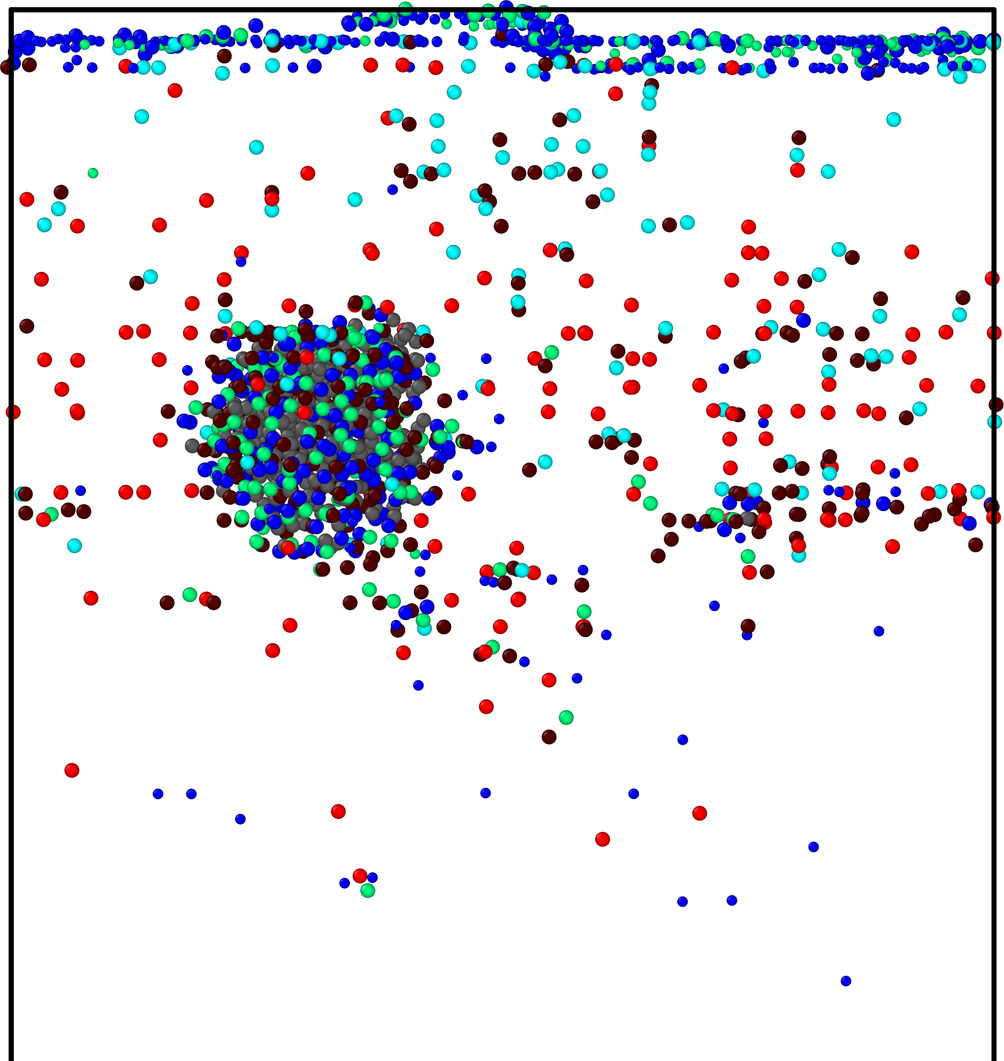}} \hfill
  \subfloat[]{\includegraphics[width=0.31\textwidth]{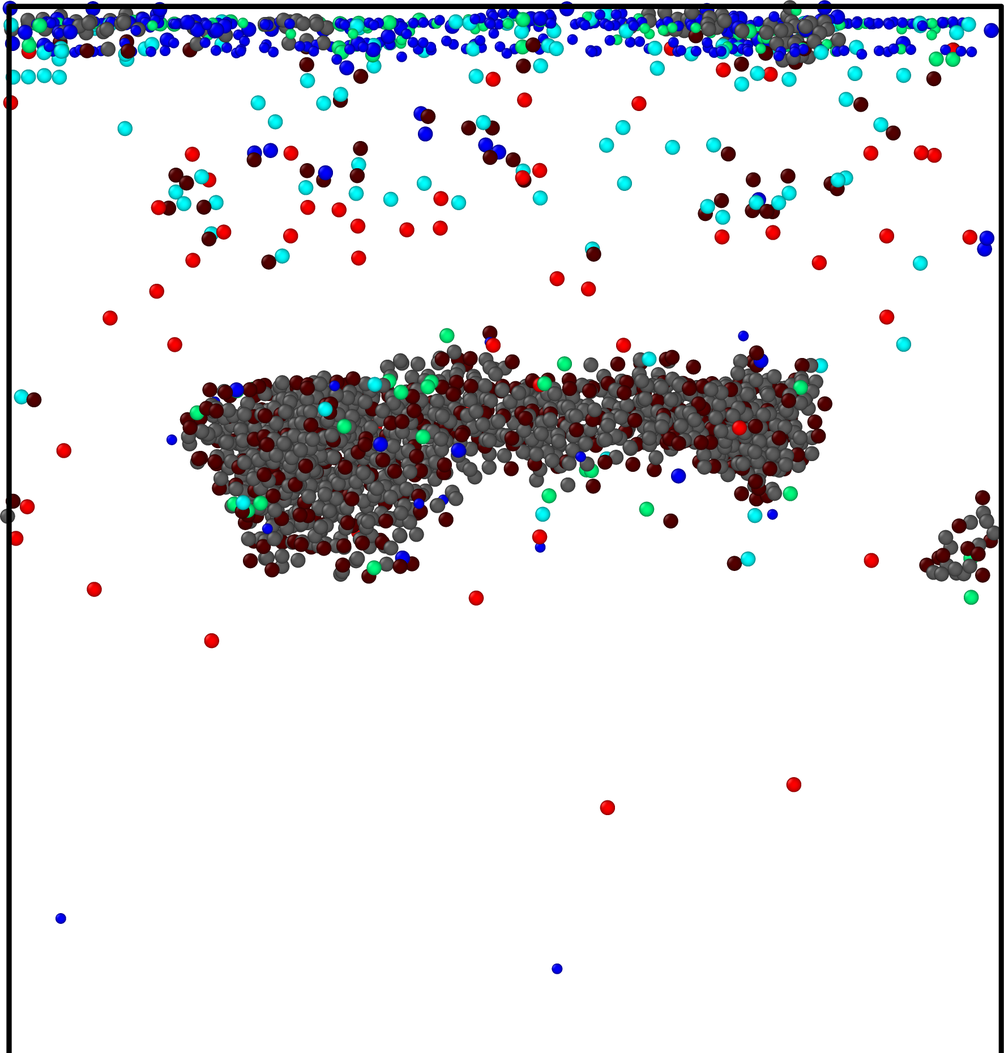}}
   \caption{Isochronal snapshots after $100$\,ns of annealing. Rows show annealing temperatures of $1500$\,K, $2000$\,K, and $2500$\,K (top to bottom), while columns compare (a) dose \mbox{$5\times10^{13}$\,cm$^{-2}$} implanted at $500$\,K, (b) dose \mbox{$7.5\times10^{14}$\,cm$^{-2}$} implanted at $500$\,K, and (c) dose \mbox{$7.5\times10^{14}$\,cm$^{-2}$} implanted at $900$\,K. For the low-dose case, clusters are still agglomerating at $1500$\,K, fully ripened near $2000$\,K, and dissolved at $2500$\,K. At the higher dose, large ripened clusters persist, with low-temperature implantation producing a more globular morphology and high-temperature implantation favoring basal-plane extension. Colors denote C defects (blue/cyan), Si defects (green), activated Al (red), other Al-related defects (brown), amorphous pockets (gray), and vacancies (small spheres).}
  \label{fig:defect_snapshots_temp}
\end{figure}

\begin{figure}[htbp!]
\centering
   \subfloat[]{\includegraphics[width=0.49\textwidth]{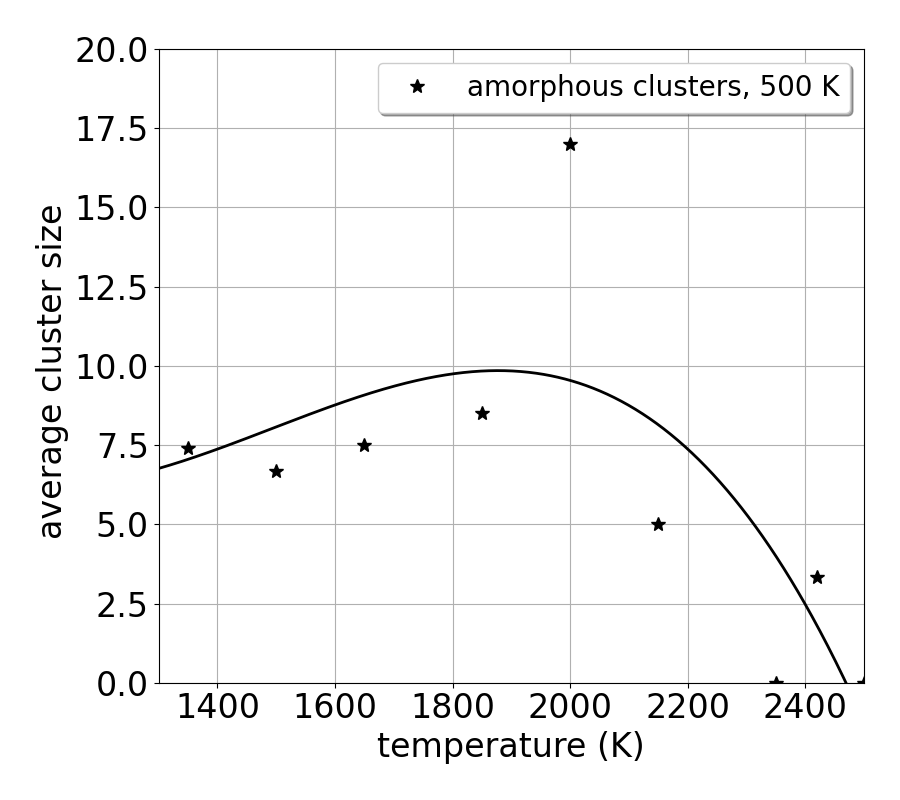}}
%  \subfloat[]{\includegraphics[width=0.3\textwidth]{figures/average_cluster_sizes_dose_8_annealing_temp.png}}
  \subfloat[]{\includegraphics[width=0.49\textwidth]{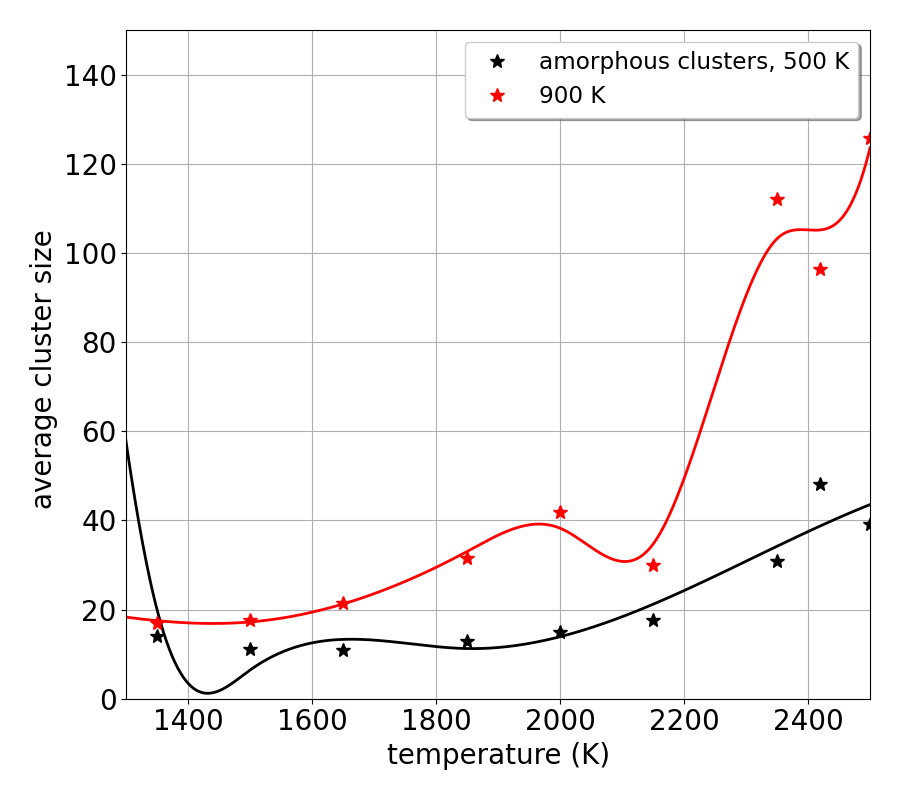}}\\
    \caption{Average cluster size at the end of annealing as a function of temperature for doses: (a) $5\times$10$^{13}$\,cm$^{-2}$ and (b) $7.5\times$10$^{14}$\,cm$^{-2}$. Lines are included as visual guides.}
  \label{fig:average_cluster_size_temp}
 \end{figure}

Each temperature represents a distinct ripening stage,
%%AH: It is not clear to me what (a) shows - which dose temperature combination
%%    It would be good to put a title above the images like in Fig.8
%DONE
%\begin{figure}[htbp!]
%\centering
\begin{figure}[htbp!]
\centering
  % Column Headers
  \hfill
  \makebox[0.32\textwidth]{\textbf{\scriptsize{$5\times 10^{13}$\,cm$^{-2}$, 500\,K}}} \hfill
  \makebox[0.32\textwidth]{\textbf{\scriptsize{$7.5\times 10^{14}$\,cm$^{-2}$, 500\,K}}} \hfill
  \makebox[0.32\textwidth]{\textbf{\scriptsize{$7.5\times 10^{14}$\,cm$^{-2}$,900\,K}}} \hfill \\
  \vspace{2pt}
  \subfloat[]{\includegraphics[width=0.33\textwidth]{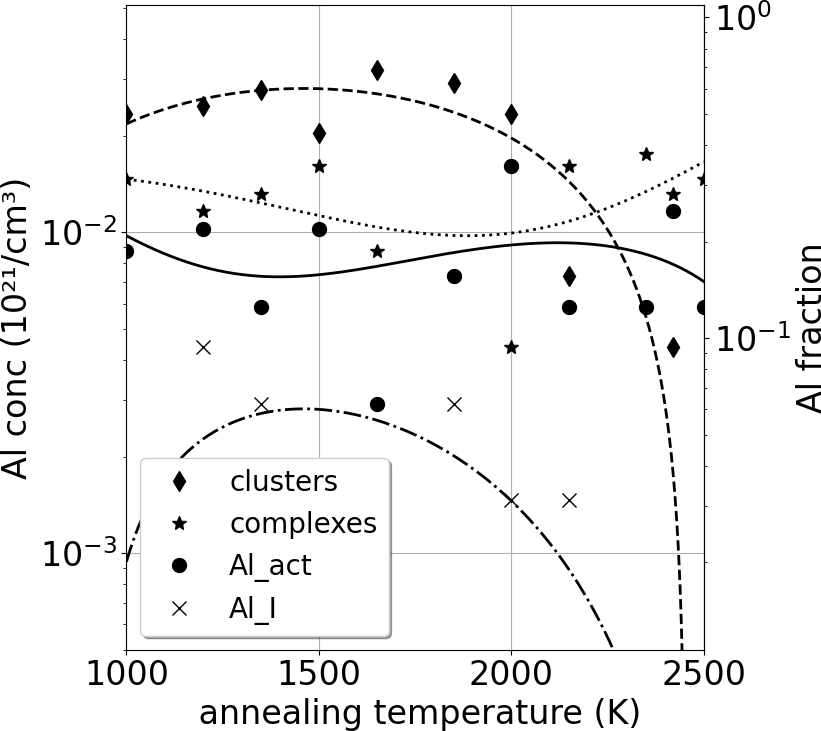}} \hfill
  \subfloat[]{\includegraphics[width=0.33\textwidth]{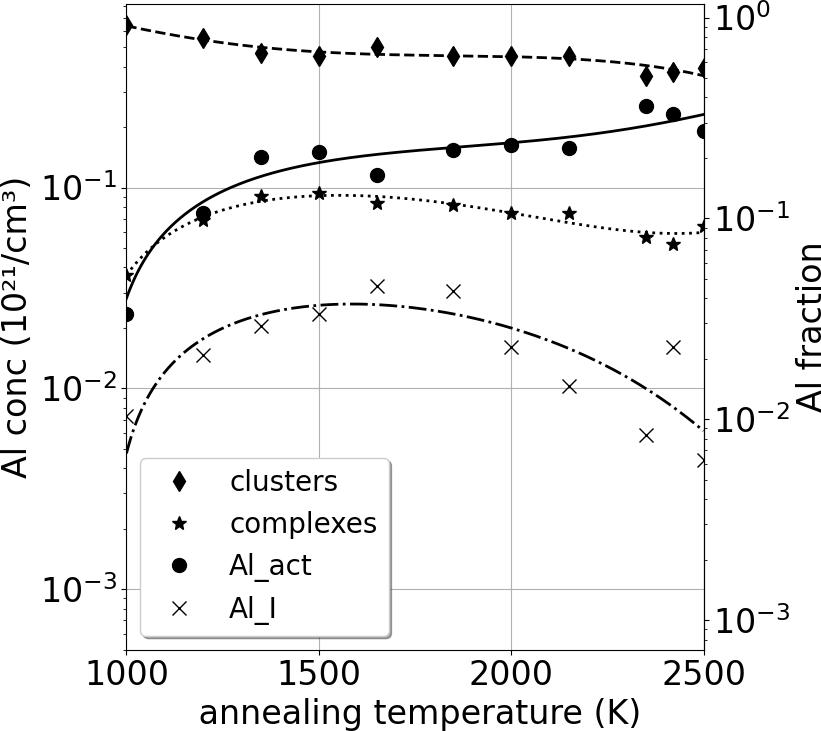}} \hfill
  \subfloat[]{\includegraphics[width=0.33\textwidth]{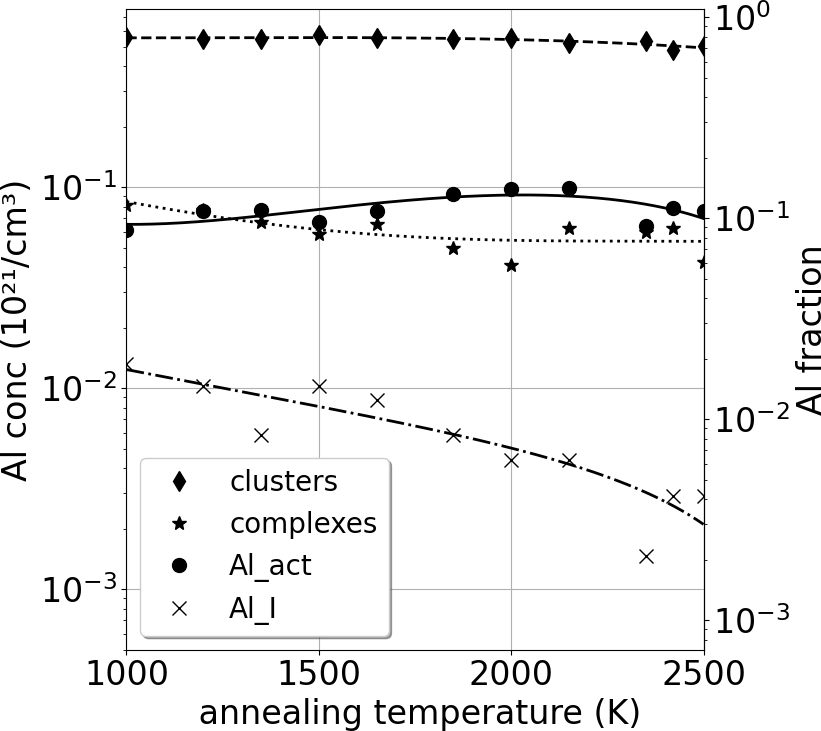}}
  \caption{Concentrations of Al species (interstitials dashed-dot, Al activated solid, complexes dotted, clusters dashed) as a function of annealing temperature for: (a) dose $5\times 10^{13}$\,cm$^{-2}$, and dose $7.5\times 10^{14}$\,cm$^{-2}$ at implantation temperatures of (b) $500$\,K and (c) $900$\,K. Lines provide visual guides.}
  \label{fig:Al_conc_temp}
\end{figure}
where isochronal snapshots of selected ripening stages at $1500$\,K, $2000$\,K, and $2500$\,K are shown in Figure~\ref{fig:defect_snapshots_temp}. At $1500$\,K, after $100$\,ns, the system with a dose of $5\times 10^{13}$\,cm$^{-2}$ is in a stage where small to medium-sized clusters are still agglomerating, shown in Figure~\ref{fig:defect_snapshots_temp}(a). At $2000$\,K, the clusters are fully ripened, and at $2500$\,K, the clusters have completely dissolved. The results in Figure~\ref{fig:Al_conc_temp}(a) and Figure~\ref{fig:average_cluster_size_temp}(a) indicate that Ostwald ripening is complete above $2150$\,K, reaching a maximum chemical Al activation of $25\,\%$ and an (Al-C) complex (sizes $2-5$) concentration of $1.4\times 10^{19}$\,cm$^{-3}$ at $2420$\,K.

For the system with implant dose of $7.5\times 10^{14}$\,cm$^{-2}$, ripening up to the maximum cluster size within $100$\,ns can be observed in Figure~\ref{fig:Al_conc_temp}(b,c) for temperatures higher than $2150$\,K. 
At the low implantation temperature of $500$\,K, recrystallization is marked by a rise in Al activation to approximately $20\,\%$ at around $1300$\,K (corresponding to the recrystallization temperature~\cite{leroch2024}), which further increases with increasing annealing temperature to about $40\,\%$ at $2350$\,K, due to the reduction of Al in complexes and compact clusters. Conversely, at elevated implantation temperatures, the activation remains almost unchanged at approximately $10\,\%$ because roughly $70\,\%$ of the Al is trapped in clusters, as illustrated by the snapshots at various temperatures in Figure~\ref{fig:defect_snapshots_temp}.
In addition, Figure~\ref{fig:average_cluster_size_temp} shows the average cluster sizes at the end of annealing. Panel (a) displays the progression of the ripening process for a dose of $5\times 10^{13}$\,cm$^{-2}$, which reaches a maximum cluster size at $2000$\,K; the data indicates that all clusters have dissolved above $2150$\,K. In Figure~\ref{fig:average_cluster_size_temp}(b), the average cluster size increases sharply starting from $2350$\,K for implantation at $900$\,K, corresponding to the formation of an extended defect. Prior to this transition, the two curves for different implantation temperatures differ only slightly.
 
 The difference in ripening stages as a function of implantation temperature at $1500$\,K, $2000$\,K, and $2500$\,K is clear from Figure~\ref{fig:defect_snapshots_temp}, particularly at the highest annealing temperatures, where clusters in both systems are fully ripened. Beyond the maximum cluster size, the primary distinction between the two systems is morphology: for low implantation temperatures of $500$\,K, the large cluster remains globular, whereas for higher implantation temperatures, it extends significantly in the basal plane.

 % \newpage
 \subsubsection{Residual defects and cluster binding energy}
Figure~S9 in the supplementary material summarizes the concentrations of the most common point defects and complexes following $100$\,ns of isochronal annealing at various temperatures. In agreement with experiments and with our observations in Section~\ref{subsec:Alsuper}, Al- and carbon-related point defects and complexes exhibit the highest kinetic stability. Conversely, concentrations of silicon-related and vacancy-related complexes decrease with increasing temperature. 
To estimate the  binding energies per interstitial in larger clusters via an Arrhenius relation, Figure~S2 in the supplementary material compares Al and C diffusivities for dose $7.5\times 10^{14}$\,cm$^{-2}$ at the highest annealing temperatures between $2150$ and $2500$\,K. 
The system transitions through different regimes during annealing. Whether the system is in an amorphous state, undergoing recrystallization, or experiencing cluster growth, the activation energy varies over time.
%as summarized in Table~S4 in the supplementary material.
While the activation energy generally results from the superposition of multiple processes, one process usually dominates within a specific regime. The time dependent activation energies for interstitial migration in the temperature range of $2150$\,K-$2500$\,K are summarized in Table~S4 in the supplementary material. During the first nanosecond, the activation energy to initiate Al diffusion in the $500$\,K system is approximately $1.2$\,eV, close to the migration barrier of $1.1$\,eV for free Al~\cite{leroch2026_JMCC}. By $30$\,ns, the energy rises to approximately $2.0$\,eV, which is close to the kick-out barrier of $1.6$\,eV for activated Al in the neutral state, driven by increased Si interstitial concentration during ripening. For the system with $900$\,K, the activation energy increases further to $2.7$\,eV by the end of annealing, correlating to the energy required to release a single Al interstitial from large interstitial clusters.
Previous studies by Ko \textit{et al.}~\cite{Ko2017} found binding energies for Si and C interstitials in pure SiC clusters of around $1.5$\,eV interstitial. Mattausch \textit{et al.}~\cite{mattausch2004} and Gali \textit{et al.}~\cite{gali2007} reported that emission of point defects from interstitial clusters in 4H-SiC can occur at energies as low as $2$\,eV. This aligns with our simulation results if the migration barrier of $1.1$\,eV for free Al is deducted from the activation energy of $2.7$\,eV~\cite{leroch2026_JMCC}.

To summarize this section: The quantitative findings as function of annealing temperature are clearly consistent with the temporal defect evolution observed for the $5\times 10^{13}$\,cm$^{-2}$ and $5\times 10^{14}$\,cm$^{-2}$ systems studied in 
Section~\ref{sec:time}.
A common feature among the systems investigated is that activation initially increases until full cluster ripening is achieved, then decreases during the subsequent dissolution process due to additional (Al-C) complex formation. This decrease depends primarily on the quantity of interstitials released from the clusters and their proximity to the surface. At Al supersaturation for $900$\,K implantation, significantly more defects agglomerate at the edges of extended stacking faults that align in the basal plane, whereas for $500$\,K, the average cluster size remains low and even larger clusters keep a globular shape.
Binding energies per interstitial of $1.6$\,eV for large clusters are in reasonably good agreement with literature values \cite{Ko2017}. It can therefore be expected that at $900$\,K, as large interstitial agglomerates dissolve, significantly more thermodynamically stable (Al-C) complexes are generated than at $500$\,K. The increase in the concentration of small (Al-C) complexes which could be secondary defects associated with the high concentration of stacking faults may explain the experimentally observed decrease in hole density over time at Al supersaturation.

\subsection{Extended defects at Al supersaturation as a function of temperature}
\begin{figure}[htbp!]
\centering
 \subfloat[]{\includegraphics[width=0.325\textwidth]{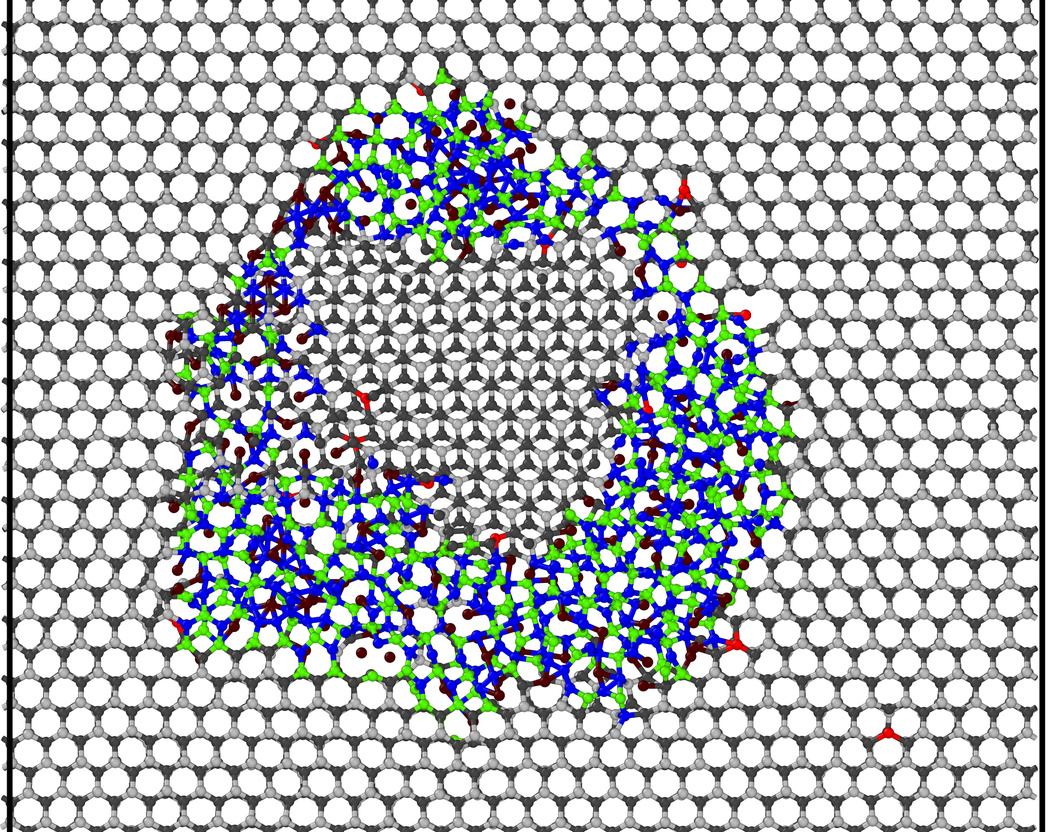}}
 \hspace{1pt}
   \subfloat[]{\includegraphics[width=0.325\textwidth]{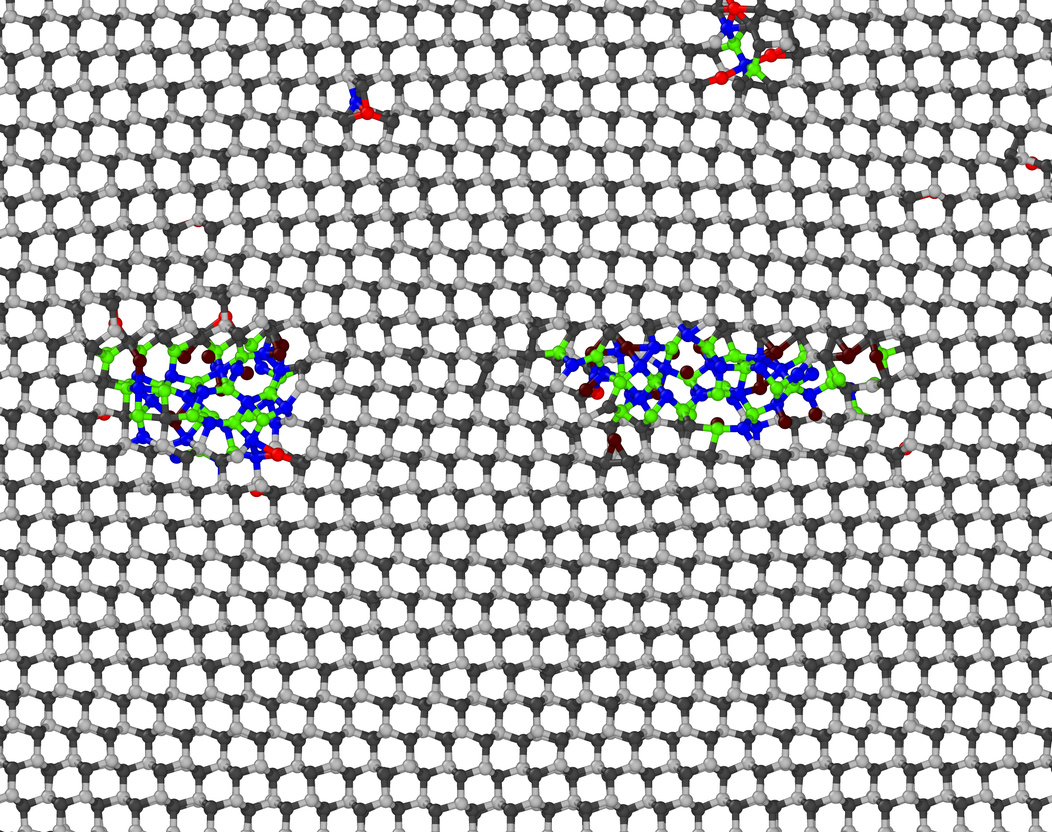}}
 \hspace{1pt}
  \subfloat[]{\includegraphics[width=0.325\textwidth]{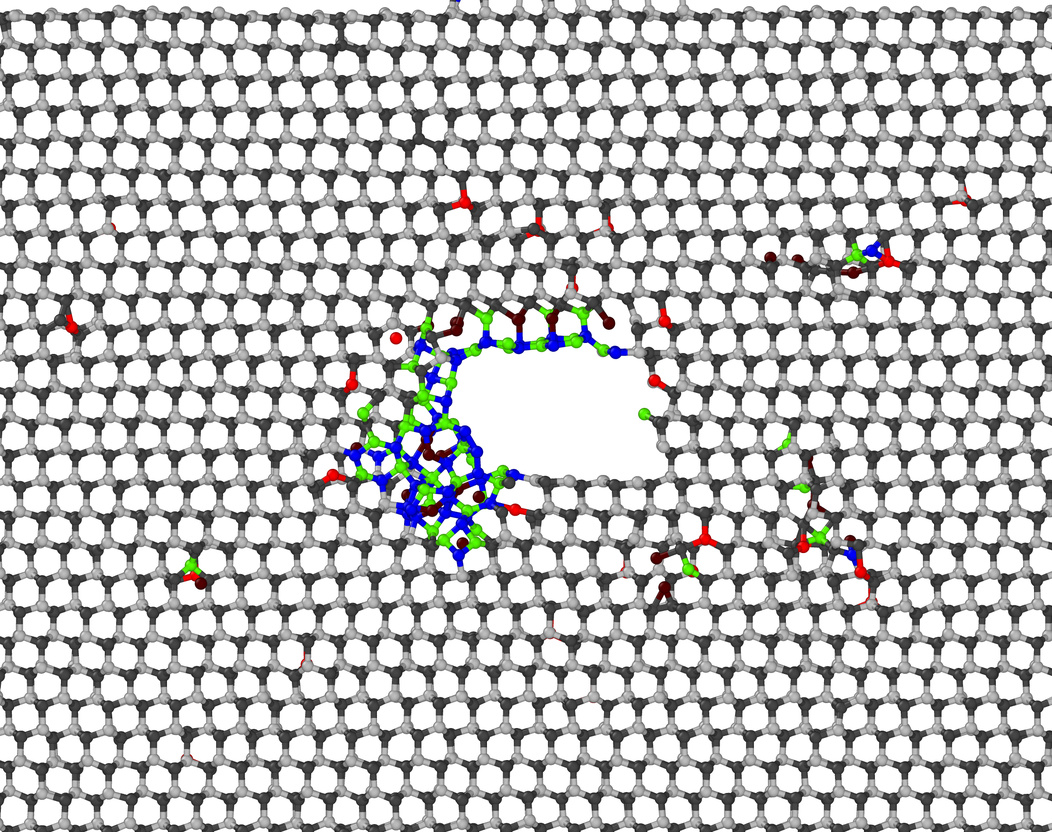}}
   \caption{Extended defect structures for a dose of $7.5\times 10^{14}$\,cm$^{-2}$ annealed at $2350$\,K after implantation at a temperature of $900$\,K: (a) cross-section through the basal plane, (b) cross-section through the $(11\overline{2}0)$ plane, and (c) void formation after implantation at $500$\,K.}
   \label{fig:dislocation}
\end{figure}
Examples of extended defects formed at Al supersaturation for the $500$\,K and $900$\,K implantations are shown in  Figure~\ref{fig:dislocation}.

At $900$\,K faulted loops, as depicted in parts (a) and (b) of the figure, are formed by the introduction of an additional layer in the stacking sequence of the SiC crystal. This layer essentially consists of excess interstitials and is free of Al, in agreement with experimental findings~\cite{Nipoti2018-2}. 
The stacking fault is encompassed in the basal plane by an amorphous torus, marking the dislocation line, which is composed of (Al-Si-C) precipitates.

At $500$\,K, instead, voids with diameters of a few nanometers as shown in Figure~\ref{fig:dislocation}(c), remain after recrystallization. These voids are decorated with under-coordinated Si atoms. 
Voids are generally not considered as critical as dislocation loops, except when dangling bonds have  occupied states in the band gap, which can act as hole traps.
Moreover, for the system implanted with the same dose but at a low temperature ($500$\,K), small intrinsic stacking faults enclosed by Shockley partials formed within $100$\,ns only at temperatures below $2100$\,K. At higher annealing temperatures, no faulted loops were observed, although the largest defect cluster at $2500$\,K reached approximately $1000$ defects. This is comparable to the $1500$ defects measured in clusters at $900$\,K. The critical factor for dislocation-loop formation is therefore not solely cluster size, but also cluster shape. The different behavior as a function of implantation temperature can be attributed to the formation history of the clusters. While the globular cluster formed during recrystallization has retained a nearly amorphous core with many vacancies the planar, almost pure interstitial clusters formed at $900$\,K resulted from the anisotropy of interstitial diffusion.  In globular clusters, lattice stresses are distributed almost isotropically, whereas in flattened clusters, residual shear stresses in the basal plane built up. At annealing temperatures above $2000$\,K, shear stresses of only a few MPa are sufficient to trigger the insertion of an extrinsic crystalline SiC layer into the basal plane~\cite{li2022}.

\subsubsection{Nucleation and growth of faulted loops}
\begin{figure}[htbp!]
\centering
  % Row 1: DXA
  \includegraphics[width=0.242\textwidth]{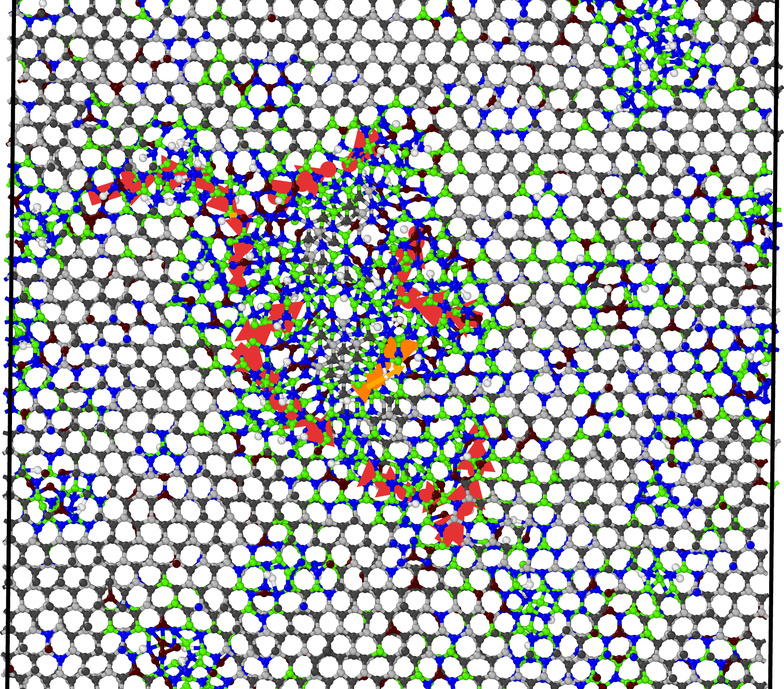} \hfill
  \includegraphics[width=0.242\textwidth]{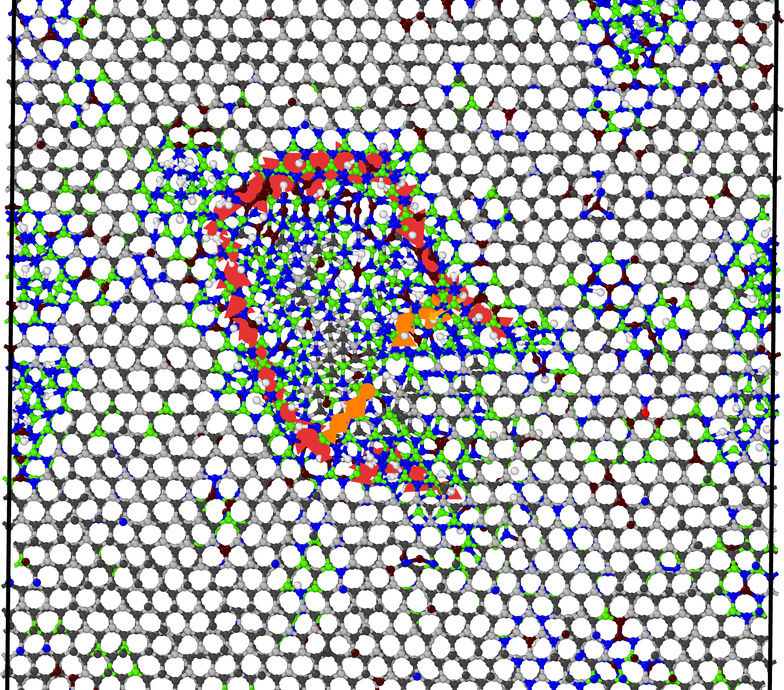} \hfill
  \includegraphics[width=0.242\textwidth]{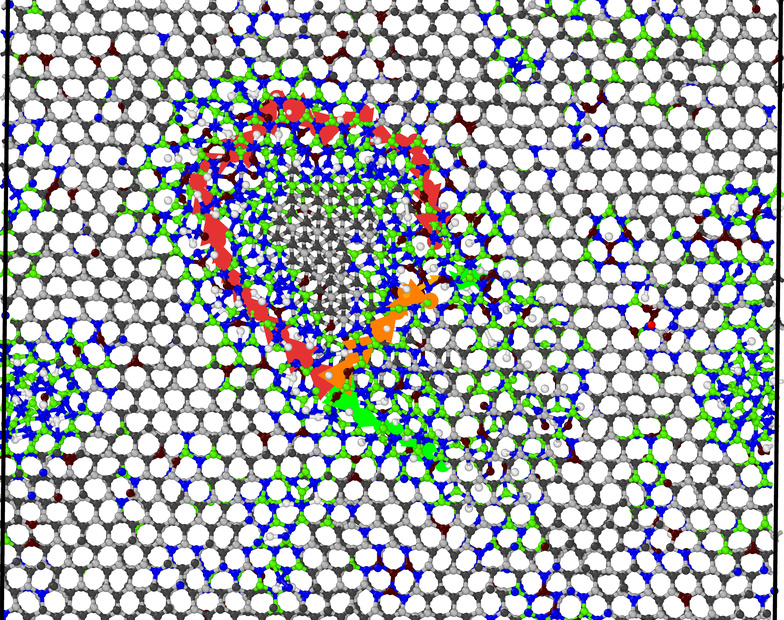} \hfill
  \includegraphics[width=0.242\textwidth]{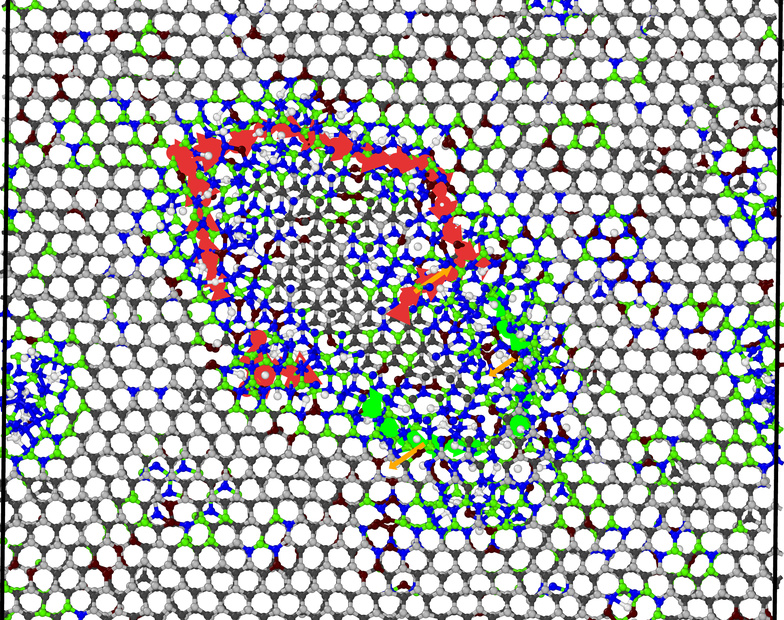} \\
  \vspace{1mm}
  % Row 2: Stacking Fault
  \subfloat[]{\includegraphics[width=0.242\textwidth]{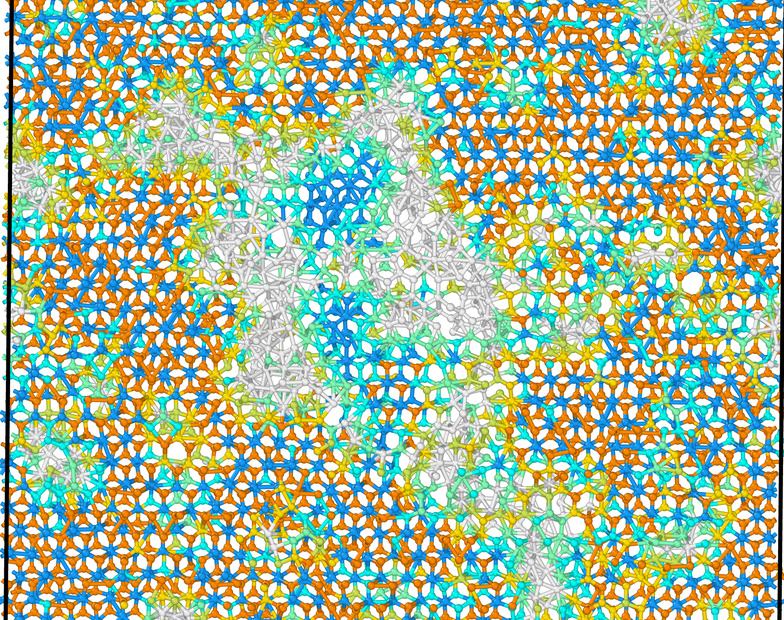}} \hfill
  \subfloat[]{\includegraphics[width=0.242\textwidth]{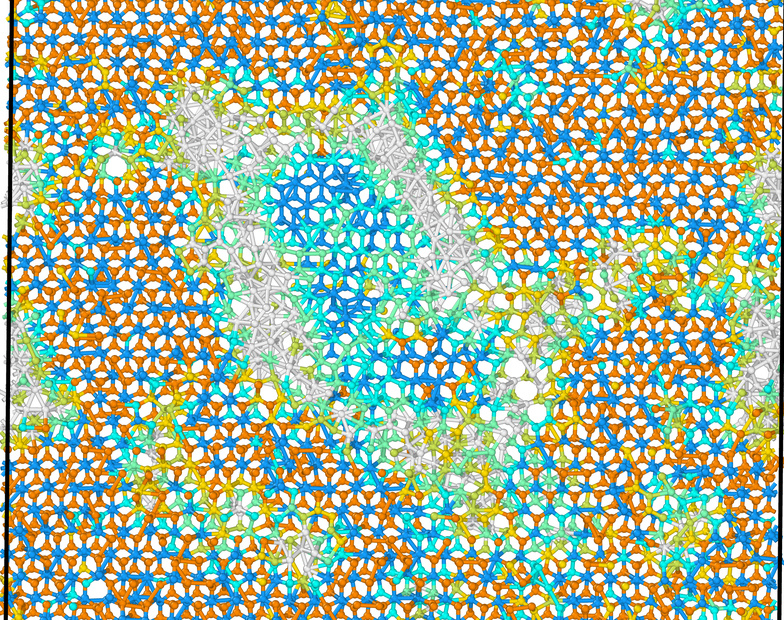}} \hfill
  \subfloat[]{\includegraphics[width=0.242\textwidth]{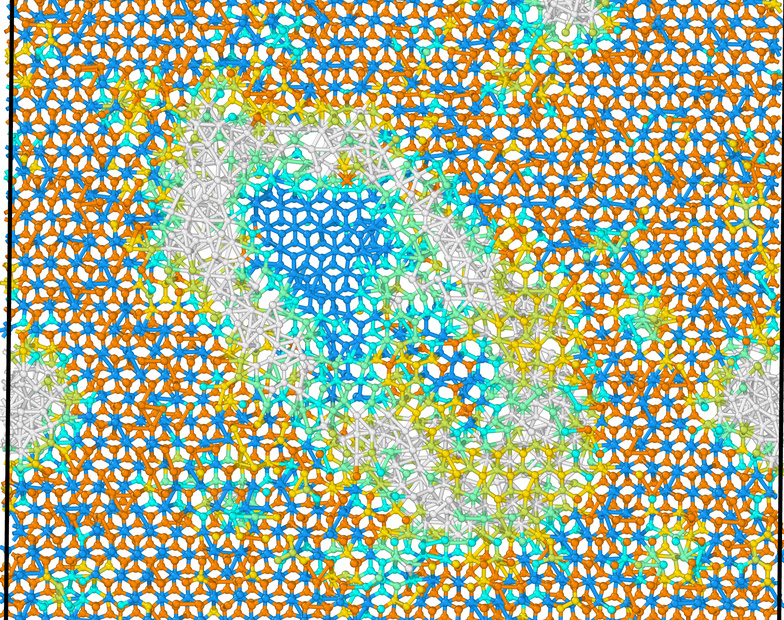}} \hfill
  \subfloat[]{\includegraphics[width=0.242\textwidth]{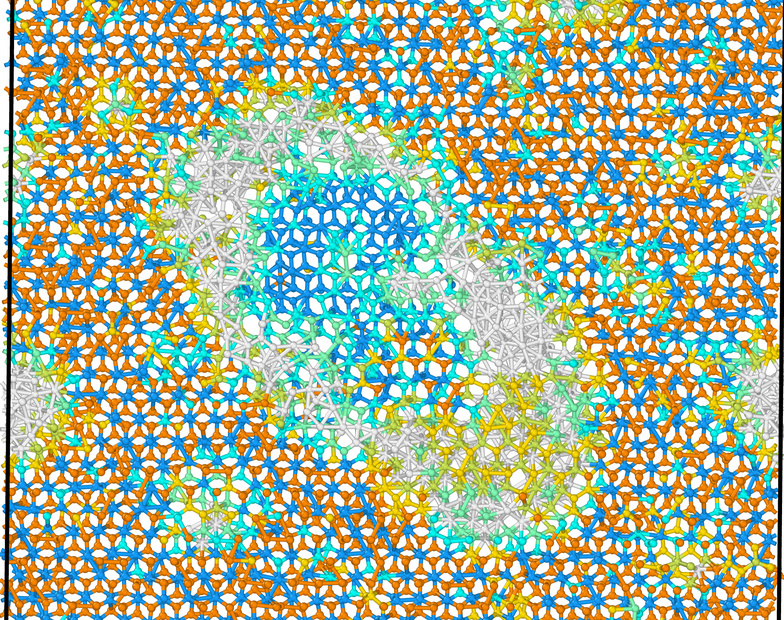}} 
  \caption{Evolution of partial dislocations and stacking faults at $2150$\,K after: (a) $1$\,ns, (b) $20$\,ns, (c) $60$\,ns, and (d) $90$\,ns. The top row shows DXA extraction of dislocation lines, and the bottom row shows stacking-fault identification.}
  \label{fig:dislocation_lines_2150K}
\end{figure}

\begin{figure}[htbp!]
\centering
  \subfloat[]{\includegraphics[width=0.48\textwidth]{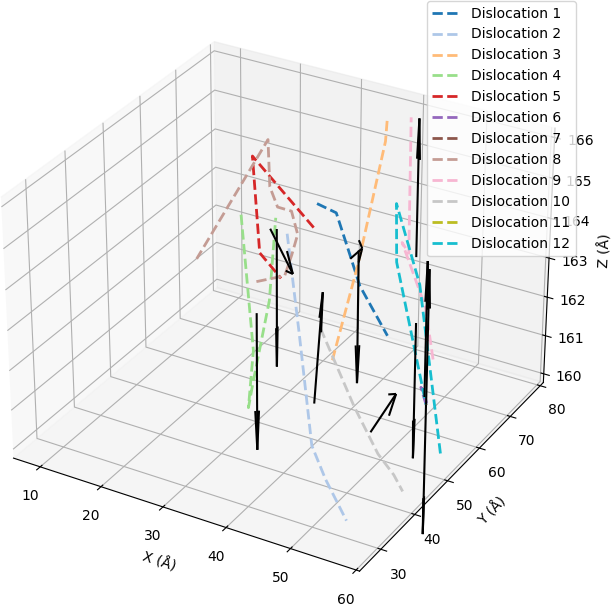}} \hfill
  \subfloat[]{\includegraphics[width=0.48\textwidth]{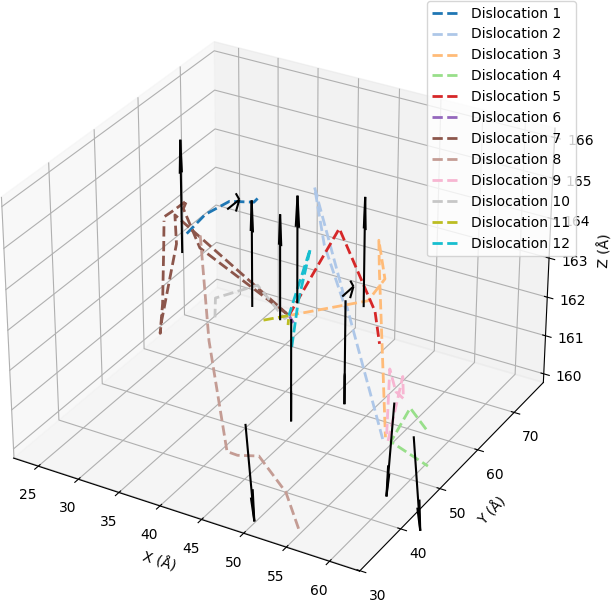}} \\
  \subfloat[]{\includegraphics[width=0.48\textwidth]{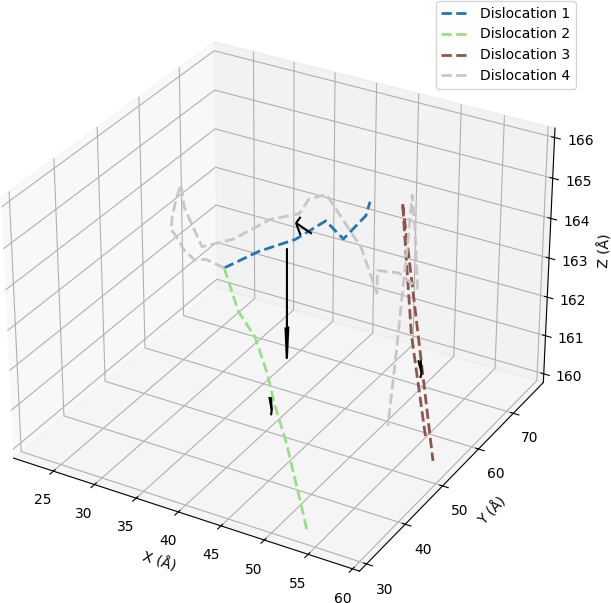}} \hfill
  \subfloat[]{\includegraphics[width=0.48\textwidth]{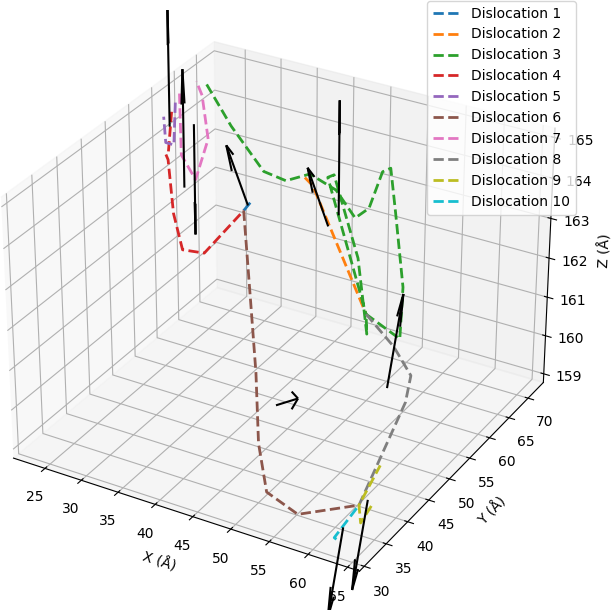}}
  \caption{Burgers-vector classification maps inlcuding partials at $2150$\,K after: (a) $1$\,ns, (b) $20$\,ns, (c) $60$\,ns, and (d) $90$\,ns.}
  \label{fig:burgers_2150K}
\end{figure}

Figure~\ref{fig:dislocation_lines_2150K} shows the temporal evolution of a faulted loop for the system with a dose of $7.5\times 10^{14}$\,cm$^{-2}$ at $900$\,K and annealing temperature of $2150$\,K. The top row shows the location of the dislocation lines as determined by OVITO's dislocation extraction algorithm (DXA). The middle row shows the stacking fault in the center of the faulted loop, visualized by coloring the atoms using OVITO's IDS algorithm to distinguish cubic (blue) and hexagonal (orange) layers. The corresponding Burgers-vector classification maps are shown separately in Figure~\ref{fig:burgers_2150K}.

After a few hundred picoseconds of annealing, when planar clusters or subdomains in irregular larger defects exceed a  critical size of around $60$ interstitials as discussed in the supplementary material and displayed in Figure~S6, the transition into stacking faults can be triggered. First tiny stacking faults at the edges of the clusters appear which grow and eventually merge. Within approximately $1$\,ns of annealing their advancing partials form ring-shaped inclusions around a continuous extrinsic stacking fault of cubically arranged atoms (in blue) as can be seen in Figure~\ref{fig:dislocation_lines_2150K}(a). As shown in Figure~\ref{fig:burgers_2150K} the partials are mainly located in the basal plane, while their Burgers vectors run mostly along the $c$-axis ($c$-type), indicating Frank-type partials typically found in implanted 4H-SiC~\cite{Persson2003}. Additionally, a few $a$-type and mixed-type partials are observed, especially at the beginning of annealing, as seen in Figure~\ref{fig:burgers_2150K}. The $a$-type dislocations shown in orange ($1/3\langle1\overline100\rangle$) and green ($1/3\langle1\overline210\rangle$) represent Shockley partials, which are characterized by dislocation lines and Burgers-vector directions in the basal plane, while the dislocation lines in red correspond to mixed and Frank-type  ($\langle0001\rangle$) partials. 

\begin{figure}[htbp!]
\centering
  % Row 1
  \includegraphics[width=0.242\textwidth]{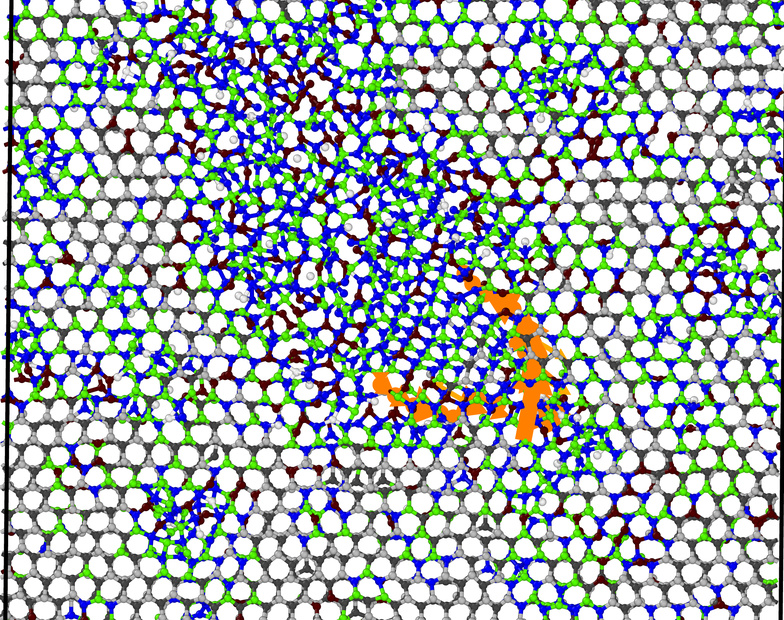} \hfill
  \includegraphics[width=0.242\textwidth]{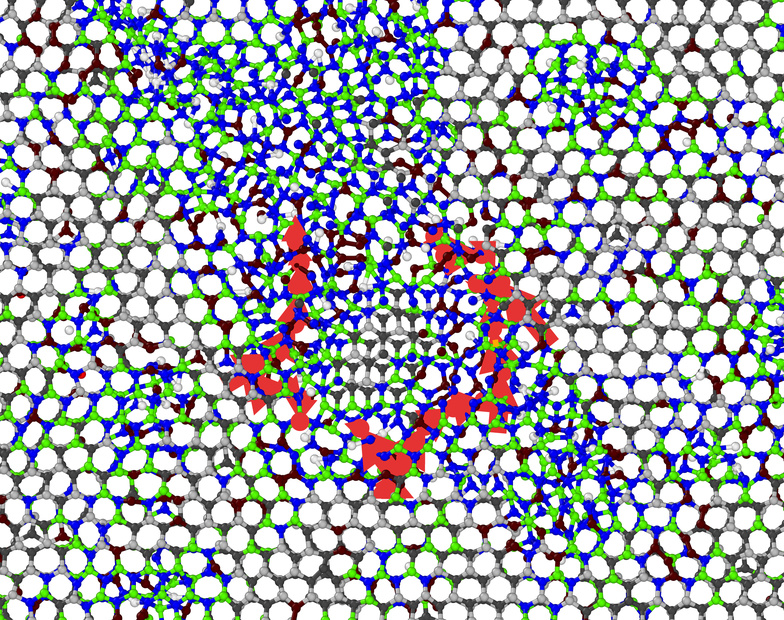} \hfill
  \includegraphics[width=0.242\textwidth]{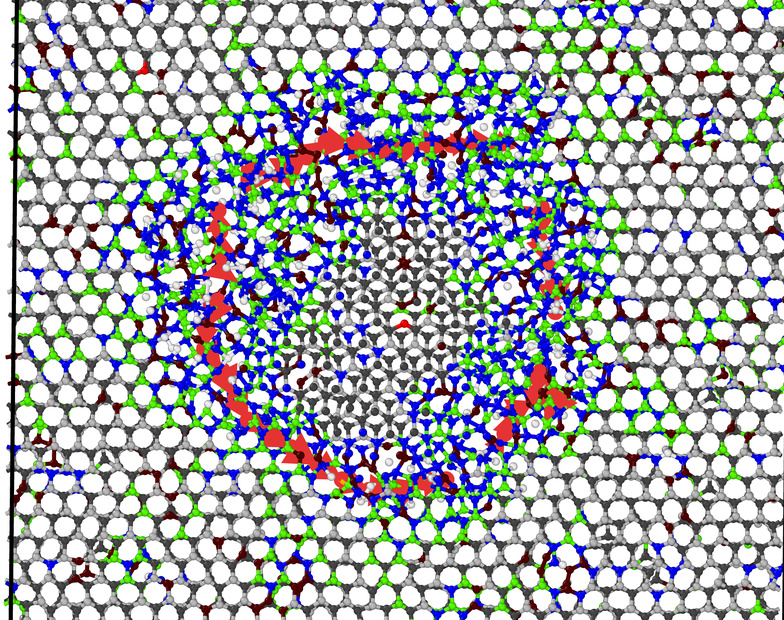} \hfill
  \includegraphics[width=0.242\textwidth]{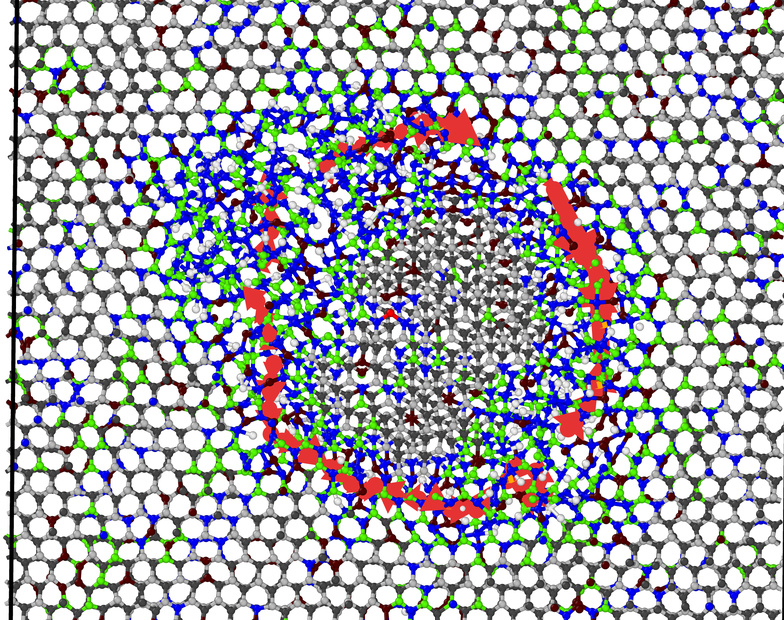} \\
  \vspace{1mm}
  % Row 2
  \subfloat[]{\includegraphics[width=0.242\textwidth]{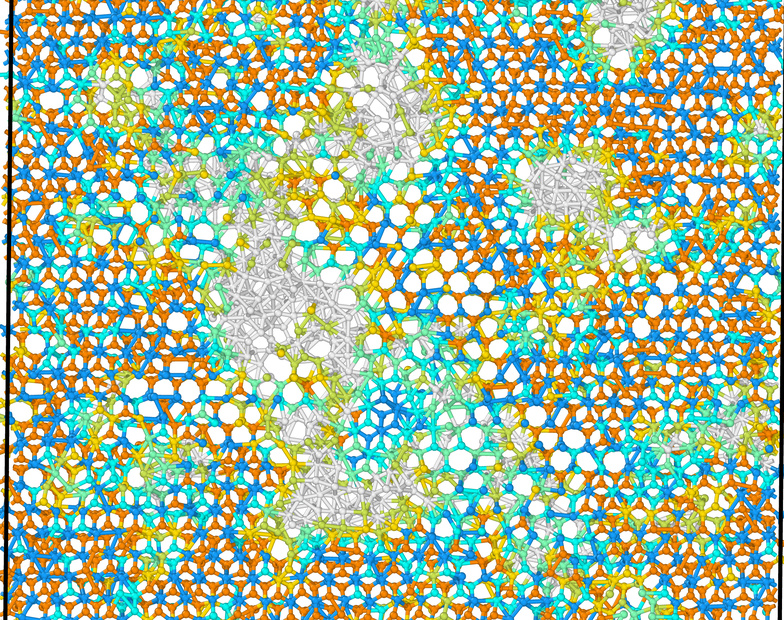}} \hfill
  \subfloat[]{\includegraphics[width=0.242\textwidth]{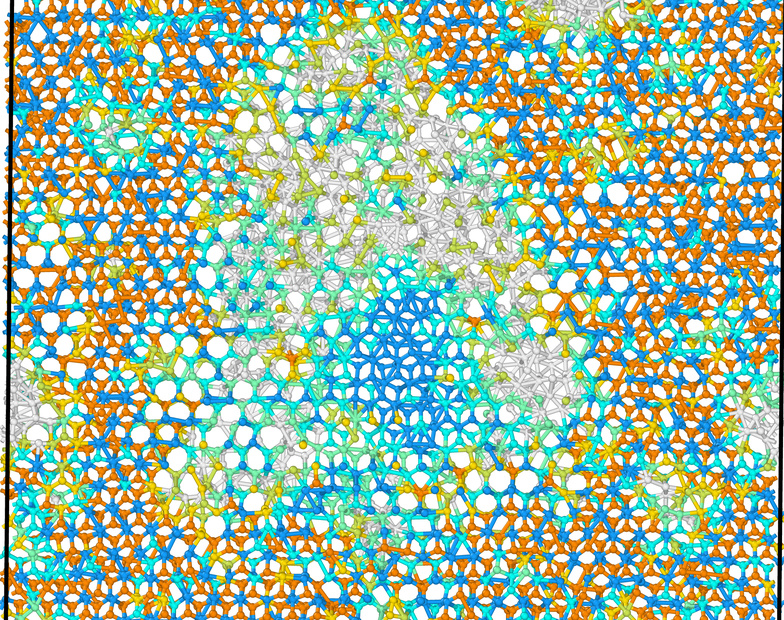}} \hfill
  \subfloat[]{\includegraphics[width=0.242\textwidth]{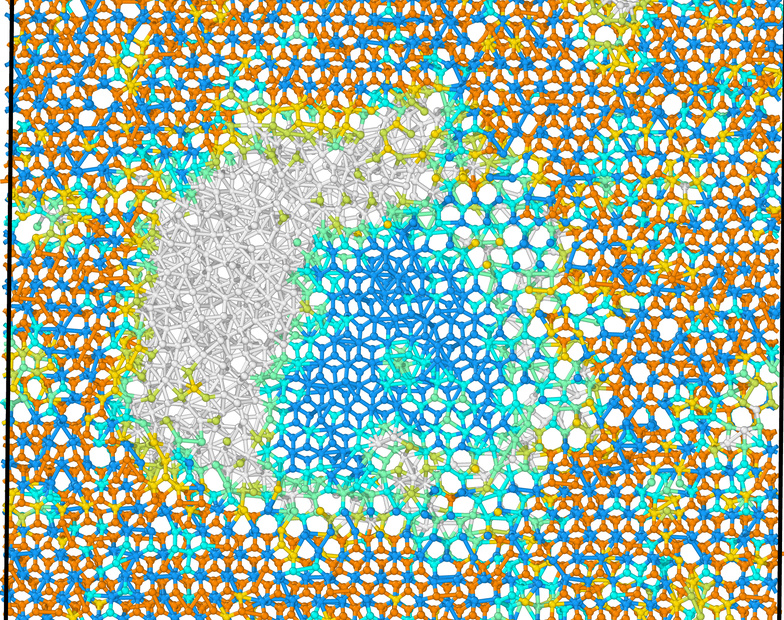}} \hfill
  \subfloat[]{\includegraphics[width=0.242\textwidth]{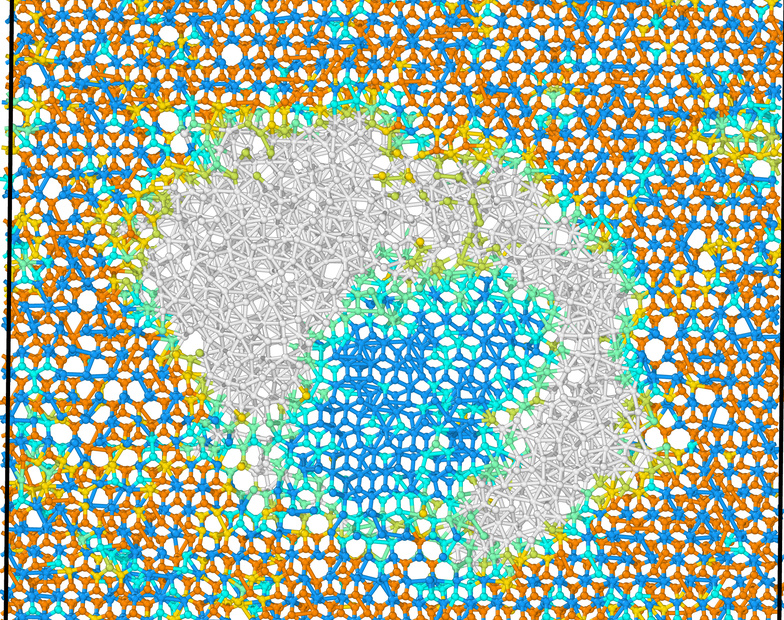}}
  \caption{Evolution of partial dislocations and stacking faults at $2500$\,K after: (a) $150$\,ps, (b) $500$\,ps, (c) $30$\,ns, and (d) $80$\,ns. The top row shows DXA extraction of dislocation lines, and the bottom row shows stacking-fault identification.}
  \label{fig:dislocation_lines_2500K}
\end{figure}

\begin{figure}[htbp!]
\centering
  \subfloat[]{\includegraphics[width=0.48\textwidth]{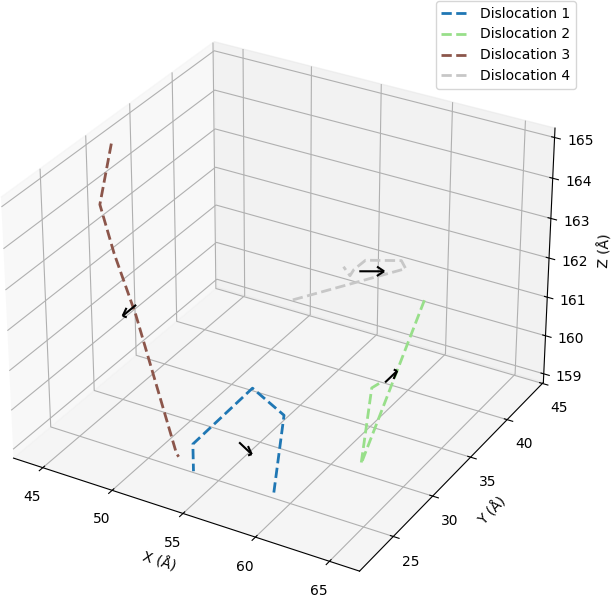}} \hfill
  \subfloat[]{\includegraphics[width=0.48\textwidth]{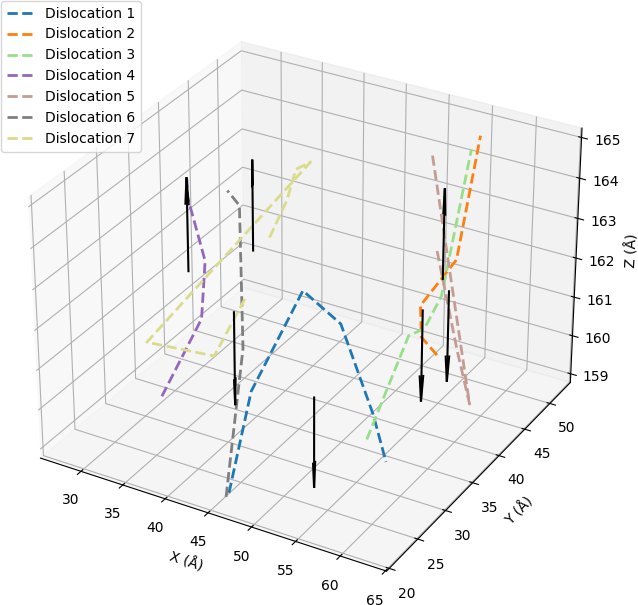}} \\
  \subfloat[]{\includegraphics[width=0.48\textwidth]{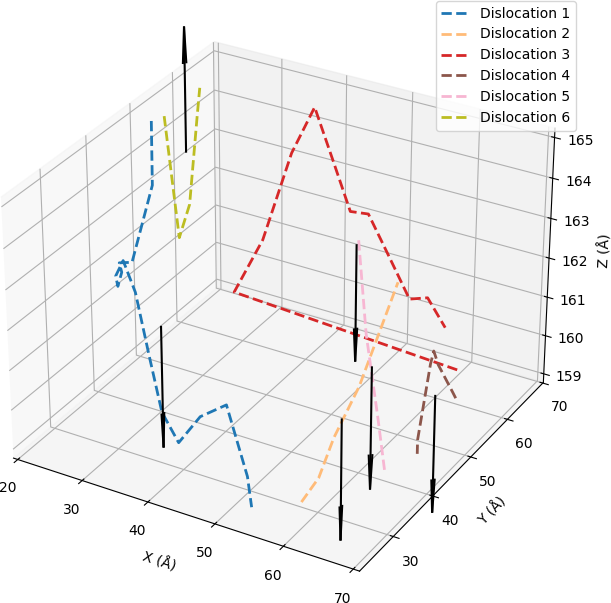}} \hfill
  \subfloat[]{\includegraphics[width=0.48\textwidth]{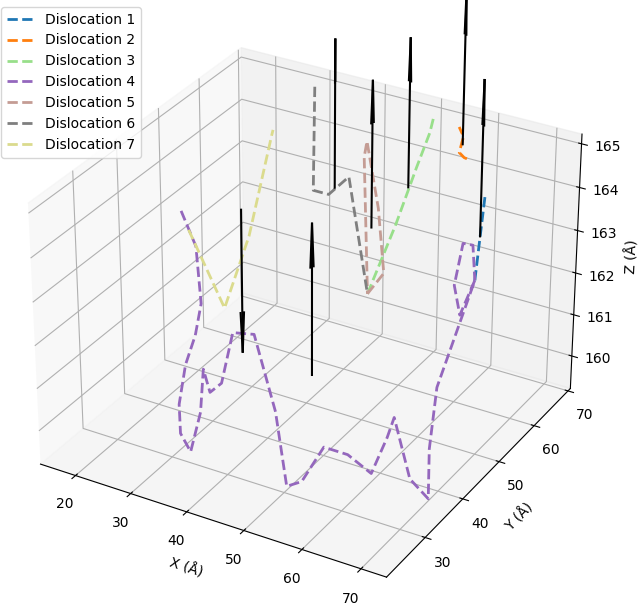}}
  \caption{Burgers-vector classification maps inlcuding partials at $2500$\,K after: (a) $150$\,ps, (b) $500$\,ps, (c) $30$\,ns, and (d) $80$\,ns.}
  \label{fig:burgers_2500K}
\end{figure}

Figure~\ref{fig:dislocation_lines_2500K} illustrates the dislocation loop formation at an annealing temperature of $2500$\,K. As in the $2150$\,K case, Shockley partials form within $150$\,ps and then transform into Frank partials within $1$\,ns as can be seen in Figure~\ref{fig:burgers_2500K}. The stacking fault grows much more rapidly by capturing further diffusing interstitials and, toward the end of annealing, extends over most of the cross-sectional area near the damage peak. The corresponding Burgers-vector classification is shown in Figure~\ref{fig:burgers_2500K}. At these high temperatures, the mobile Shockley partials become active and annihilate, leaving only stable Frank-type loops.
The loop is not yet fully developed in Figure~\ref{fig:burgers_2500K}(d); it does not consist of a continuous loop, but is composed of numerous partials that lie within a monolayer of approximately $5$\,\AA{} in thickness. 

\subsubsection{Activation energy for nucleation and growth}
The time evolution for dislocation length and stacking fault area at different annealing temperatures is shown in Figure~S7 of the supplementary material. The time required for the nucleation of a $3$\,nm long dislocation line (sum of all the partials), triggered by the shear stress of irregular planar clusters starting with sizes of about $60$ defects at temperatures between $1500$\,K and $2500$\,K, is shown in an Arrhenius plot in Figure~S7(c). The nucleation rate has an activation energy of approximately $1.1$\,eV and an attempt frequency of $6.7\times 10^{11}$s$^{-1}$, in good agreement with a proposed dislocation model~\cite{SUN2013216}. Dislocation length and stacking fault area initially rise sharply and then increase linearly before reaching saturation at approximately $50$\,ns. The growth rate, estimated via an Arrhenius relation, yields a range of activation energies from $1.5$ to $2.7$\,eV, which is within the range of values reported for dislocation-loop growth in steel~\cite{yang2011}, and matches well with the activation energy for interstitial migration calculated in Table~S5 of the SI.

To summarize the findings of this section: The nucleation of faulted loops is characterized by the simultaneous appearance of (i) an inserted basal layer, (ii) a local 3C-like stacking sequence, and (iii) partial dislocation segments bounding the transformed region. The defect subsequently grows by expansion and coalescence of the partial segments rather than by the formation of an isolated stacking fault followed by later dislocation nucleation, as described in textbooks. The nucleation of small stacking fault domains proceeds rapidly with an activation energy of $1$\,eV, consistent with estimates based on a macroscopic model \cite{SUN2013216}. The seeds for dislocation nucleation are medium- to large sized planar clusters/cluster segments in the size range of $60$ to several hundreds of interstitials which serve as reservoirs for stacking faults however, without entirely transforming into stacking faults.

\section{Conclusion}

In this work, MD simulations were applied to resolve how implantation temperature and implanted Al dose jointly govern defect evolution and dopant activation in 4H-SiC during annealing. The stable defects identified after annealing are in good agreement with experiments~\cite{Kumar2024} and DFT predictions~\cite{gali2007,hornos2008}. Residual damage under all implantation conditions includes V$_\mathrm{C}$ and small Al- and C-related  complexes, where (Al-C) dimers and trimers remain the most common compensating defects. Some of them are thermodynamically very stable, others are persistent because of high kinetic barriers such as the  Al$_\mathrm{Si}$C$_\mathrm{I}$ which acts as a dynamic trapping state: it is continuously formed by capture of C interstitials and dissolved by thermally activated release, 
which explains the abundance of the dimers at the end of annealing on MD-accessible timescales~\cite{leroch2026_JMCC} despite its low binding energy of $0.5$\,eV.

In contrast, as shown in our previous publication~\cite{leroch2026_JMCC}, the trimers consisting of two carbon antisites and one Al interstitial are particularly thermodynamically stable, with binding energies in the neutral state of up to $5$\,eV. In addition, some of the trimers identified \cite{leroch2026_JMCC} exhibit shallow levels between $0.2$\,eV and $0.5$\,eV above the VBM, which could explain the measured DLTS signals \cite{gali2007}.
A detailed study of (Al-C) complexes of varying sizes, composition and orientation would reveal additional stable structures. However, the Morse potential used here does not provide the necessary accuracy; rather, the focus of this study is on the temporal evolution of faulted loops and its impact on Al activation and the formation of compensating secondary defects as postulated on the basis of experiments~\cite{Negoro2004}. The formation of faulted loops has never been tracked before in this way using MD simulations. A previous study~\cite{leroch2026_JMCC} has already demonstrated that the Gao-Weber-Morse potential describes entropy-driven processes, such as the formation of extended defects as a function of the implantation conditions, in phenomenological agreement with experiment~\cite{WANG2023,ZANG2025}.

With increasing Al supersaturation, larger (Al-Si-C) agglomerates, stacking faults, and basal plane faulted loops emerge. 
The reported Al solubility-related concentration varies between $0.5$ and $2\times10^{20}$\,cm$^{-3}$ in annealed SiC~\cite{Nipoti2018-2}.
The key kinetic finding is that, at Al supersaturation, the implantation temperature controls  whether the system is mainly dominated by large compact interstitial clusters, or by planar extended defects. At elevated implantation temperature ($900$\,K), interstitial-rich planar clustering is promoted already during implantation and acts as a precursor for extended defects during annealing. At lower implantation temperature ($500$\,K), stronger local disorder is retained and subsequently consumed by epitaxial regrowth, which favors substitutional Al incorporation. 
For extended-defect formation, our atomistic trajectories resolve spontaneous loop nucleation and growth. Once a critical planar cluster size is reached, tiny ($\sim 20$ atoms) cubically ordered stacking faults form on timescales from a few hundred picoseconds to $10$\,ns depending on the annealing temperature. Atomistic simulations indicate that the transformation proceeds through the nucleation of several spatially separated 3C-like stacking-fault embryos, each bounded by partial dislocation segments. These embryos subsequently grow and coalesce into a continuous faulted loop. Loop formation therefore does not follow the textbook picture of a single coherent disk that nucleates and then expands.  Instead, several locally transformed regions within one connected cluster appear first and only later merge into a single fault. As the stacking fault advances, interstitials are incorporated into the inserted basal layer, the local cluster loses atoms,
so that the measured cluster may shrink. 
In addition, diffusing interstitials continue to be trapped at the dislocation lines, leading to renewed growth of planar clusters around the stacking faults, which, fed by the interstitials, also continue to grow.
The extracted activation energy for dislocation nucleation is about $1.1$\,eV (for a summed partial length of $\sim3$\,nm, corresponding to a stacking fault of about $20$ atoms). The nucleation is a thermally activated process, triggered by shear stresses of medium- to large sized planar clusters.  Stacking-fault growth follows an Arrhenius behavior with an activation energy of about $2.1$\,eV~\cite{SUN2013216}. The dominant extended defect structure  at $900$\,K is faulted interstitial loops of  Frank-type, whereas transient Shockley partials are mainly observed at early stages or at lower temperatures, consistent with experiment~\cite{Persson2003,li2022}. The estimated Al-interstitial binding energy in larger Al-Si-C clusters is about $1.6$\,eV per interstitial, in good agreement with  values reported for pure SiC clusters~\cite{Ko2017}.

When large clusters begin to dissociate, the electrical activation of Al decreases because, due to the large number of interstitials released, Al acceptors are bound into stable (Al-C) complexes, which act as compensating centers in p-type semiconductors. Since the concentration of the formed (Al-C) complexes depends on the size of the initial defect clusters, higher Al-complex concentrations are expected at $900$\,K where large numbers of interstitials agglomerate around stacking faults, than at $500$\,K. Compensating (Al-C) complexes could therefore be the cause of the decline in electrical Al activation over time that was first observed experimentally at Al supersaturation by Negoro \emph{et al.}~\cite{Negoro2004}. Moreover, it is in line with the spectroscopic experiments of Kumar \emph{et al.}~\cite{Kumar2024} who detected defect signals indicating the presence of Al- and C-related complexes in implanted 4H-SiC.

The atomistic rates, cluster binding energies  and mechanisms studied here could provide direct input for, and improve the understanding of, the modeling of faulted loops in  kinetic Monte Carlo simulations, which is needed for upscaling toward experimentally relevant times and dimensions.

\section*{CRediT authorship contribution statement}
\textbf{S. Leroch}: Methodology, Investigation, Formal analysis, Data curation, Writing - original draft
\textbf{R. Stella}: Investigation, Writing - review $\&$ editing
\textbf{A. Hössinger}: Conceptualization, Writing - review $\&$ editing
\textbf{L. Filipovic}: Conceptualization, Supervision, Funding acquisition, Resources, Writing - review $\&$ editing
\section*{Data availability}
All original data and characterization files generated or analyzed during this study are included in this published article and its supplementary material files.
\section*{Acknowledgements}
% \vspace{-5pt}
Financial support by the Federal Ministry of Labour and Economy, the National Foundation for Research, Technology and Development and the Christian Doppler Research Association is gratefully acknowledged. The computational results presented have been achieved using the Vienna Scientific Cluster (VSC).The authors acknowledge TU Wien Bibliothek for financial support through its Open Access Funding Program.

%% ---- Bibliography ---------------------------------------------------
\bibliographystyle{elsarticle-num}
\bibliography{SiC_update_with_DOIs}

%% =====================================================================
%%  SUPPLEMENTARY MATERIAL
%% =====================================================================
\startsupplement

\begin{center}
  {\Large\bfseries Supplementary Material\par}
  \vspace{0.8em}
  {\large Faulted loop nucleation and dopant activation in Al-implanted 4H-SiC\par}
  \vspace{0.8em}
  {Sabine Leroch, Robert Stella, Andreas Hössinger, Lado Filipovic\par}
\end{center}
\vspace{1.5em}

\noindent Supplementary figures, tables, and analyses supporting the main manuscript.
\vspace{1em}

\section{Additional Results}

\subsection{Analysis Definitions and Dose-to-Concentration Mapping}

Table~\ref{tab:definitions} summarizes the operational definitions used throughout the manuscript for activation, defect classes, and clustering metrics.

\begin{table}[H]
\centering
\setlength{\tabcolsep}{6pt}
\caption{Definitions used in the main manuscript and supplementary analysis.}
\begin{tabular}{p{0.32\linewidth} p{0.67\linewidth}}
\textbf{Quantity} & \textbf{Definition used in this work} \\ \hline
Chemically activated Al & Al atom on a substitutional Si site: Al$_\mathrm{Si}$. \\
Perfect activation & Isolated Al$_\mathrm{Si}$ (not bound in a defect complex/cluster). \\
General activation & All substitutional Al$_\mathrm{Si}$, including Al$_\mathrm{Si}$ bound in complexes/clusters. \\
Cluster size $s$ & Number of connected defects in a cluster (connectivity criterion based on the first minimum of the partial radial pair distribution function). \\
Cluster-size classes & Small: $3 \le s < 10$; medium: $10 \le s < 100$; large: $s \ge 100$. \\
Dislocation nucleation criterion & First appearance of a dislocation network with total extracted DXA line length of approximately $3$\,nm (corresponding to a critical planar cluster size of about $20$--$50$ defects). \\
Stacking-fault area & Convex-hull area of the largest connected cubic layer enclosed by dislocation lines (DXA + IDS-based workflow). \\
\hline
\end{tabular}
\label{tab:definitions}
\end{table}

To connect areal dose (cm$^{-2}$) and concentration (cm$^{-3}$) scales, we use the same geometric estimate as in the main text. Assuming an effective implanted-region thickness of $d\approx9$\,nm, the characteristic concentration is approximated as
\begin{equation}
  C_{\mathrm{eff}} \approx \frac{\Phi}{d},
\end{equation}
where $\Phi$ is the implantation dose. The resulting mapping is summarized in Table~\ref{tab:dose_mapping}.

\begin{table}[H]
\centering
\setlength{\tabcolsep}{10pt}
\caption{Approximate conversion between areal dose and concentration using $d\approx9$\,nm ($=9\times10^{-7}$\,cm).}
\begin{tabular}{cc}
\textbf{Dose $\Phi$ (cm$^{-2}$)} & \textbf{$C_{\mathrm{eff}}\approx \Phi/d$ (cm$^{-3}$)} \\ \hline
$5\times10^{13}$ & $5.6\times10^{19}$ \\
$1\times10^{14}$ & $1.1\times10^{20}$ \\
$2\times10^{14}$ & $2.2\times10^{20}$ \\
$5\times10^{14}$ & $5.6\times10^{20}$ \\
$7.5\times10^{14}$ & $8.3\times10^{20}$ \\
\hline
\end{tabular}
\label{tab:dose_mapping}
\end{table}
The number of Al ions implanted at different doses according to the implantation volume of $10\times9\times21$\,nm$^{-3}$ corresponding to approximately $190000$ atoms is given in Table~\ref{tab:dose_numbers}
\begin{table}[H]
\centering
\setlength{\tabcolsep}{10pt}
\caption{Number of implanted Al ions for each implantation dose.}
\begin{tabular}{cc}
\textbf{Dose $\Phi$ (cm$^{-2}$)} & \textbf{Al ions} \\ \hline
$5\times10^{13}$ & $32$ \\
$1\times10^{14}$ & $64$ \\
$2\times10^{14}$ & $128$ \\
$5\times10^{14}$ & $256$ \\
$7.5\times10^{14}$ & $480$ \\
\hline
\end{tabular}
\label{tab:dose_numbers}
\end{table}
\newpage
\subsection{Kick-in Process}
This subsection summarizes the DFT-NEB pathway and barrier sequence for Al kick-in at a carbon antisite in the neutral state.
\begin{figure}[htbp]
\centering
\subfloat[]{\fbox{\includegraphics[width=.7\linewidth]{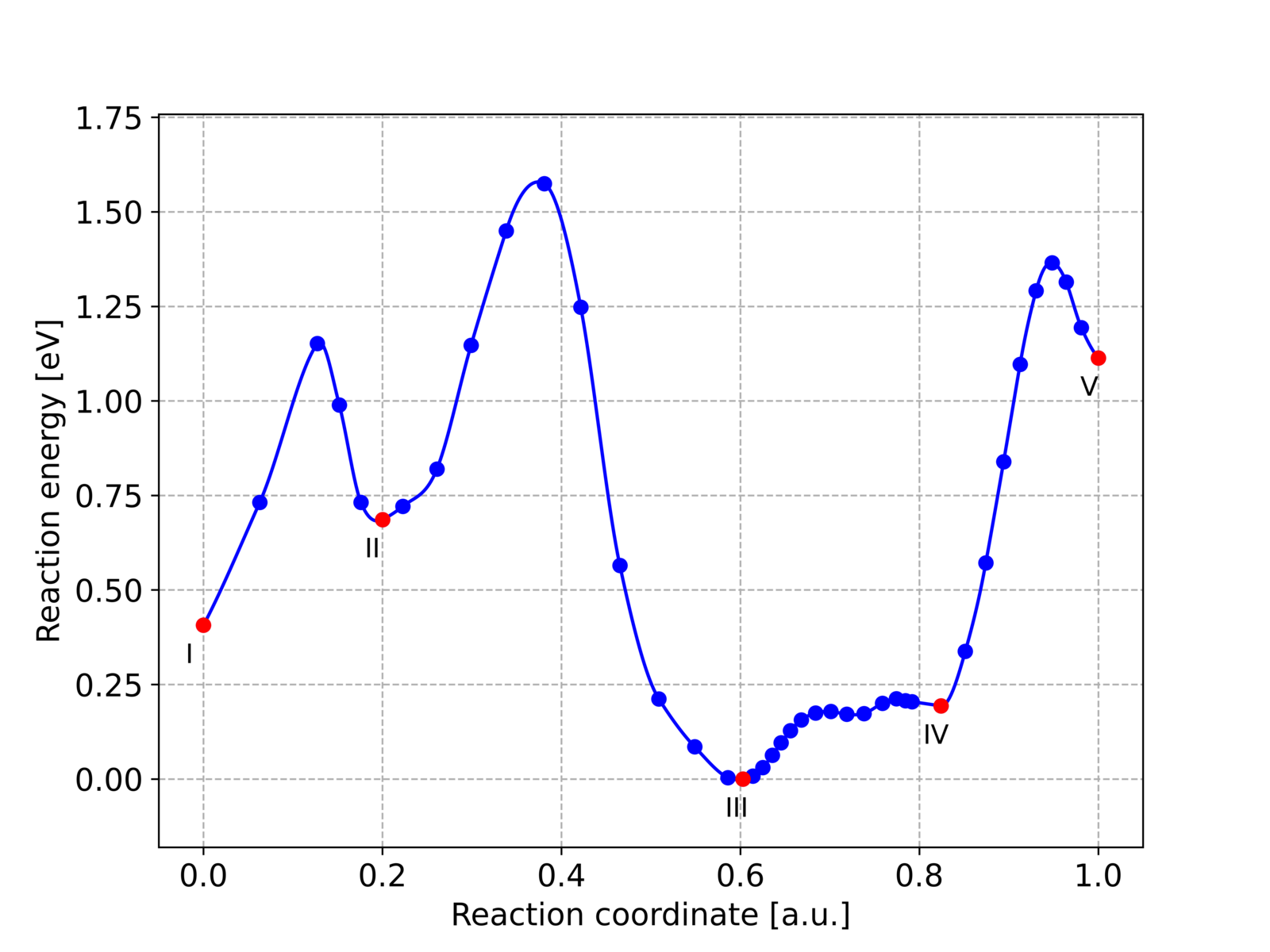}}}
 \caption{DFT kick-in barrier of the Al interstitial to the carbon antisite in the neutral state. The kick-in starts with the formation of an (Al-C)$_{Si}$ split interstitial at point V. Proceeds to the stable Al$_{Si}$C$_I$ complex at III and IV,  detaches into isolated defects at II, with the carbon interstitial diffusing away at I.}
\label{fig:kick-in}
\end{figure}
\begin{table}[ht!]

\centering
\setlength{\tabcolsep}{20pt}   
\begin{tabular}{lc}
\textbf{Process} & Barrier height \\ \hline
$ \text{I}\rightarrow \text{II}$ & $0.74$\,eV ($0.42$\,eV)\\
$ \text{III}\rightarrow \text{II}$ &  $1.57$\,eV ($0.85$\,eV)\\
$ \text{III}\rightarrow \text{IV}$ & $0.21$\,eV ($0.01$\,eV)\\
$ \text{IV}\rightarrow \text{V}$ & $1.16$\,eV ($0.25$\,eV)\\
\hline
\end{tabular}
\caption{Migration barrier heights for the $\mathrm{Al}_\text{Si,h}\mathrm{C}_\text{I,c}$ in Figure \ref{fig:kick-in}. Values in parenthesis denote the reverse direction.}
\label{tab:barriers_DFT_CI_dissociat}
\end{table}
\subsection{Activation energy of diffusion}
\subsubsection{Diffusion in dependence of implantation temperature/damage}
 \begin{figure}[hbtp]
\centering
  \subfloat[]{\fbox{\includegraphics[width=0.48\textwidth]{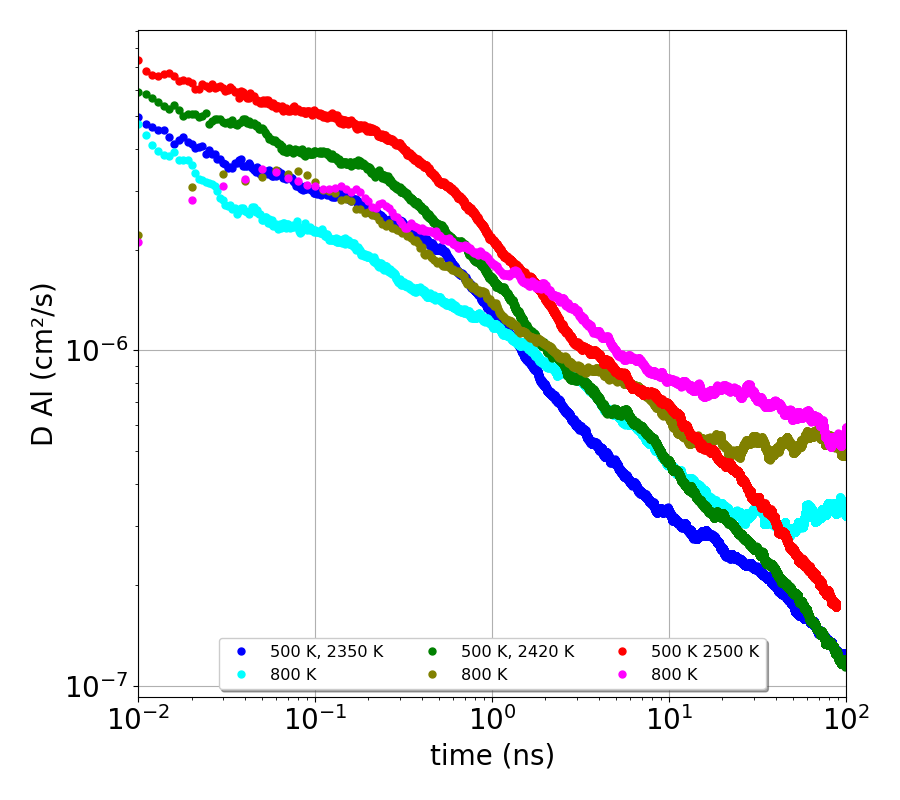}}}\hfill
   \subfloat[]{\fbox{\includegraphics[width=0.48\textwidth]{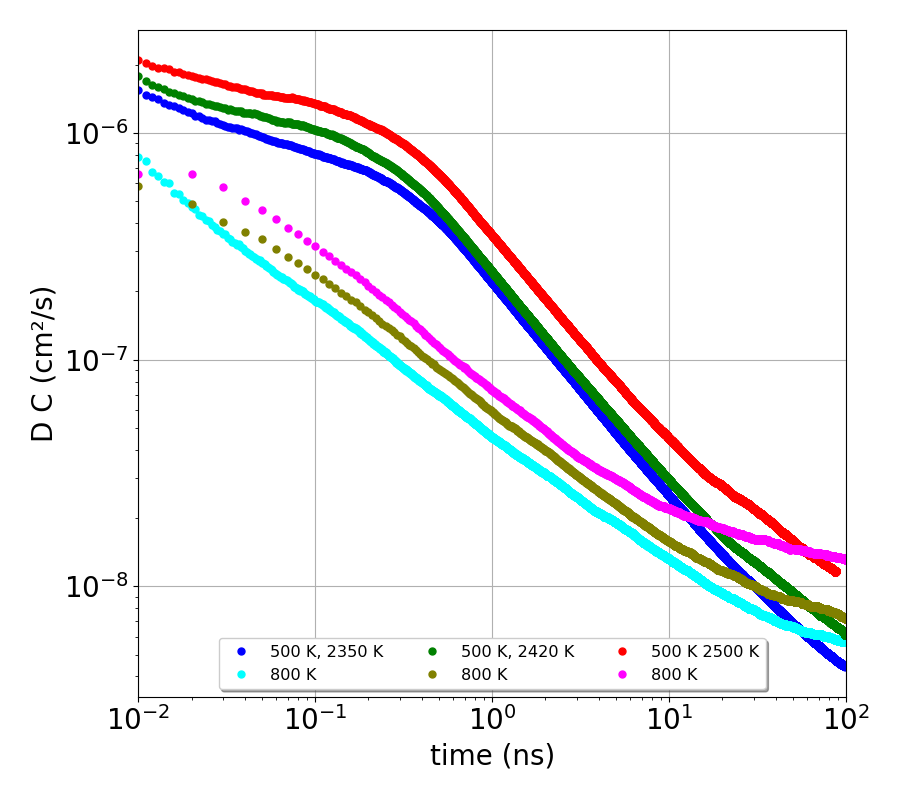}}}
   \caption{Diffusivity of Al (a) and C (b) at high annealing temperatures between $2350$-$2500$\,K for dose $7.5\times$10$^{14}$\,cm$^{-2}$ and implantation temperatures of $500$ and $900$\,K.}
\label{fig:Diff_dose_10}
\end{figure} 
In the system with low implantation temperature, recrystallization within the first nanosecond appears as a plateau in Al diffusivity. A secondary, weaker plateau, coinciding with large-cluster ripening, appears between $5$ and $40$\,ns depending on temperature. In the system with higher implantation temperature, this second plateau is much more dominant and persists from $10$\,ns until the end of annealing, indicating significant mass transfer during the formation of extended defects.
\begin{table}[ht!]
\centering
\setlength{\tabcolsep}{9pt}   % adjust once; applies to all columns equally
\caption{Time-dependent diffusion activation energies for Al, C, and Si interstitials for dose $7.5\times$10$^{14}$\,cm$^{-2}$ and implantation temperatures $500$ and $900$\,K at high annealing temperatures between $2100$ and $2500$\,K.}
\begin{tabular}{cc|ccc}%\hline
\textbf{Temperature} (K) & \textbf{time} (ns) & E$_{Al}$ (eV)& E$_{Si}$ (eV) & E$_{C}$ (eV) \\ \hline\hline
500 & 0.1  & 1.2 & 1.4 & 1.4\\
900 & 0.1  & 1.5 & 1.3 & 1.3\\
\hline
500 & 5    & 1.5 & 1.2 & 1.2 \\
900 & 5    & 1.4 & 1.3 & 1.3\\
\hline
500 & 30   & 1.9 & 2.0 & 2.0\\
900 & 30   & 2.2 & 2.1 & 2.1\\
\hline
500 & 100  & 1.9 & 2.9 & 2.7\\
900 & 100  & 2.7 & 2.7 & 2.5\\
% \hline
\end{tabular}
\label{tab:activation_energy_diff}
\end{table}
\subsubsection{Al-C complex binding energies}
The transient diffusivities shown below support the kinetic interpretation of Al$_{Si}$C$_I$ complex dissolution, discussed in the main text.
\begin{figure}[htbp]
\centering
\subfloat[]{\fbox{\includegraphics[width=.48\linewidth]{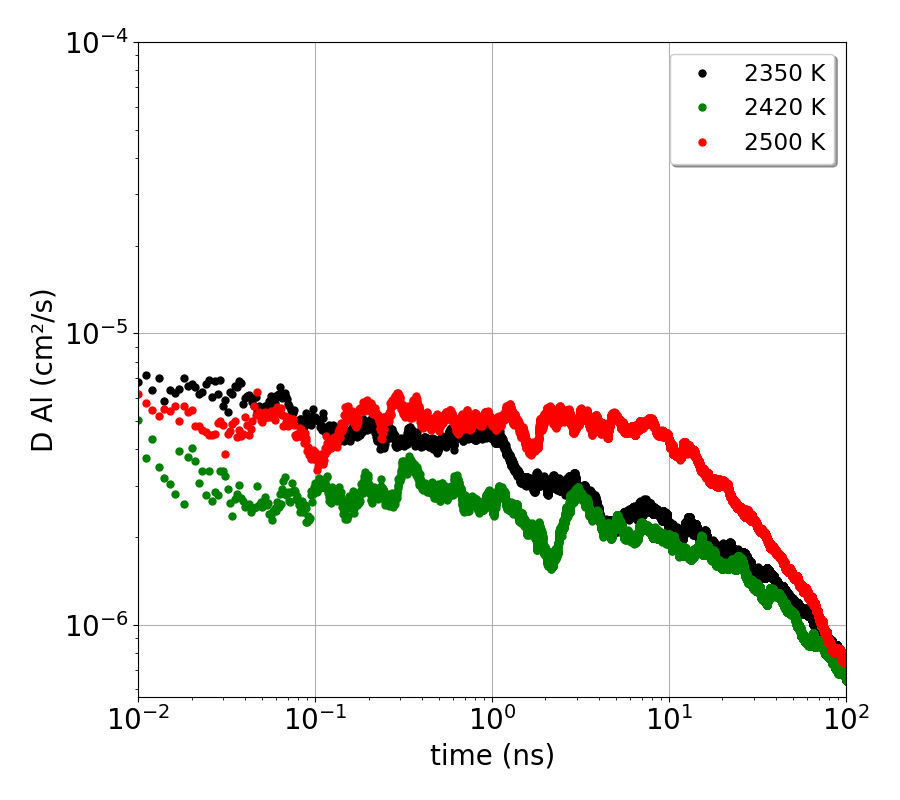}}}
\hfill
\subfloat[]{\fbox{\includegraphics[width=.48\linewidth]{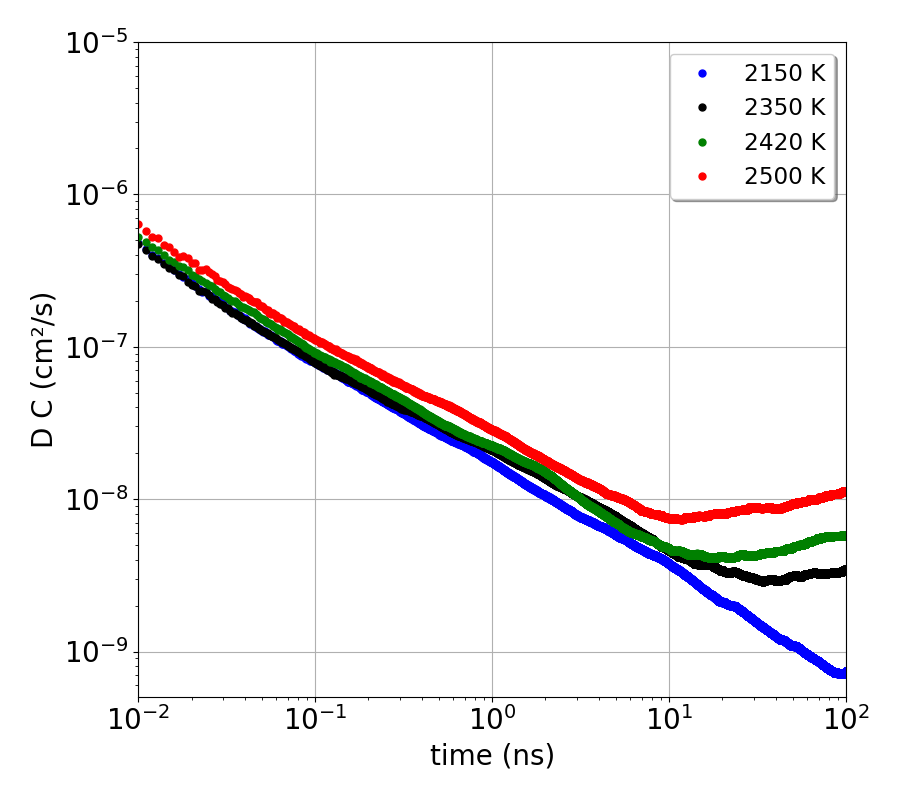}}}
%\subfloat[]{\fbox{\includegraphics[width=.3\linewidth]{figures/Ahrrenius_diffusion_C_paper.png}}}
  \caption{Transient enhanced diffusion of Al a) and C b) during cluster growth and dissolution at high annealing temperatures and dose \mbox{$5\times10^{13}$}\,cm$^{-2}$. %c) Arrhenius plot for the carbon interstitial after $100$\,ns annealing with activation energy for migration of $3.2$\,eV and pre-factor $2.8$$\times$10$^{-2}$cm$^2$/s.
  }
\label{fig:diffusion}
\end{figure}
\newpage
\subsection{Transition from planar clusters to faulted loops}
\subsubsection{Cluster shape}
Planar clusters in the basal plane play a role in the formation of stacking faults. Stacking faults arise from shear stresses caused by the incorporation of planar defects into the crystal lattice.  Shear stresses along the basal plane can cause a change in the sequence of the crystalline layers, leading to the formation of intrinsic stacking faults, or, at higher stresses and temperatures, insert an additional layer into the basal plane, which corresponds to an extrinsic stacking fault. A so-called dislocation line forms at the edges of the stacking fault, which contains the residual defects of the clusters that generated it. 

The factor $S$ to define the shape of an interstitial cluster is given as
\begin{equation}
    S= \frac{2\Delta z}{\Delta x + \Delta y}
    \label{eqn:shape}
\end{equation}
with $\Delta x$, $\Delta y$ and $\Delta z$ describing the cluster extensions in the $x$, $y$ (basal plane) and $z$ (c-axis) direction.  A shape factor of $1$ would mean a perfectly spherical cluster, while a factor of $0$ would indicate a perfect plane. In practice, a shape factor $0.5<S<1.5$ indicates a globular cluster, while $S\le0.5$ and $S\geq1.5$ describe planar clusters in the basal plane and along the c-axis, respectively.
\begin{figure}[htbp]
\centering
\subfloat[]{\fbox{\includegraphics[width=.48\linewidth]{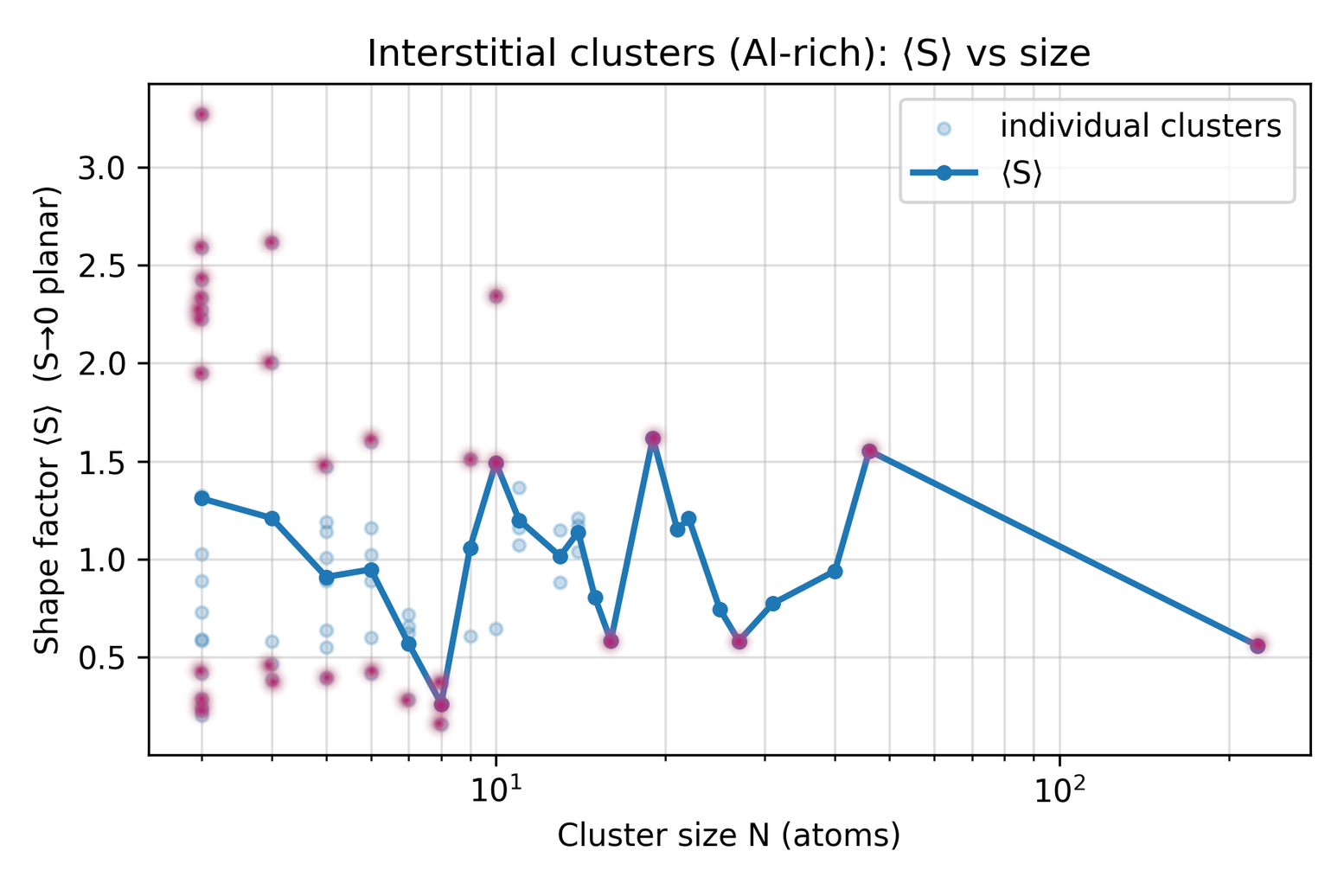}}}\hfill
\subfloat[]{\fbox{\includegraphics[width=.48\linewidth]{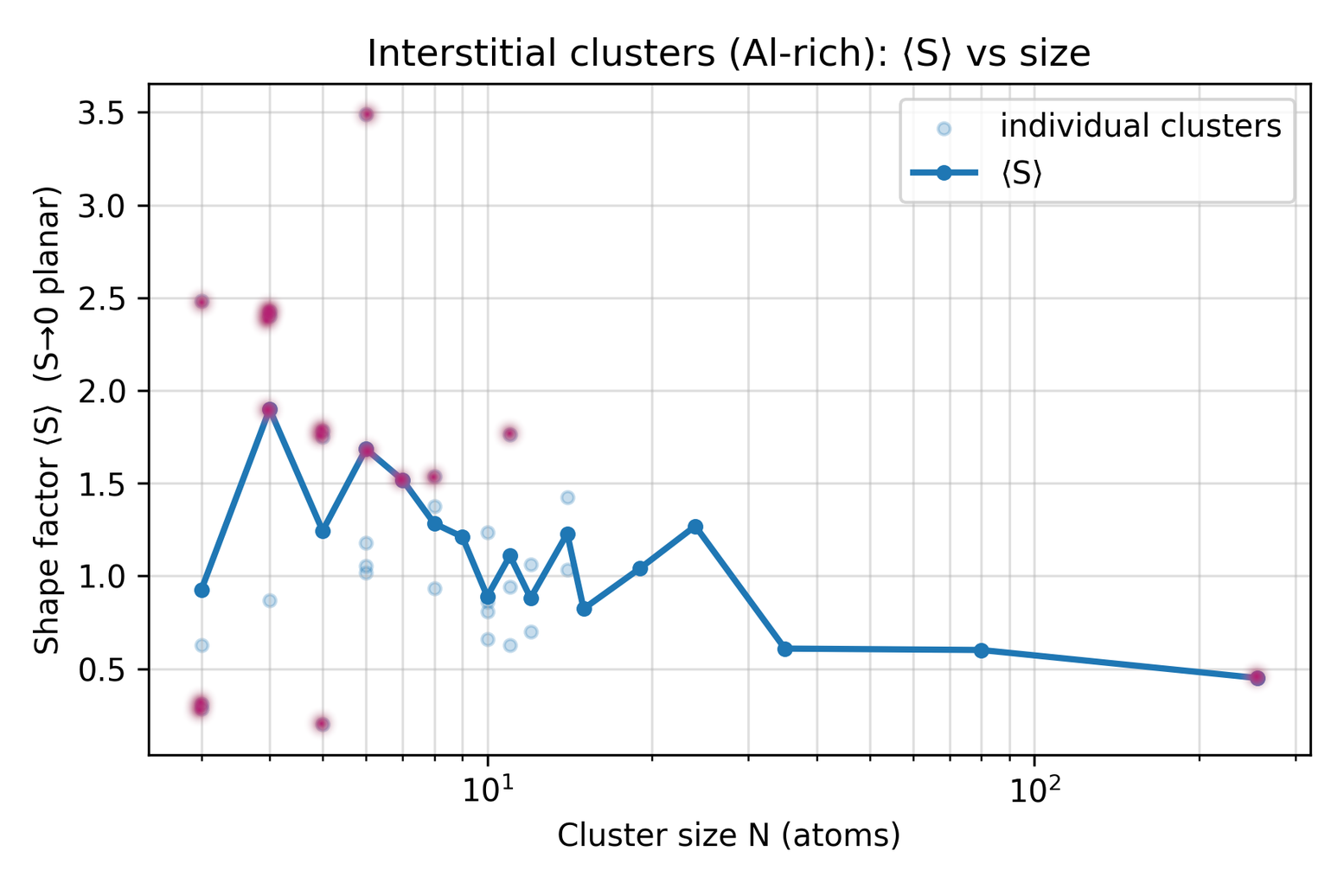}}}\\
\subfloat[]{\fbox{\includegraphics[width=.48\linewidth]{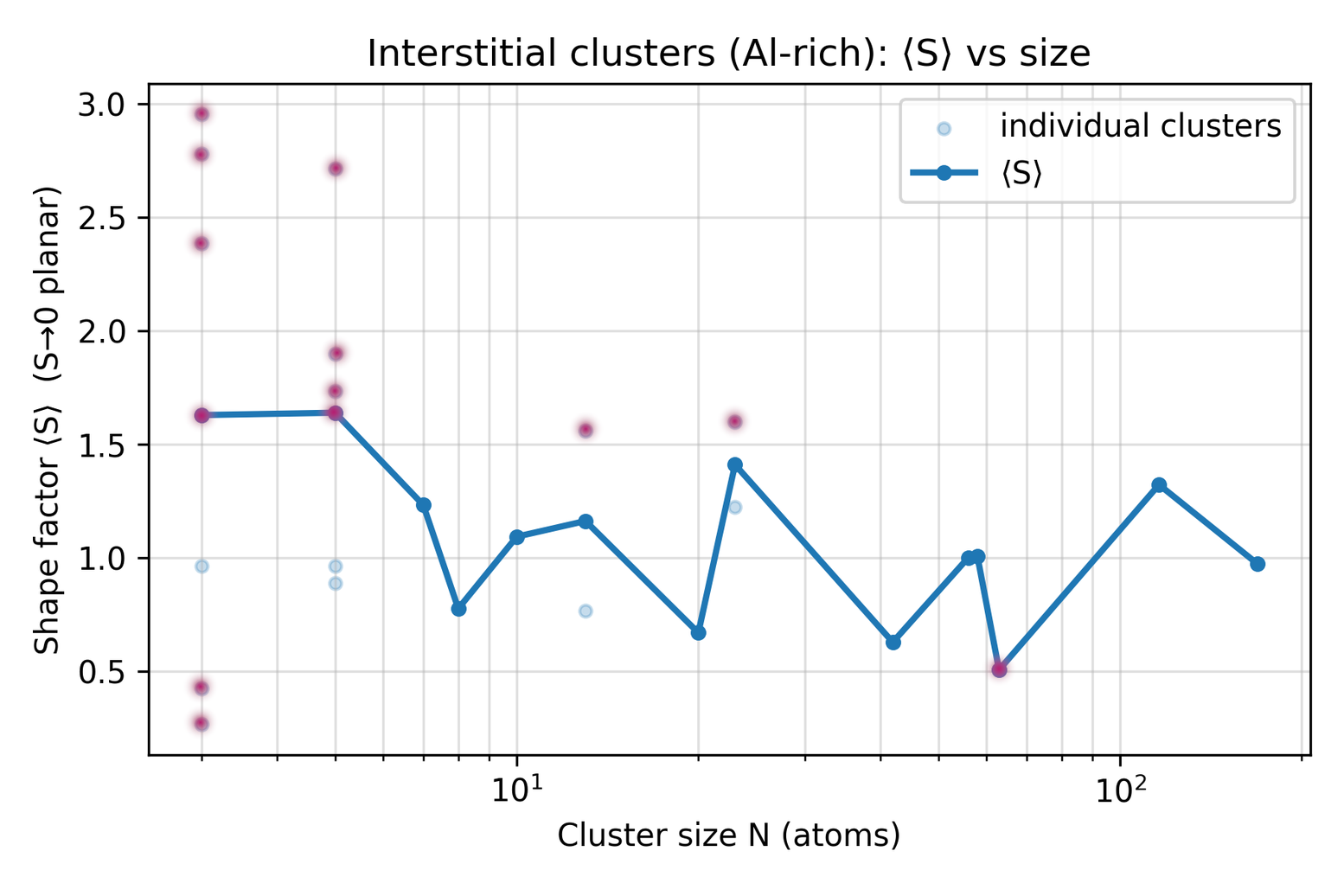}}}\hfill
\subfloat[]{\fbox{\includegraphics[width=.48\linewidth]{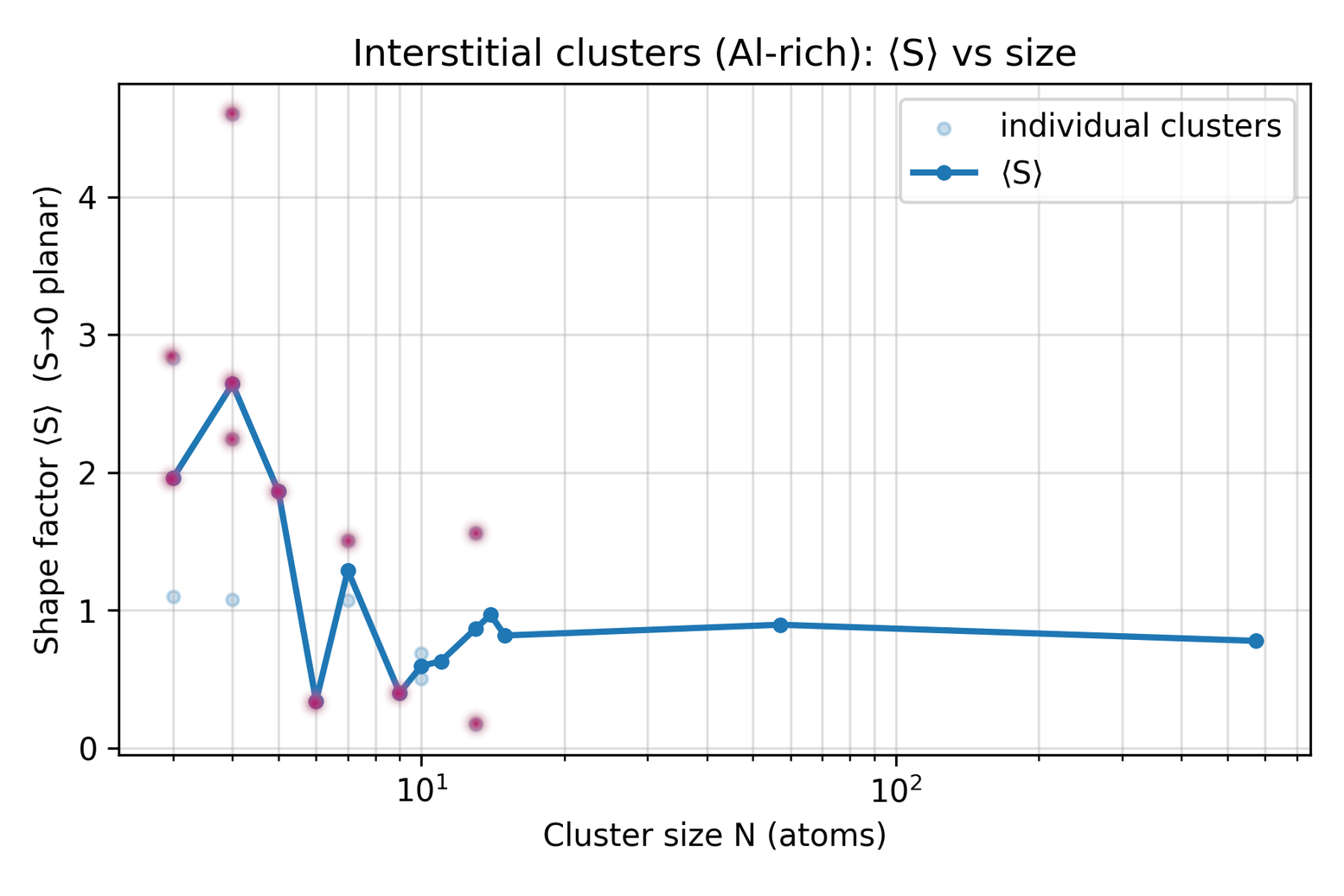}}}\\
\subfloat[]{\fbox{\includegraphics[width=.48\linewidth]{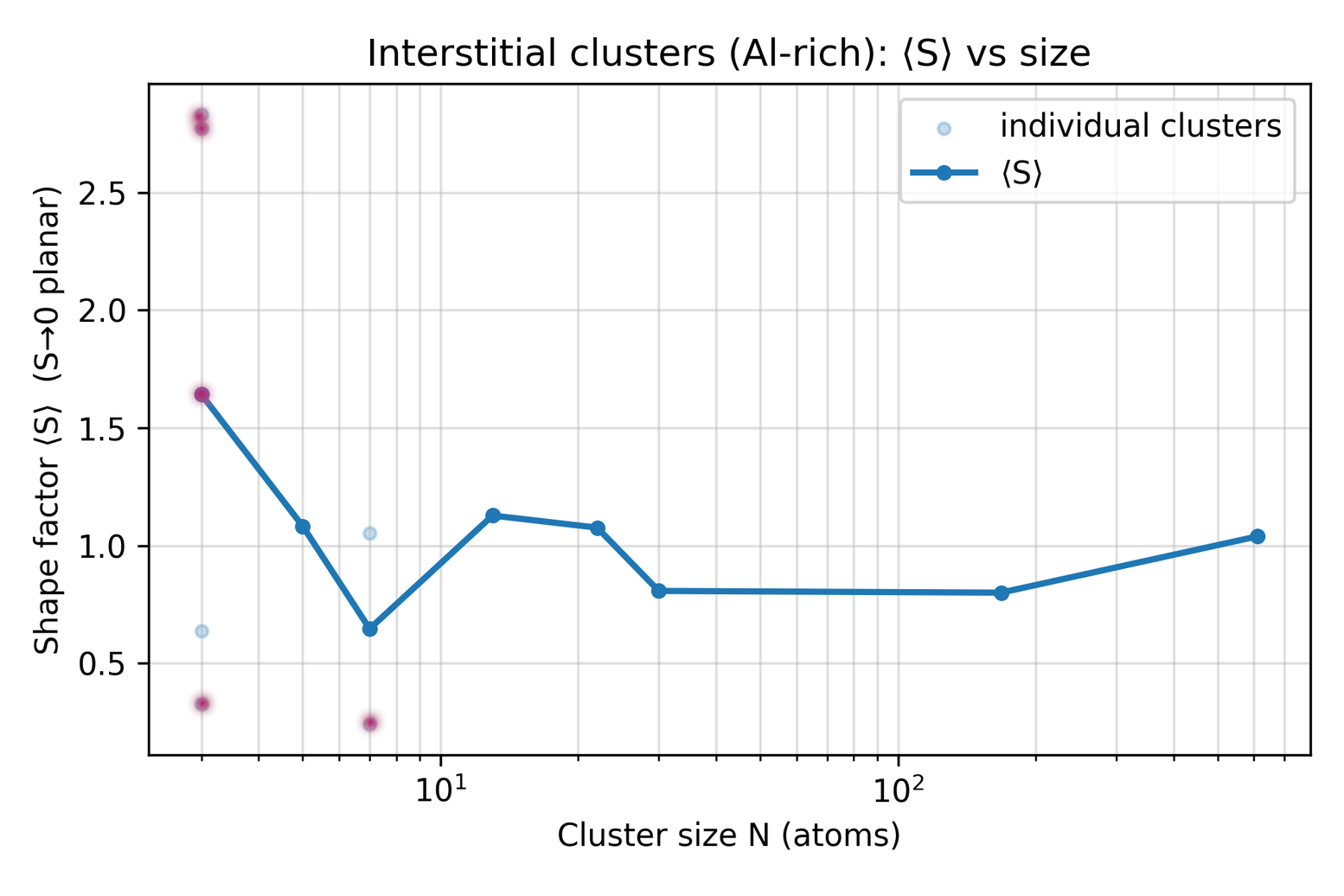}}}\hfill
\subfloat[]{\fbox{\includegraphics[width=.48\linewidth]{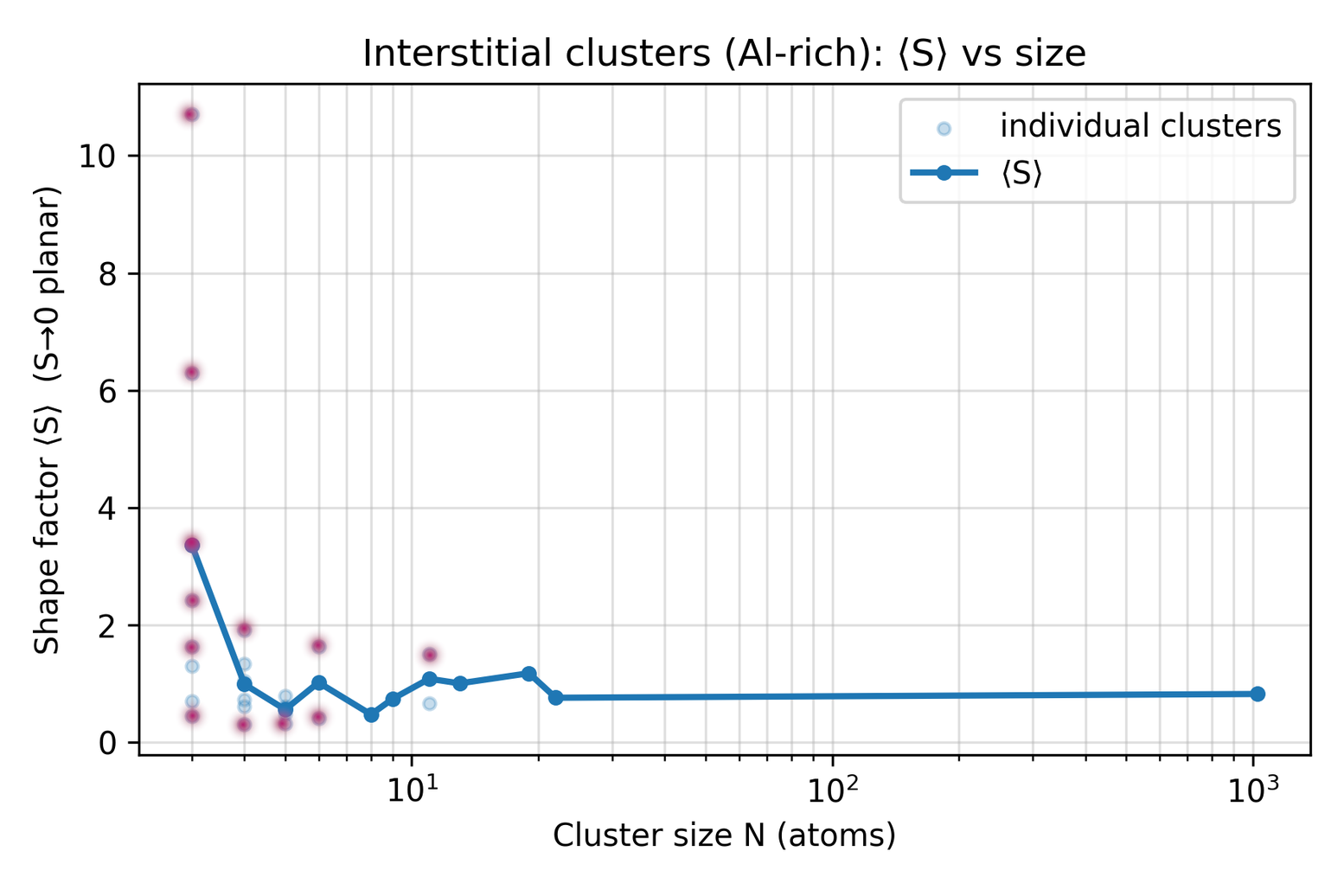}}}
  \caption{Shape factors $S$ for dose \mbox{$7.5\times10^{14}$\,cm$^{-2}$} and implantation temperature $500$\,K at annealing temperatures a) $1500$\,K, b) $2000$\,K, c) $2150$\,K, d) $2350$\,K, e) $2420$\,K, and f) $2500$\,K. Shown are the shape factors of the individual clusters after $100$\,ns of annealing, where the line provides the average of $S$ for the chosen cluster size. Shape factors in magenta indicate planar shaped clusters orientated along the c-axis (S>1.5) or lying in the basal plane (S<0.5), shape factors in blue indicate globular shaped clusters.}
\label{fig:shape_500K_dose10}
\end{figure}
\begin{figure}[htbp]
\centering
\subfloat[]{\fbox{\includegraphics[width=.48\linewidth]{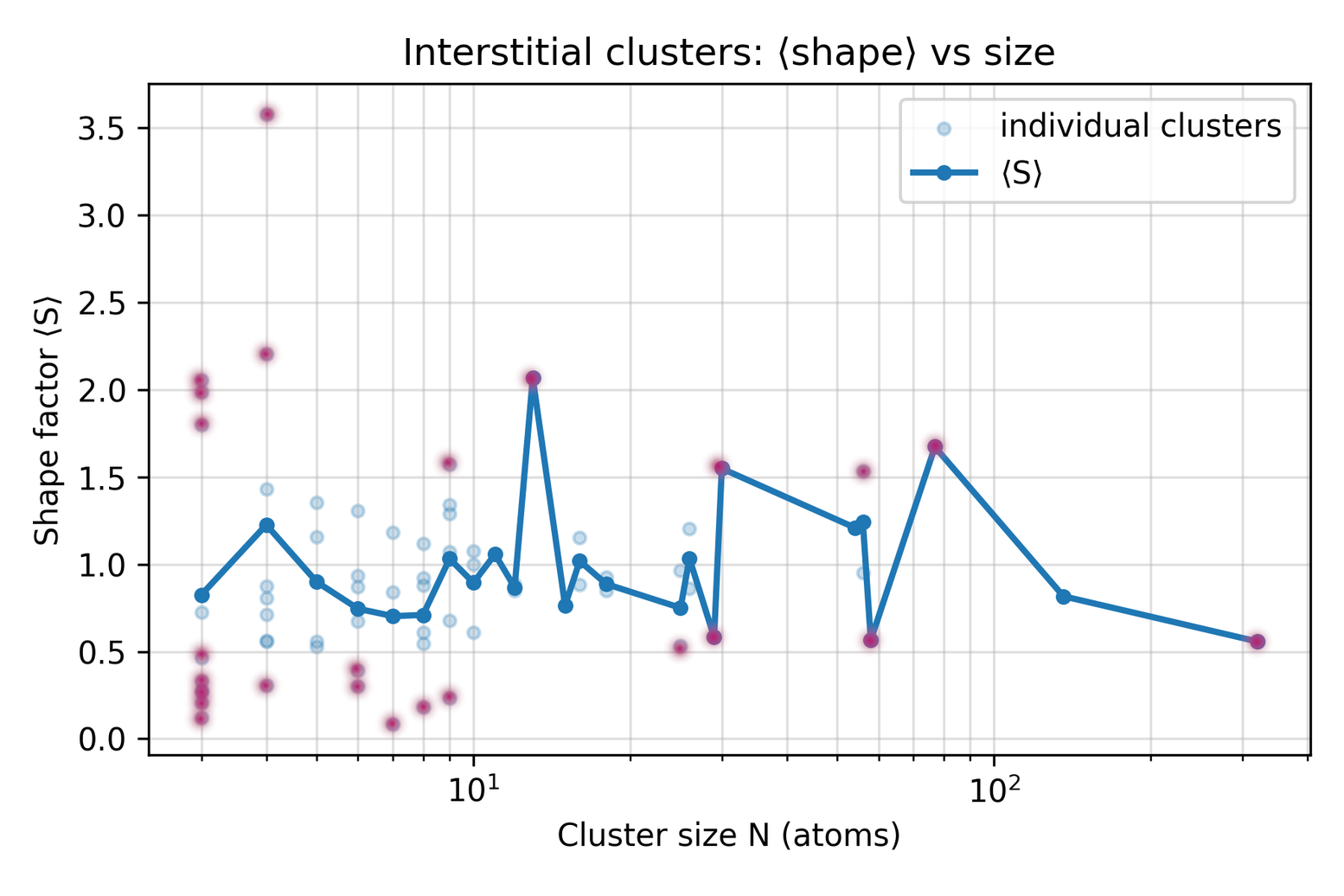}}}\hfill
\subfloat[]{\fbox{\includegraphics[width=.48\linewidth]{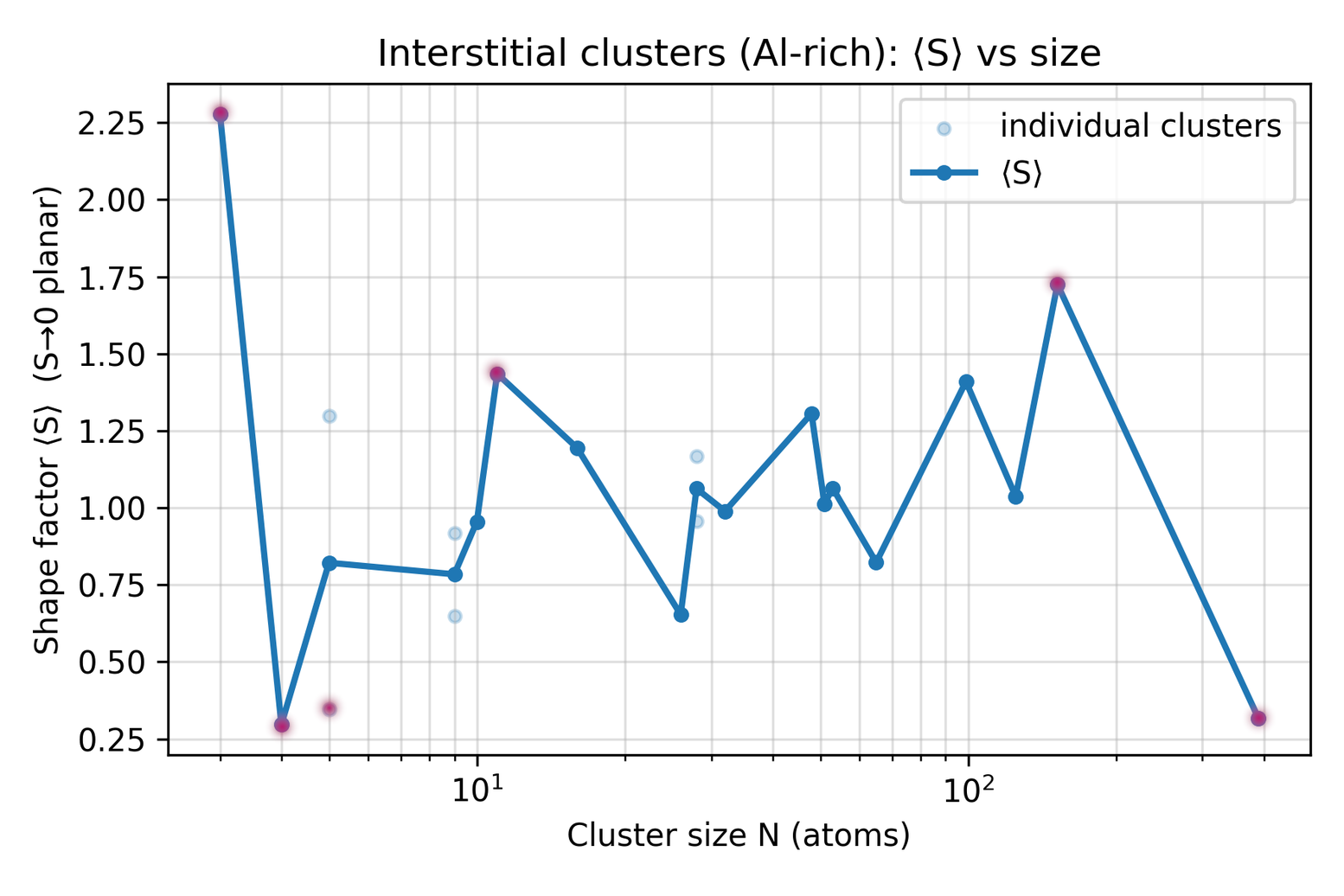}}}\\
\subfloat[]{\fbox{\includegraphics[width=.48\linewidth]{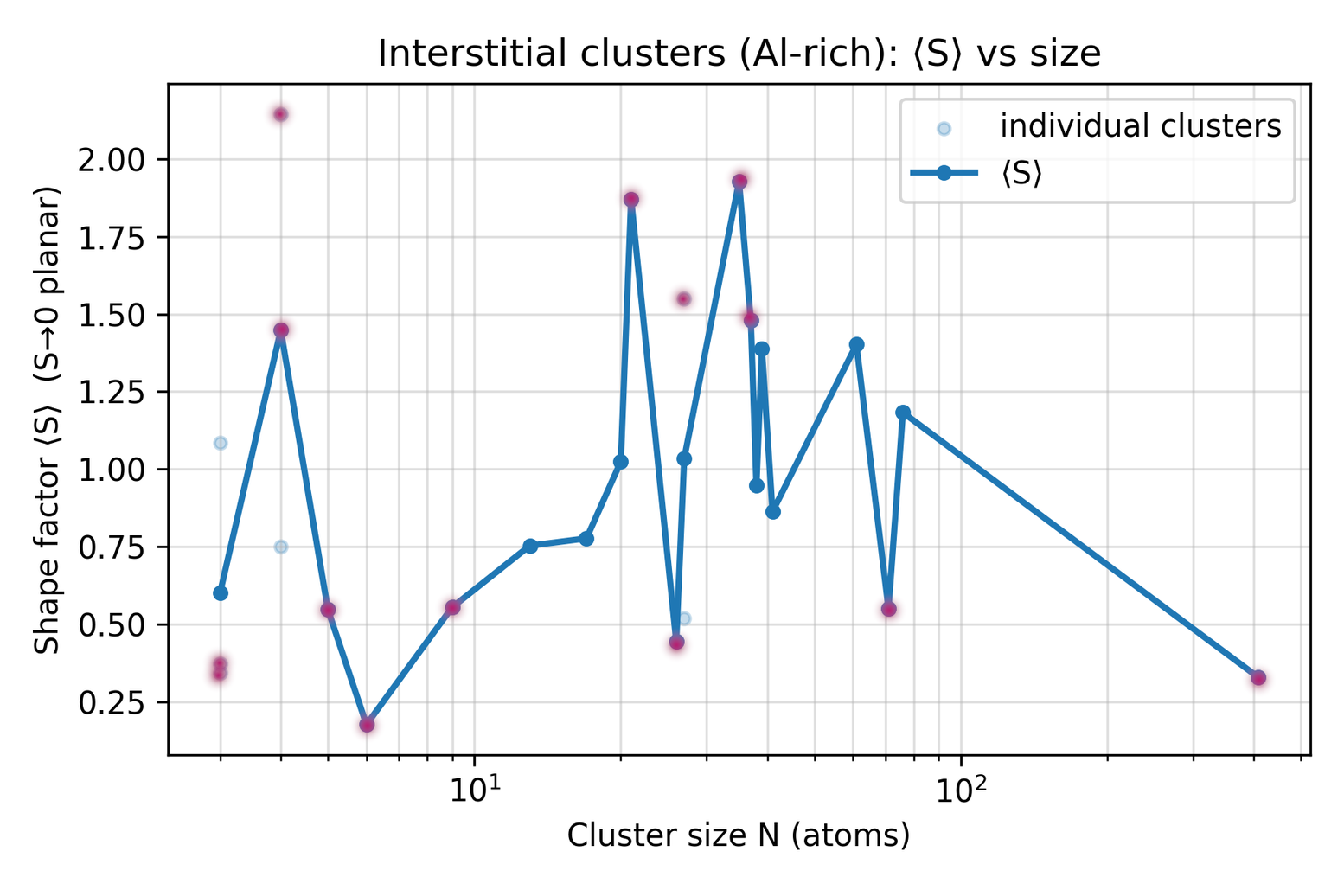}}}\hfill
\subfloat[]{\fbox{\includegraphics[width=.48\linewidth]{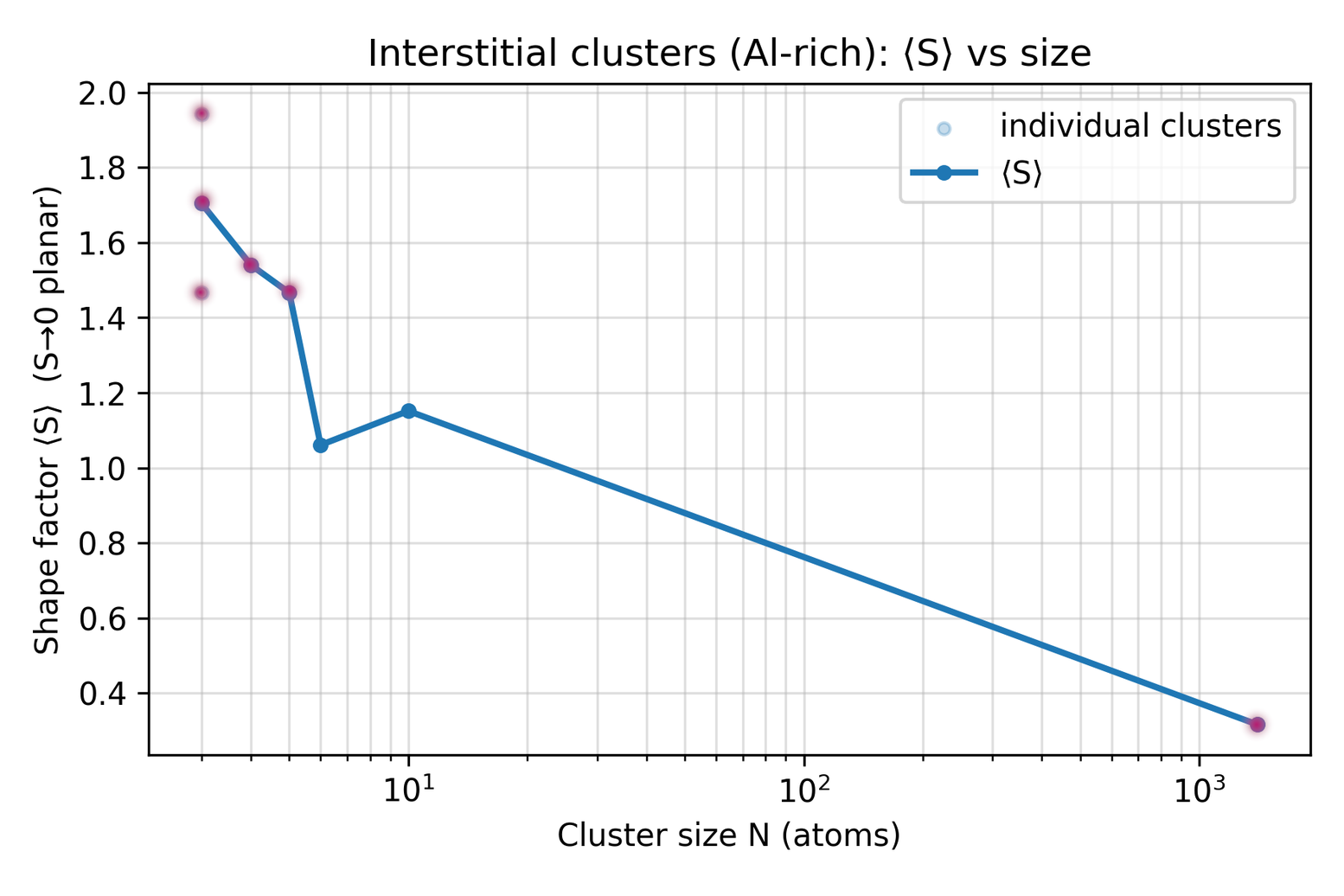}}}\\
\subfloat[]{\fbox{\includegraphics[width=.48\linewidth]{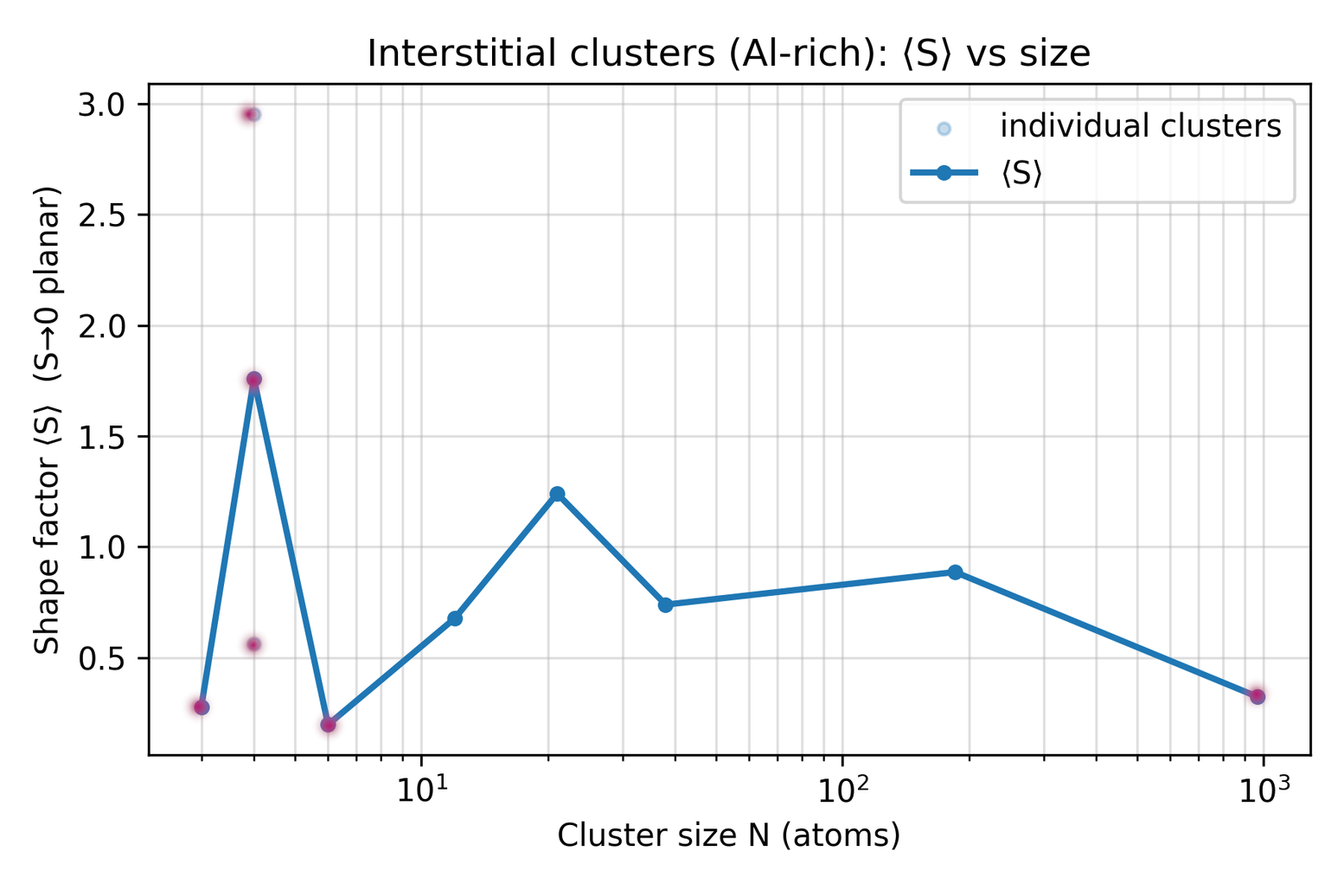}}}\hfill
\subfloat[]{\fbox{\includegraphics[width=.48\linewidth]{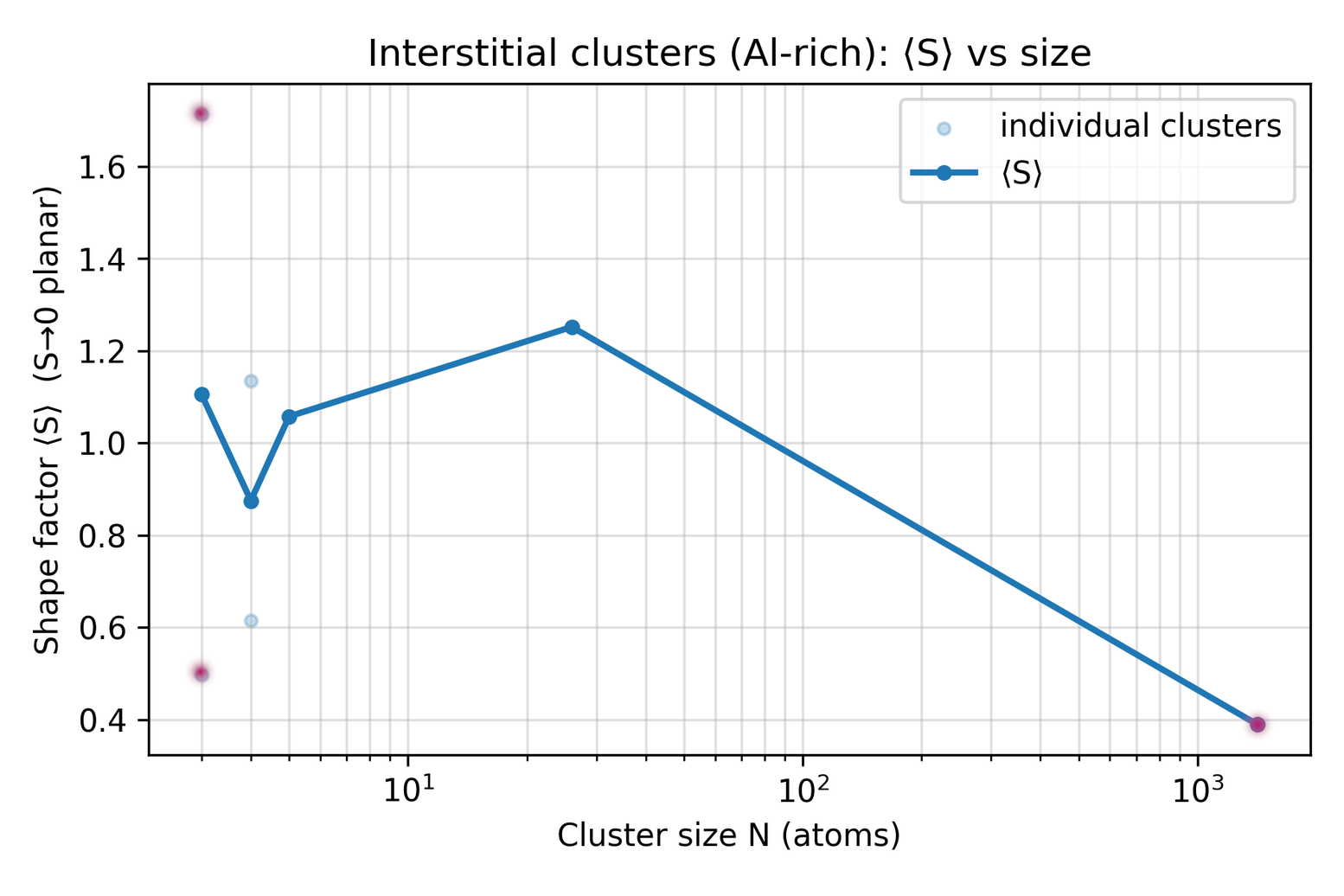}}}
\caption{Shape factors $S$ for dose \mbox{$7.5\times10^{14}$\,cm$^{-2}$} and implantation temperature $900$\,K at annealing temperatures a) $1500$\,K, b) $2000$\,K, c) $2150$\,K, d) $2350$\,K, e) $2420$\,K, and f) $2500$\,K. Shown are the shape factors of the individual clusters after $100$\,ns of annealing, where the line provides the average of the individual $S$ for the chosen cluster size. Shape factors in magenta indicate planar shaped clusters orientated along the c-axis ($S>1.5$) or lying in the basal plane ($S<0.5$), shape factors in blue indicate globular shaped clusters.}
 \label{fig:shape_900K_dose10}
\end{figure}
Figure~\ref{fig:shape_500K_dose10} and Figure~\ref{fig:shape_900K_dose10} show the shape factors of all clusters at the end of annealing, broken down by cluster size, for implantation temperatures of $500$ and $900$\,K, respectively. The annealing temperatures range from $1500$\,K to $2500$\,K in panels a-f. At low temperatures or short annealing times, the system is dominated by a large number of medium-sized clusters, which gradually diminish with rising temperature. 
Noticeably, small clusters up to a size of $10$ are preferentially arranged in planar structures either in the basal plane or along the c-axis, displayed by magenta spheres. For medium-sized clusters ($10\leq\mathrm{size}<100$), planar clusters occur only at temperatures below $2200$\,K (a-c), while at higher temperatures (d-f), the remaining clusters are globular. The planar structures occur most frequently around cluster sizes of $20$ and $60$ defects. The shapes of the large clusters ($\mathrm{size}\geq 100$) differ significantly in Figure~\ref{fig:shape_500K_dose10} and Figure~\ref{fig:shape_900K_dose10}: while even clusters with several hundred defects remain predominantly globular at low implantation temperatures, large defect clusters at $900$\,K are all planar. The only exception are clusters at annealing temperatures below $2100$\,K in Figure~\ref{fig:shape_500K_dose10}(a,b). In these systems, planar clusters with about $200$ defects can be observed even at $500$\,K.
The planar defects that form at low implantation temperatures are faulted clusters primarily representing intrinsic stacking faults enclosed by Shockley partials. Since these defects are sessile and less thermally stable, they disappear at annealing temperatures above $2000$\,K.
At high implantation temperatures, however, thermally very stable Frank-type partials form, by inserting additional cubic crystalline domains into the basal plane.

\subsubsection{Critical cluster size}
\begin{sidewaysfigure}
\centering
\subfloat[]{%
  \begin{minipage}[t]{.242\linewidth}
    \centering
    \includegraphics[width=\linewidth]{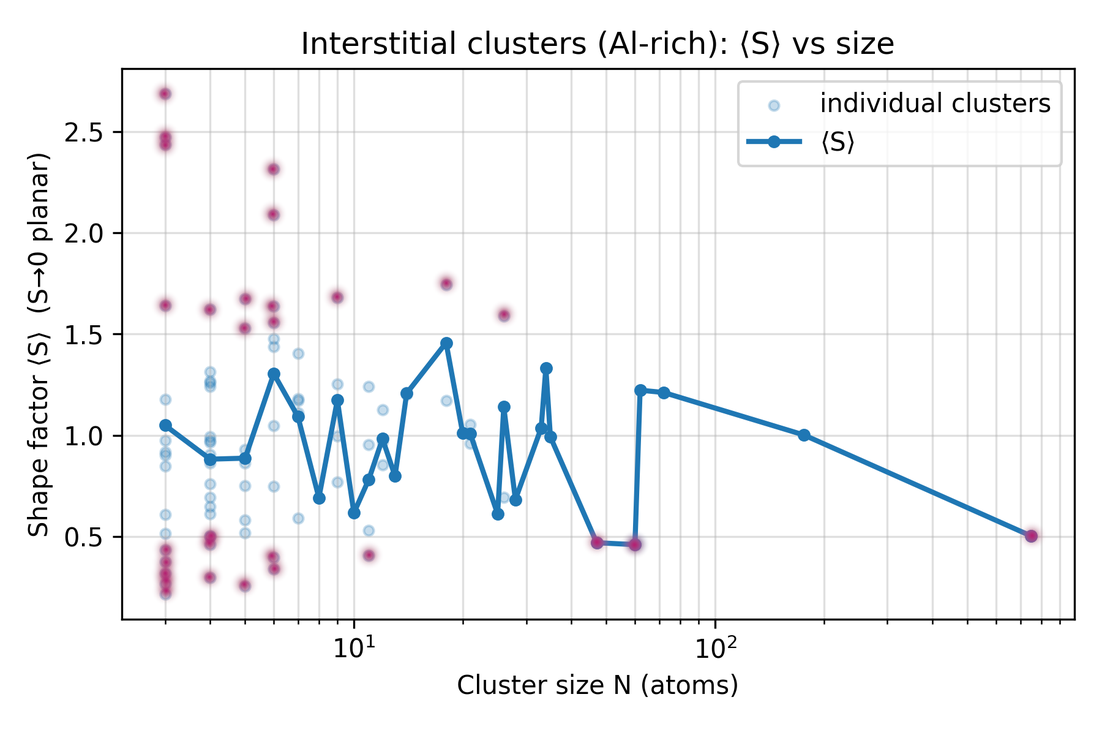}\\[1mm]
     \includegraphics[width=\linewidth]{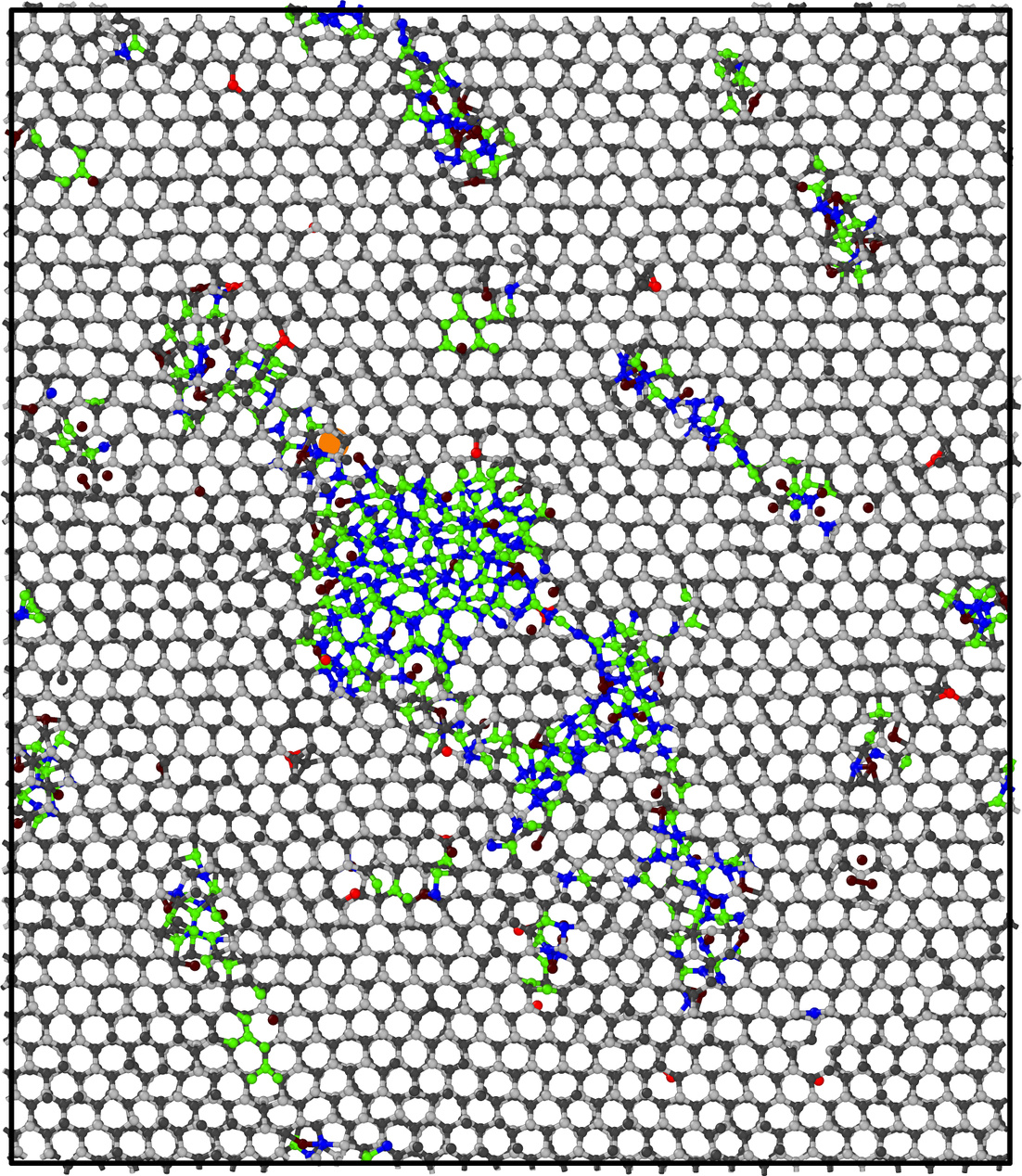}\\
    \includegraphics[width=\linewidth]{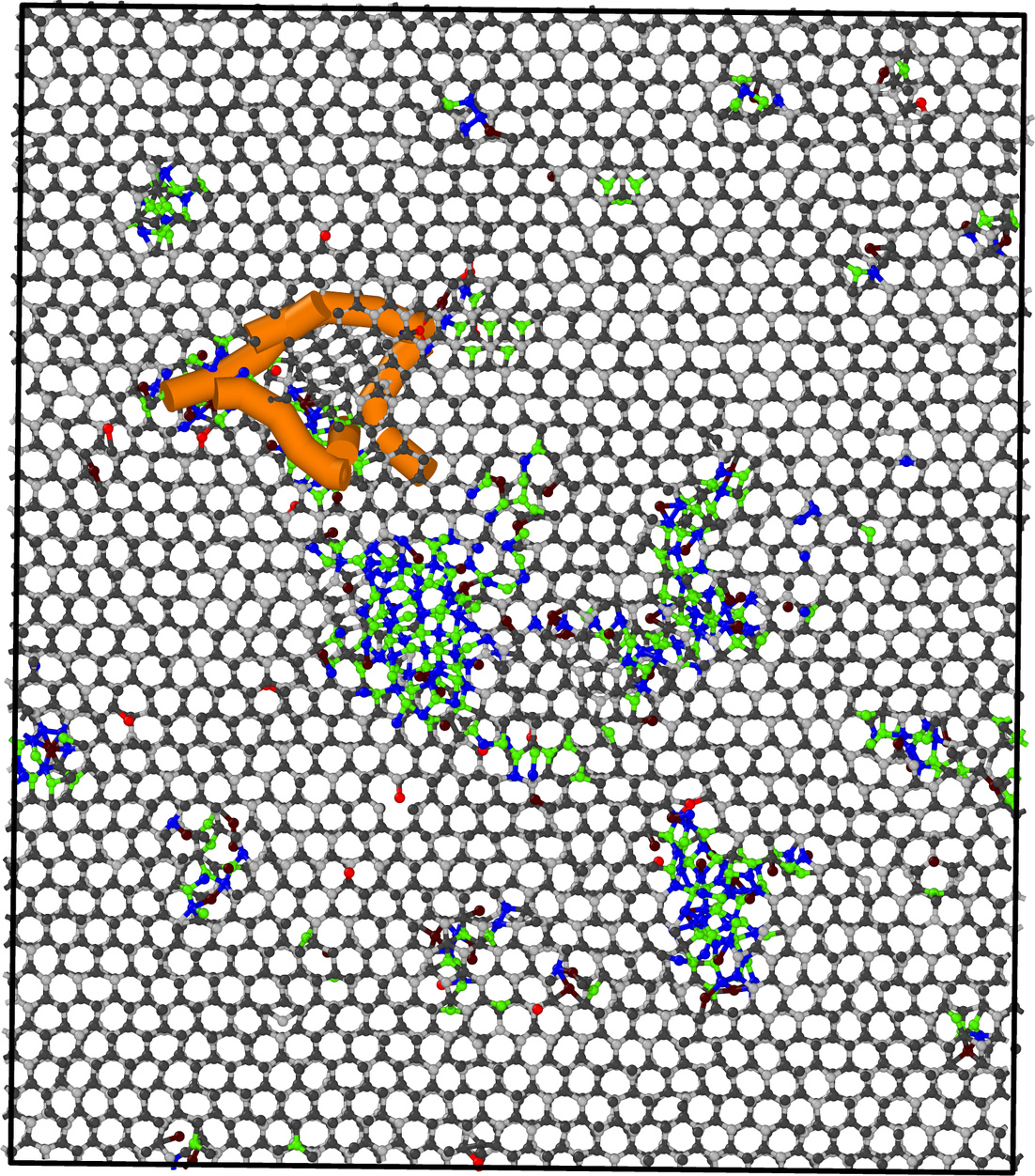}
  \end{minipage}
}\hfill
\subfloat[]{%
  \begin{minipage}[t]{.242\linewidth}
    \centering
    \includegraphics[width=\linewidth]{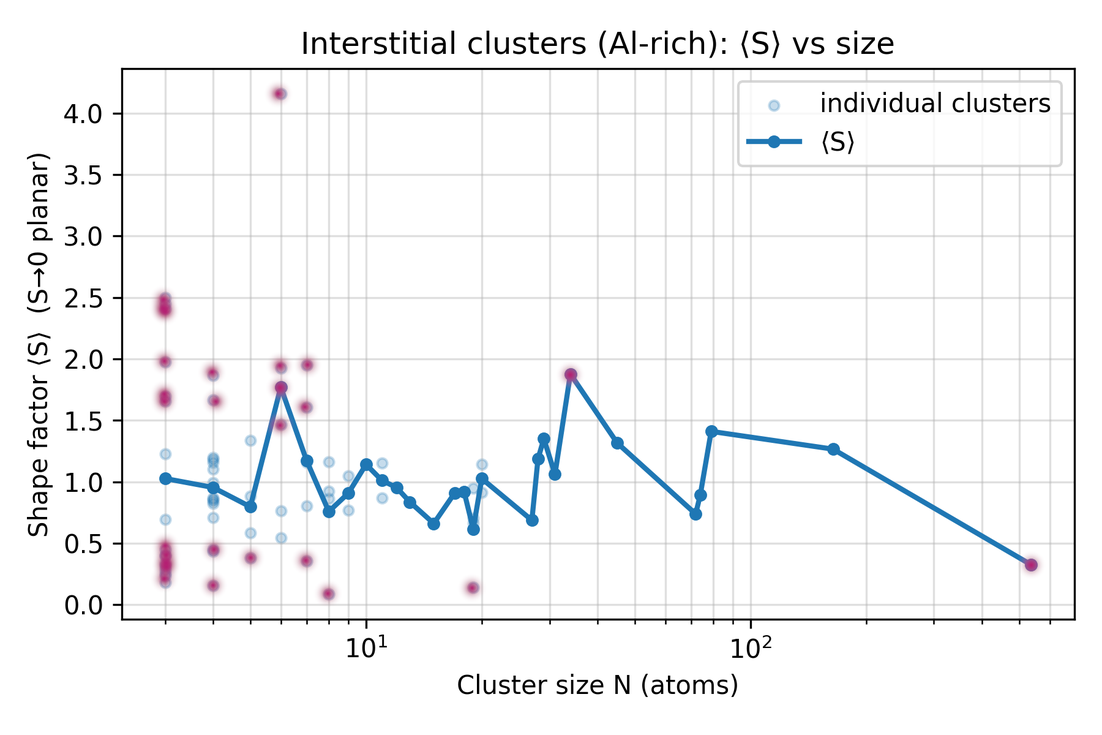}\\[1mm]
    \includegraphics[width=\linewidth]{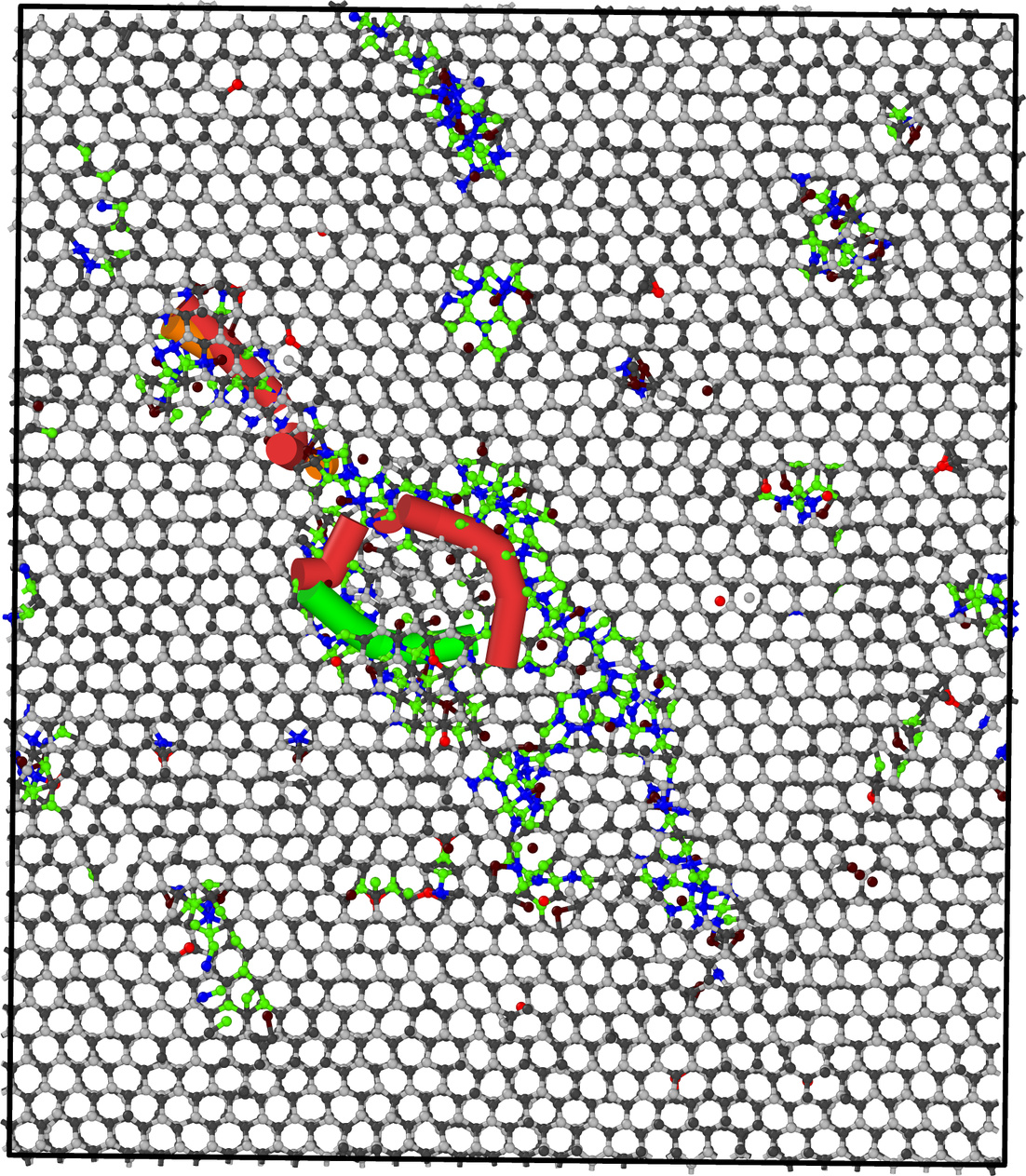}\\
     \includegraphics[width=\linewidth]{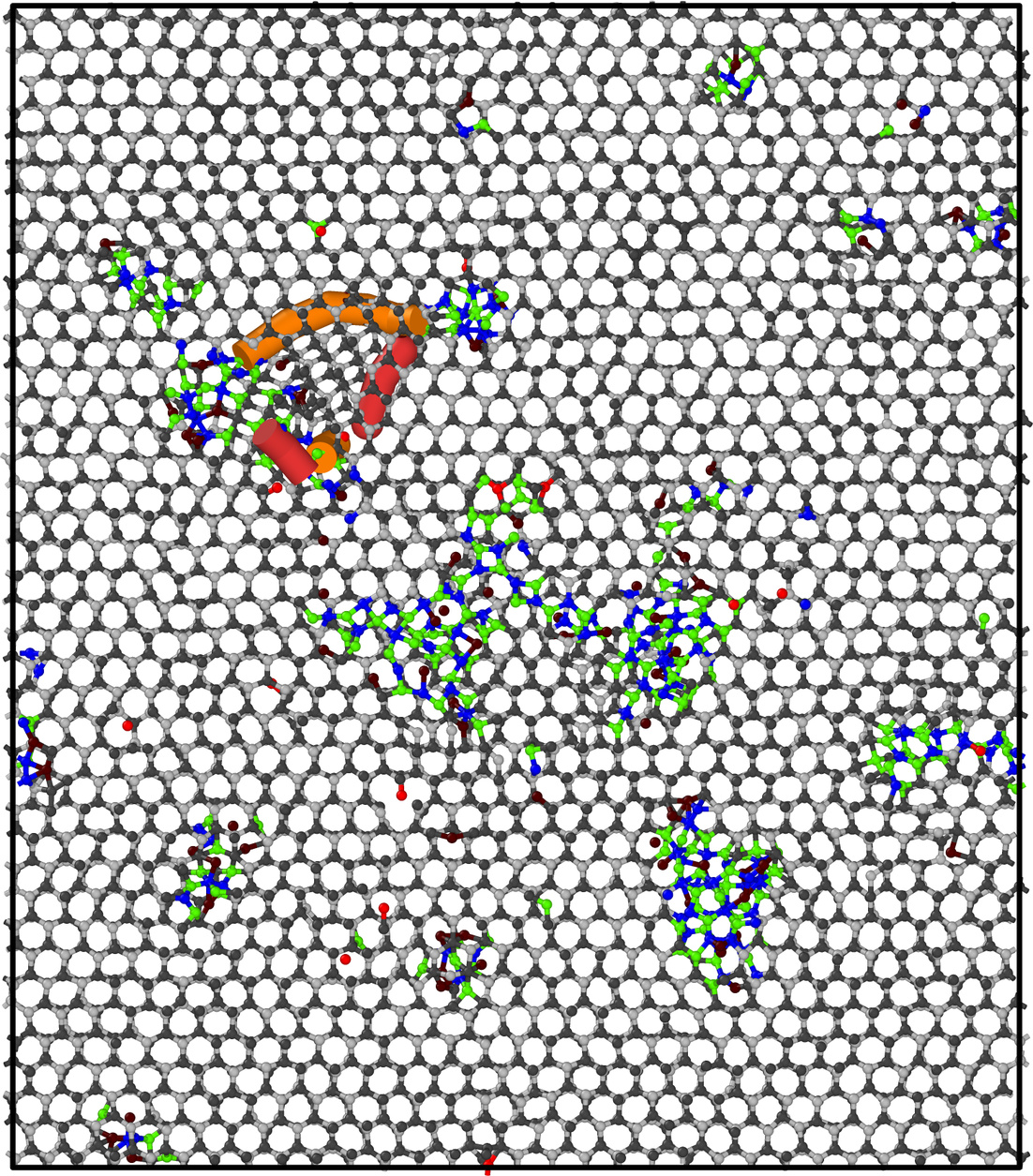}
  \end{minipage}
}\hfill
\subfloat[]{%
  \begin{minipage}[t]{.242\linewidth}
    \centering
    \includegraphics[width=\linewidth]{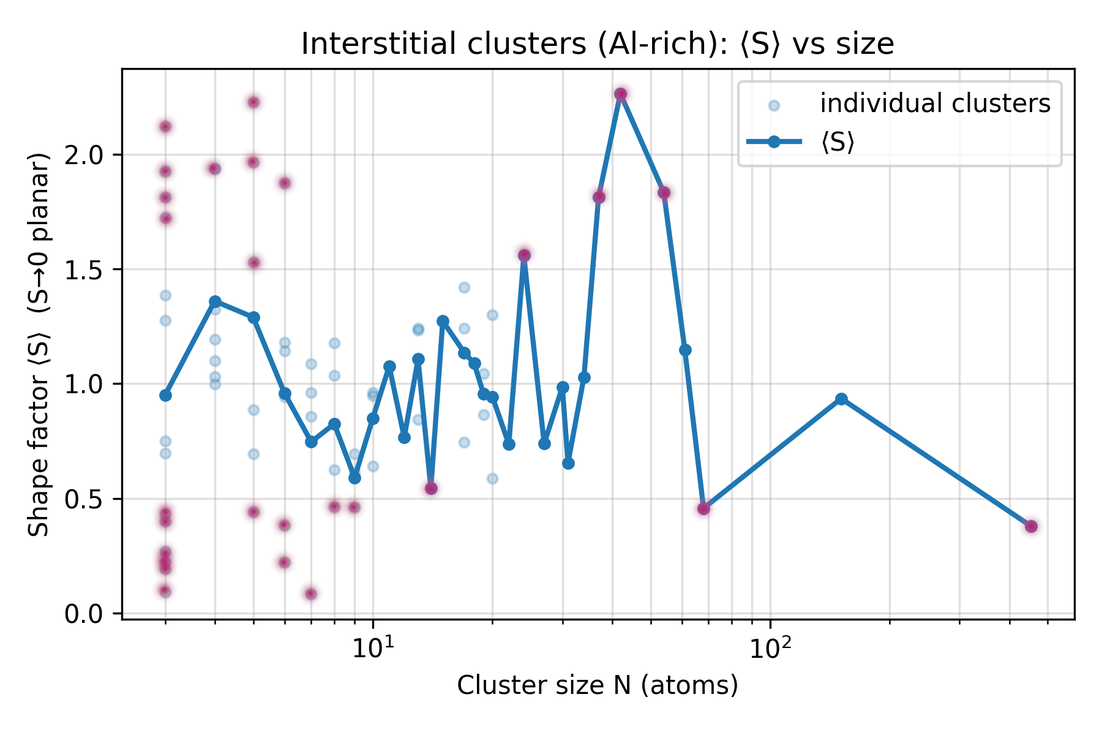}\\[1mm]
    \includegraphics[width=\linewidth]{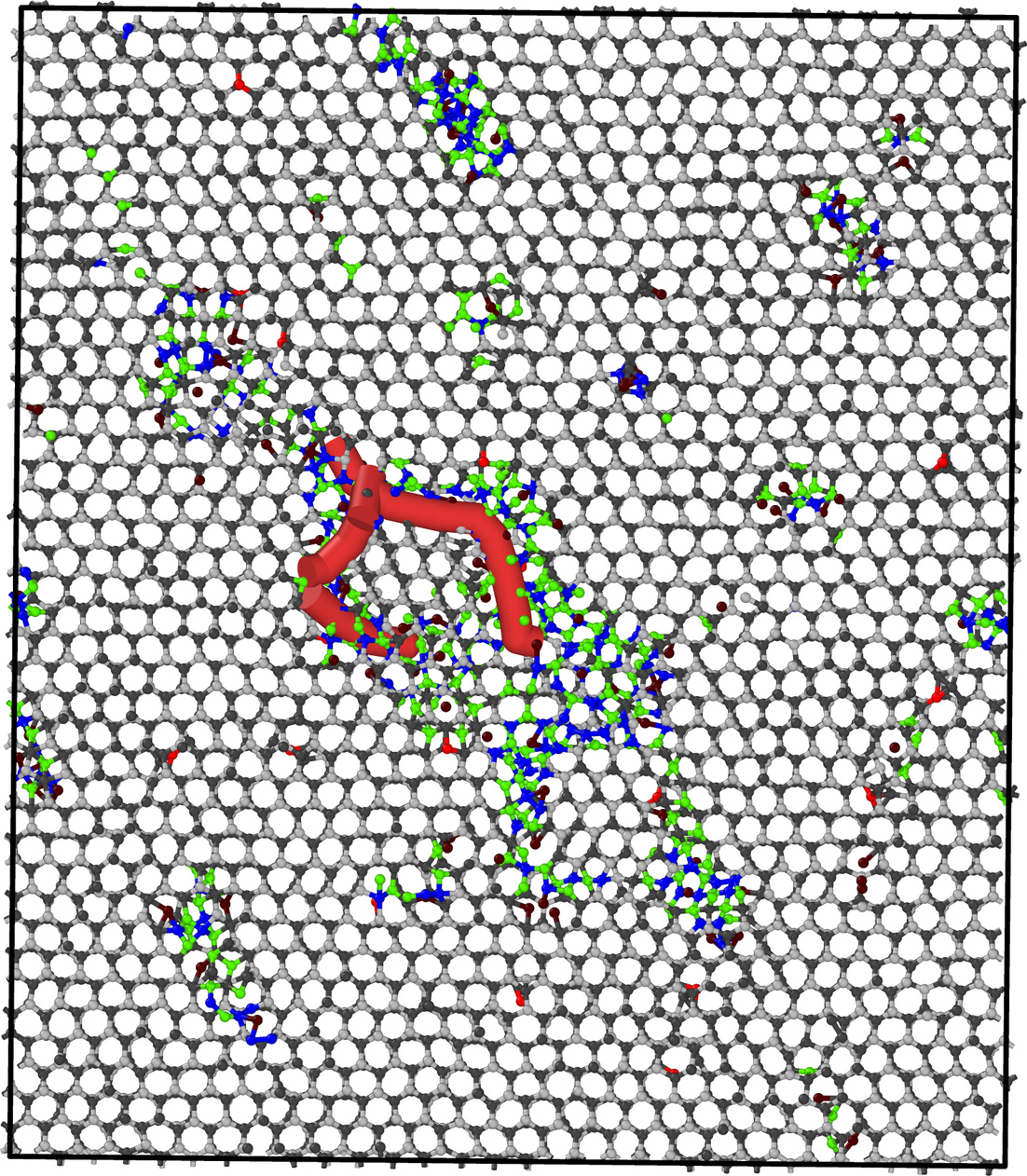}
     \includegraphics[width=\linewidth]{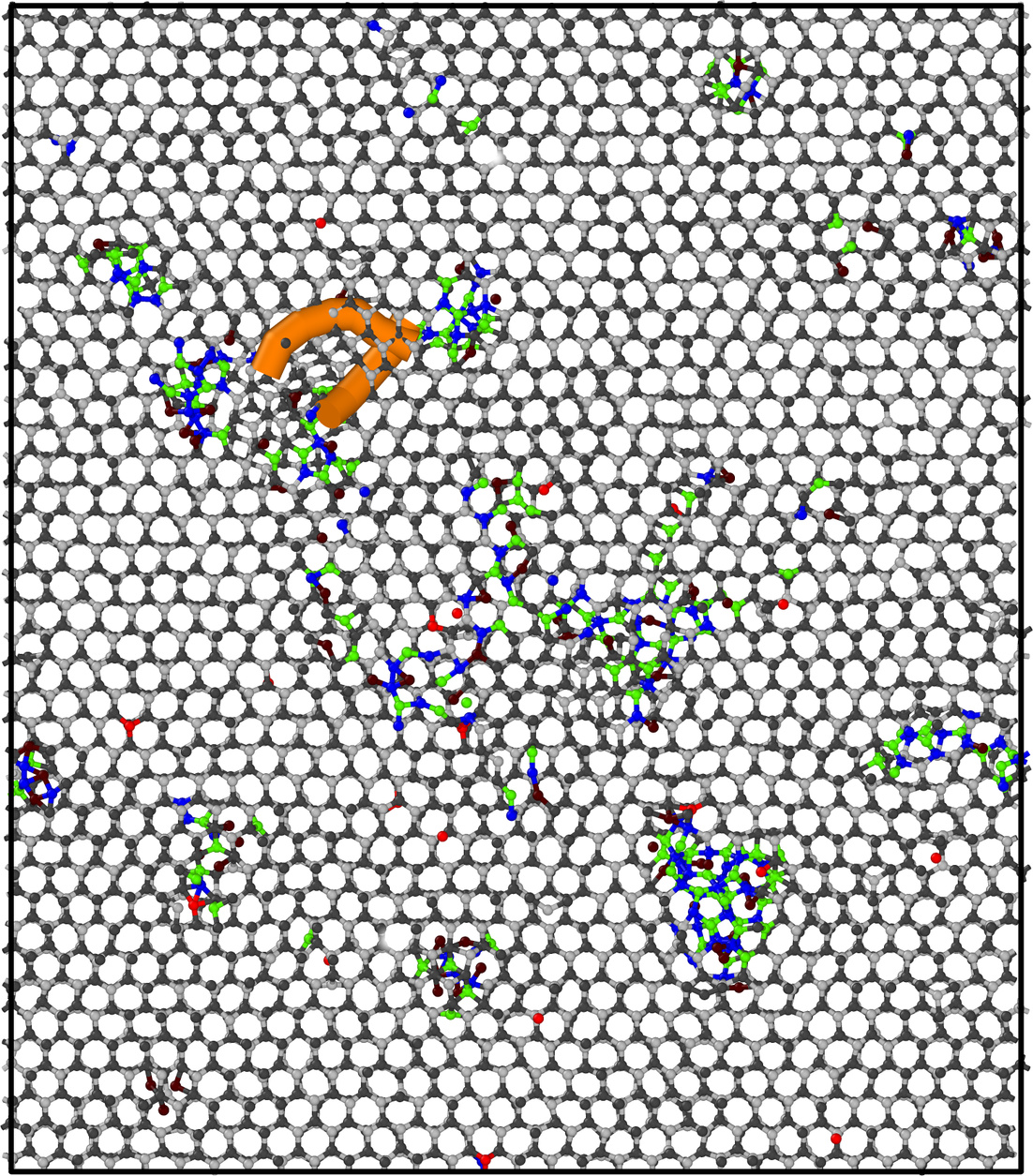}
  \end{minipage}
}\hfill
\subfloat[]{%
  \begin{minipage}[t]{.242\linewidth}
    \centering
    \includegraphics[width=\linewidth]{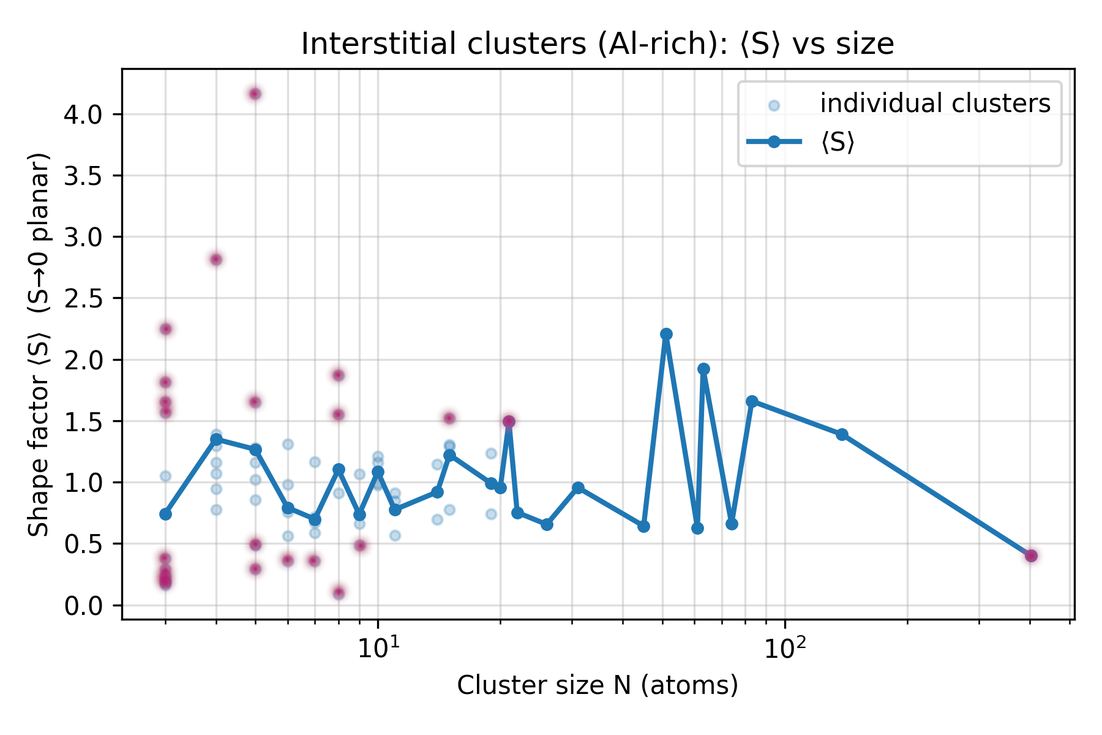}\\[1mm]
    \includegraphics[width=\linewidth]{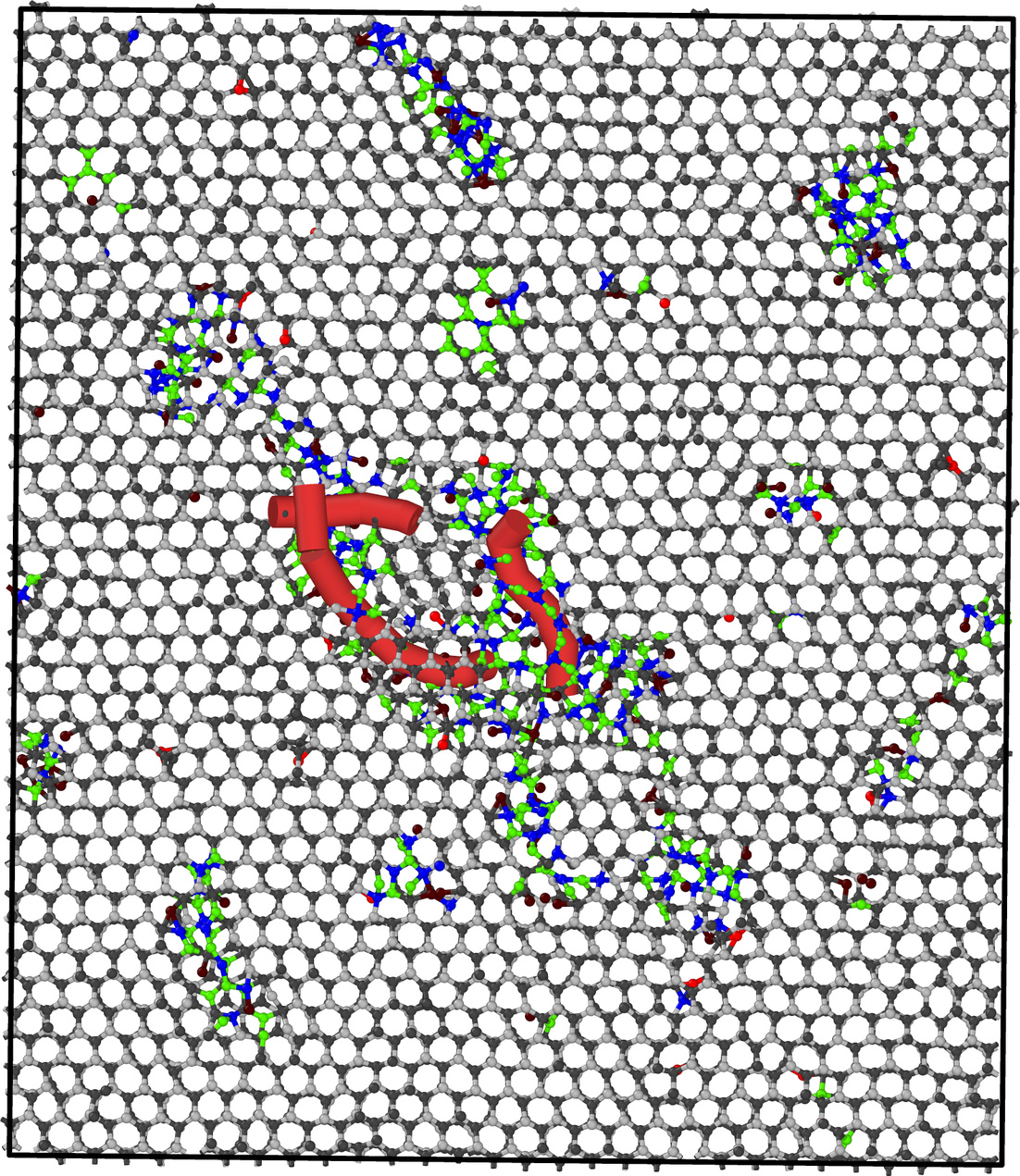}
     \includegraphics[width=\linewidth]{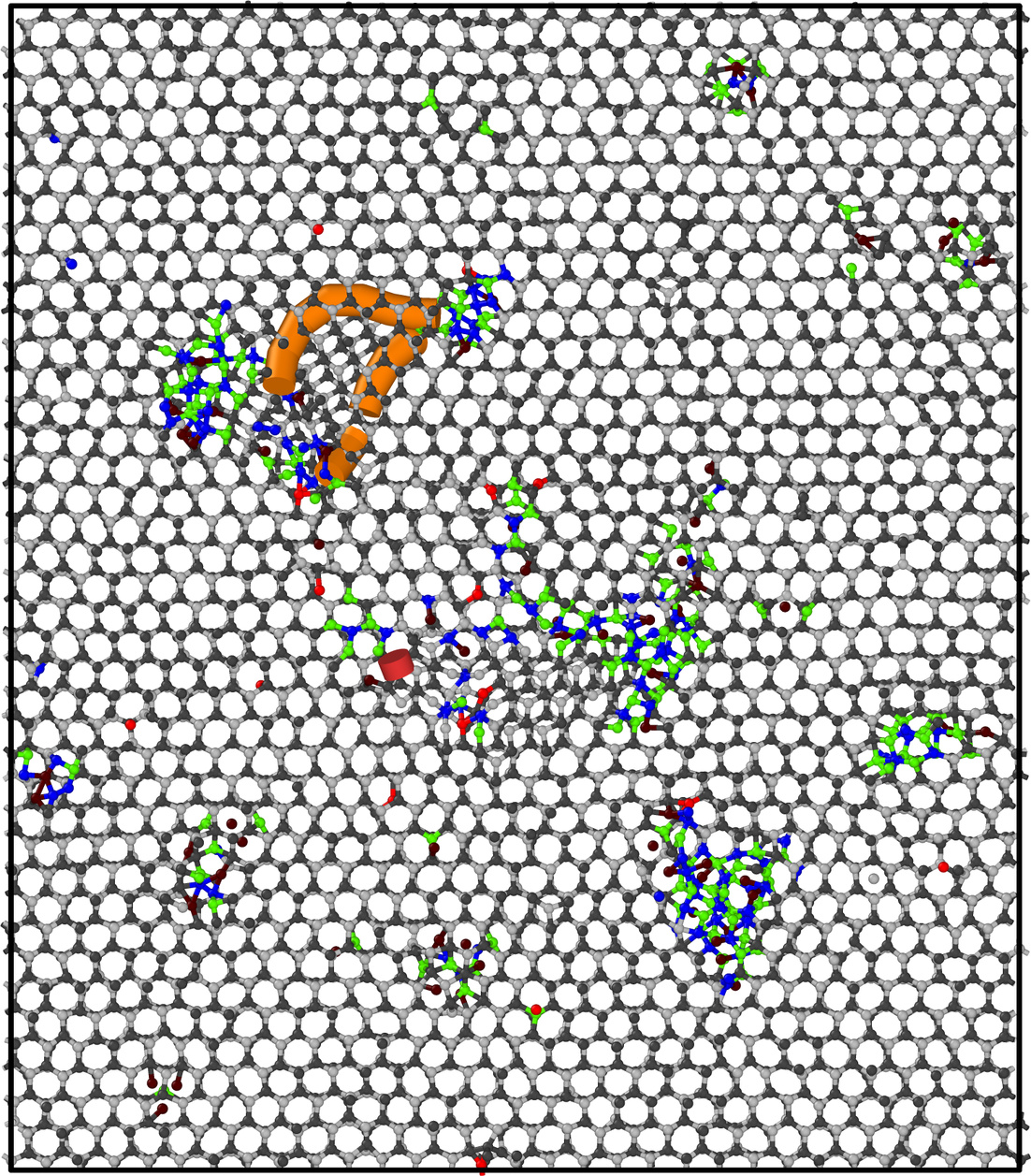}
  \end{minipage}
}
  \caption{Each subpanel (a)--(d) shows one time point for dose \mbox{$7.5\times10^{14}$\,cm$^{-2}$}, implantation temperature $900$\,K, and annealing temperature $1500$\,K: top, shape factors; bottom, corresponding dislocation-loop structure in two stacked atomic layers. Time points are (a) $10$\,ns, (b) $50$\,ns, (c) $70$\,ns, and (d) $90$\,ns of annealing. Orange ($1/3\langle 1\overline100\rangle$) and green ($1/3\langle 1\overline210\rangle$) dislocation lines indicate Shockley partials in the basal plane, while red indicates mixed dislocations.}
\label{fig:dislocation_1500K_dose10}
\end{sidewaysfigure}

Depending on the temperature, there is a critical size for the planar cluster that can build up sufficient shear stress in the basal plane to cause the appearance of stacking faults in the crystal lattice.
There is no fixed value of planar cluster size where the transition into a stacking fault instantly takes place. 
It is a stochastic/thermally activated process, as evidenced by the fact that the transformation proceeds more rapidly at higher temperatures. For the transition to occur, a barrier/activation energy must be overcome, which results from the restructuring of disordered defect clusters into ordered stacking faults. Since the transition is stochastic, the critical size of the planar cluster can vary greatly. There may be clusters in the system that have already undergone the transformation, while other clusters of the same size or even larger remain as irregular planar defects. By observing the transition at various annealing temperatures, we are attempting to narrow down the range of size values for the formation of Shockley and Frank faulted loops.

Since dislocation loops form spontaneously and grow rapidly by epitaxial regrowth which is further fed by the capture of additional defects at the edges of the loop, we chose a low annealing temperature of $1500$\,K to better track this dynamic process in Figure~\ref{fig:dislocation_1500K_dose10}. 
%This allows us to draw a correlation between planar clusters of certain size and the formation of a faulted loop.
Small clusters (size $<10$) do not cause sufficient lattice distortion to generate stacking faults. However, medium-sized clusters can already build up enough stress. The planar defects marked in magenta in part (a) of Figure~\ref{fig:dislocation_1500K_dose10} have sizes of $45$ and $60$ interstitials and are located at the edges of the Shockley partials that form after approximately $10$\,ns of annealing. The faulted loop has a stacking fault size of about $20$ atoms with a total length of all partials of $3$\,nm and persists until the end of the annealing.  After approximately $50$\,ns, another stacking fault forms in the center of the crystal. The transition takes place in the upper part of the largest planar irregular shaped cluster, consisting of about $600$ defects. As the additional crystalline layer is inserted, the planar cluster subsequently decreases to approximately $400$ defects at the end of the annealing cycle, where the planar clusters are acting as interstitial reservoirs.
However, the reduction in cluster size cannot be explained solely by the formation of stacking faults, in which only parts of the large cluster are involved. Rather, the cluster itself undergoes internal recombination reactions with the numerous vacancies that are still present at the beginning of the annealing process, causing the large cluster to shrink before the stacking fault emerges after $50$\,ns.
%The large central cluster did not lead to the formation of a dislocation loop from the beginning. This is likely because the cluster initially contained numerous vacancies, and 
Moreover, it is only after further recombination with vacancies that the necessary shear stress for the formation of stacking faults could be built up. 
This may also help to explain why, in 4H-SiC with a high damage ratio, the presence of numerous Frenkel pairs suppresses the formation of extended defects.
At other temperatures, larger clusters with $100$ or $250$ defects were also identified as precursors to stacking faults. In most cases, however, it was parts of large irregular planar structures that caused the transition. 
%To define a transition probability as a function of cluster size, one would have to multiply the nucleation rate by the cluster size distribution for each annealing temperature. However, the statistical data are insufficient to make a valid statement.
%The observed behavior suggests that the stacking fault does not nucleate as one coherent disk as assumed by the text book picture.  Instead, multiple locally transformed regions appear first. As they grow, their partials meet,
%and the individual faults merge into one. As the stacking fault advances, interstitials are incorporated into the inserted basal layer, the local cluster loses atoms,
%therefore the measured cluster shrinks. 
%A shrinking cluster before fault formation suggests that the cluster is not simply growing until it transforms.
%Instead, it may be redistributing atoms. So a shrinking cluster does not necessarily mean atoms disappear.
%It may simply mean
%the interstitials become crystallographically incorporated into the stacking fault. The planar clusters are not the final defect.
%Instead, they are probably acting as interstitial reservoirs.
\subsubsection{Dislocation nucleation rate and stacking fault growth rate}
\begin{figure*}[hbtp]
\centering
\subfloat[]{\fbox{\includegraphics[width=0.48\textwidth]{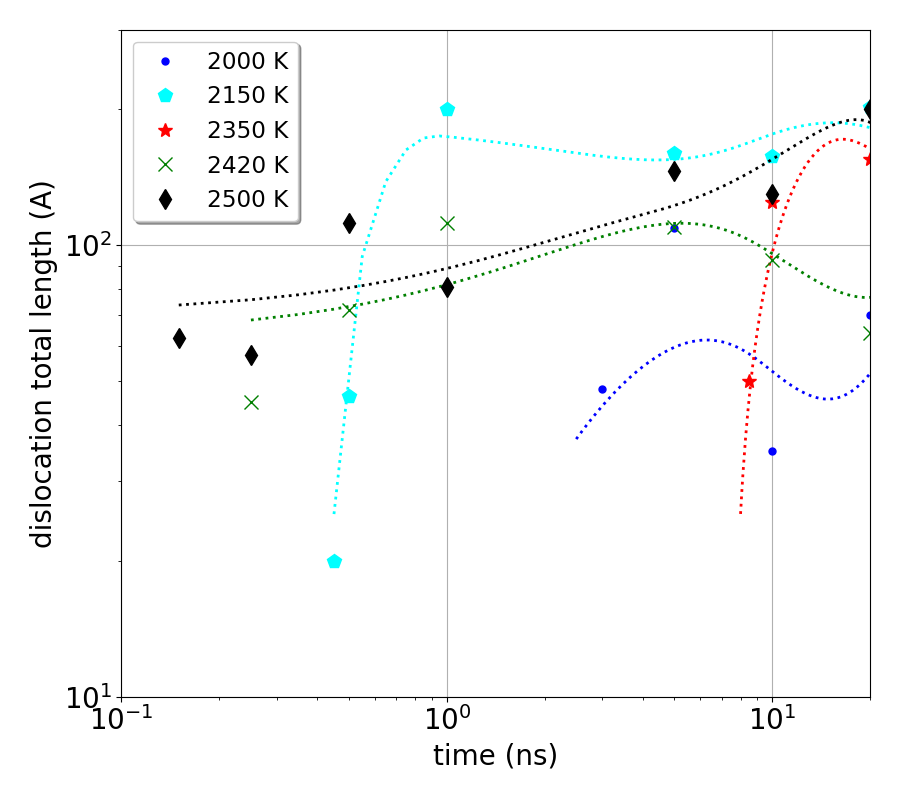}}}\hfill
 \subfloat[]{\fbox{\includegraphics[width=0.48\textwidth]{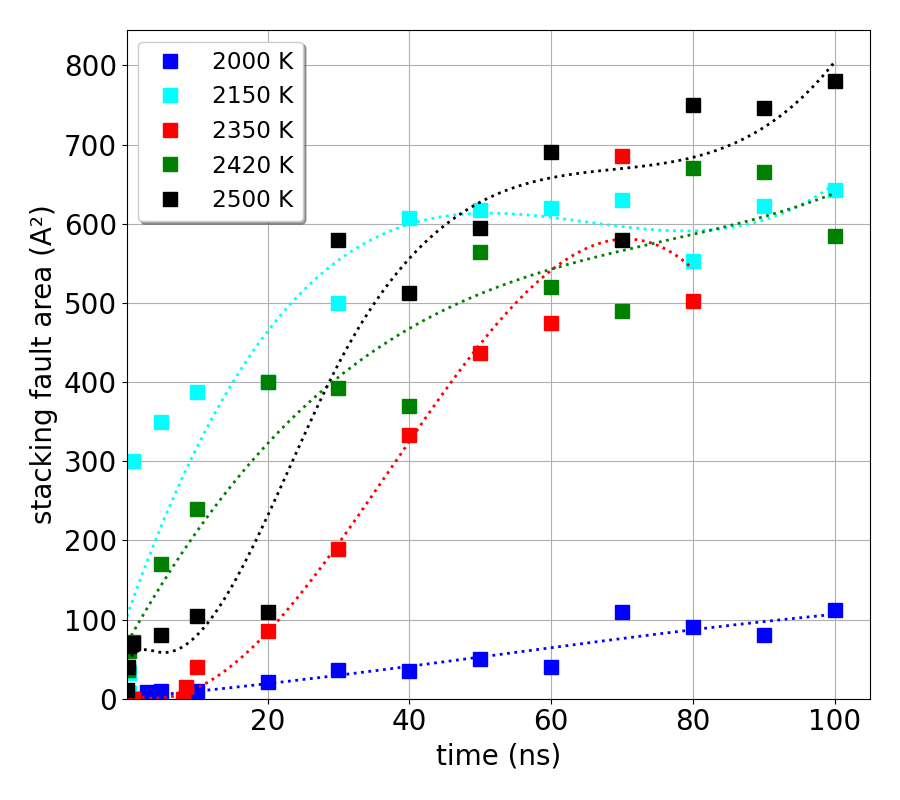}}}\\
 \subfloat[]{\fbox{\includegraphics[width=0.48\textwidth]{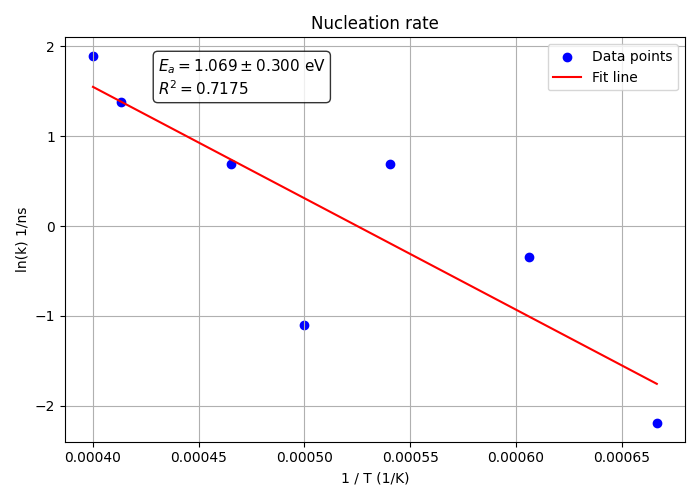}}}\hfill
 \subfloat[]{\fbox{\includegraphics[width=0.48\textwidth]{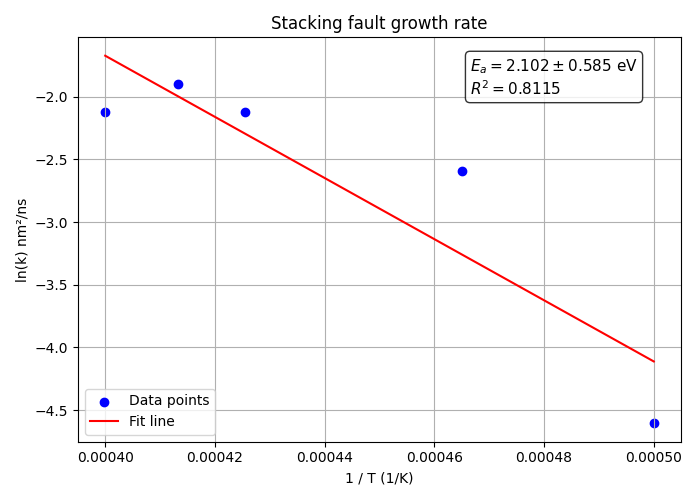}}}
  \caption{Total dislocation length (a) and stacking fault area (b) over time. Dislocation nucleation rate with pre-exponential factor of $6.7\times 10^{11}$s$^{-1}$, and activation energy E$_\mathrm{act}=1.07$\,eV (c). Stacking fault growth rate with pre-factor $3.3\times 10^{-2}$cm${^2}$/s and activation energy E$_\mathrm{act}=2.1$\,eV.}
  \label{fig:stacking_faults}
 \end{figure*}

Figure~\ref{fig:stacking_faults} shows the nucleation and growth rates of the faulted loops along with the activation energies. The time it takes for a dislocation line of approximately $3$\,nm in length to nucleate at different annealing temperatures is used to calculate the activation energy. For the growth rate, the slope shown in the Figure~\ref{fig:stacking_faults}(b) which occurs after the first spontaneous growth has finished was used. Specifically, for $2350$\,K, the slope between $10$ and $50$\,ns is used.   

In Summary: The seeds for stacking faults, consisting of about $20$ atoms encompassed by partials of $3$\,nm  length, form spontaneously with low activation energies of $1.1$\,eV in the temperature range between $1500$ and $2500$\,K. The observed behavior suggests that multiple locally faulted embryos appear along the evolving planar clusters and subsequently coalesce. Those planar clusters with critical sizes starting from $60$ to several $100$ interstitials, which trigger the transition by serving as interstitial reservoir during formation and growth of the extended planar defect, can be independent interstitial clusters (as shown in the lower panel of Figure~\ref{fig:dislocation_1500K_dose10}) as well as different parts of a single, elastically connected irregular interstitial aggregate (as shown in the center panel of Figure~\ref{fig:dislocation_1500K_dose10}). 
%\begin{figure}[htbp]
%\centering
%\fbox{\includegraphics[width=.45\linewidth]{figures/avg_shape_vs_size_interstitials_Alrich_2350K_dose_9_500K.png}}
%\fbox{\includegraphics[width=.45\linewidth]{figures/avg_shape_vs_size_interstitials_Alrich_2350K_dose_9_900K.png}}

%  \caption{Transient enhanced diffusion of Al (a) and C (b) during cluster growth and dissolution at high annealing temperatures and dose \mbox{$5\times10^{13}$}\,cm$^{-2}$.c) Arrhenius plot for the carbon interstitial after $100$\,ns with activation energy for migration of $3.2$\,eV and pre-factor $2.8$$\times$10$^{-2}$cm$^2$/s.}
%\label{fig:shape_dose9}
%\end{figure}
%\begin{figure}[htbp]
%\centering
%\fbox{\includegraphics[width=.48\linewidth]{figures/avg_shape_vs_size_interstitials_Alrich_1500K_dose_4_500K.png}}
%\fbox{\includegraphics[width=.48\linewidth]{figures/avg_shape_vs_size_interstitials_Alrich_2000K_dose_4_500K.png}}

%  \caption{Transient enhanced diffusion of Al (a) and C (b) during cluster growth and dissolution at high annealing temperatures and dose \mbox{$5\times10^{13}$}\,cm$^{-2}$.c) Arrhenius plot for the carbon interstitial after $100$\,ns with activation energy for migration of $3.2$\,eV and pre-factor $2.8$$\times$10$^{-2}$cm$^2$/s.}
%\label{fig:shape_dose4}
%\end{figure}

\newpage
\subsection{Cluster composition}
\begin{sidewaysfigure}[hbtp]
\centering
   \subfloat[]{\fbox{\includegraphics[width=0.3\textwidth]{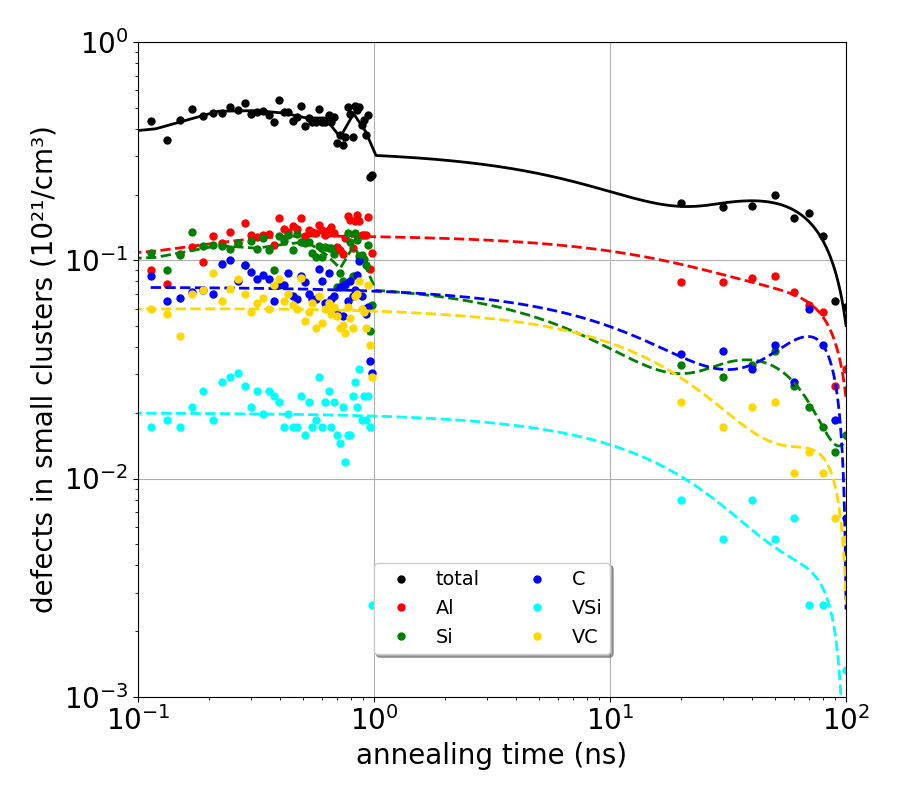}}}\hfill
   \subfloat[]{\fbox{\includegraphics[width=0.3\textwidth]{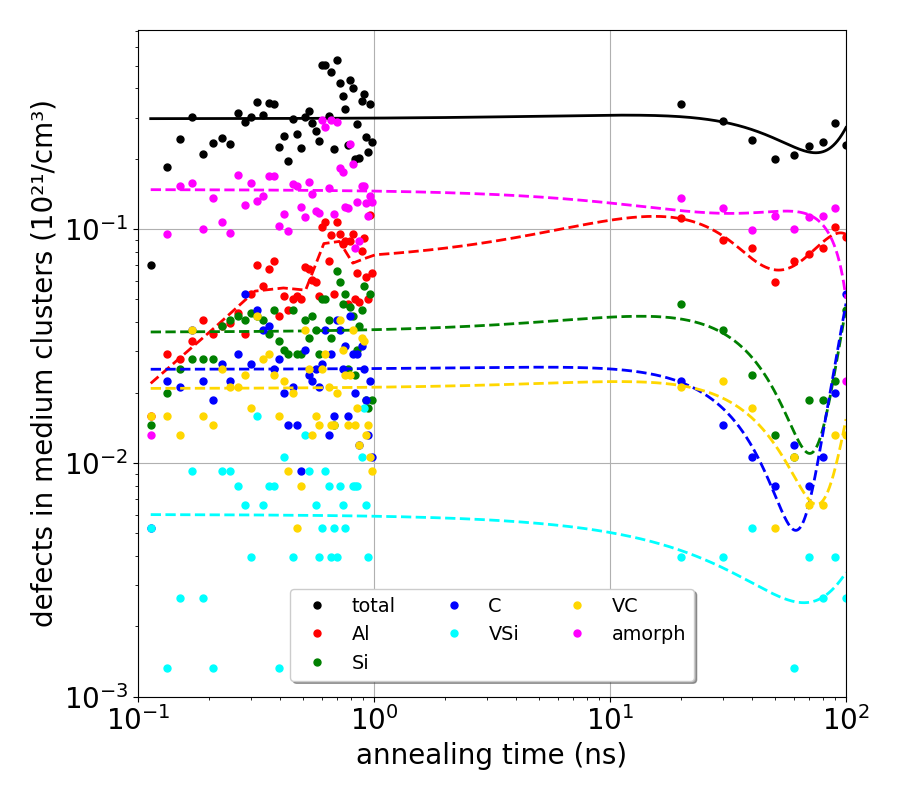}}}\hfill
   \subfloat[]{\fbox{\includegraphics[width=0.3\textwidth]{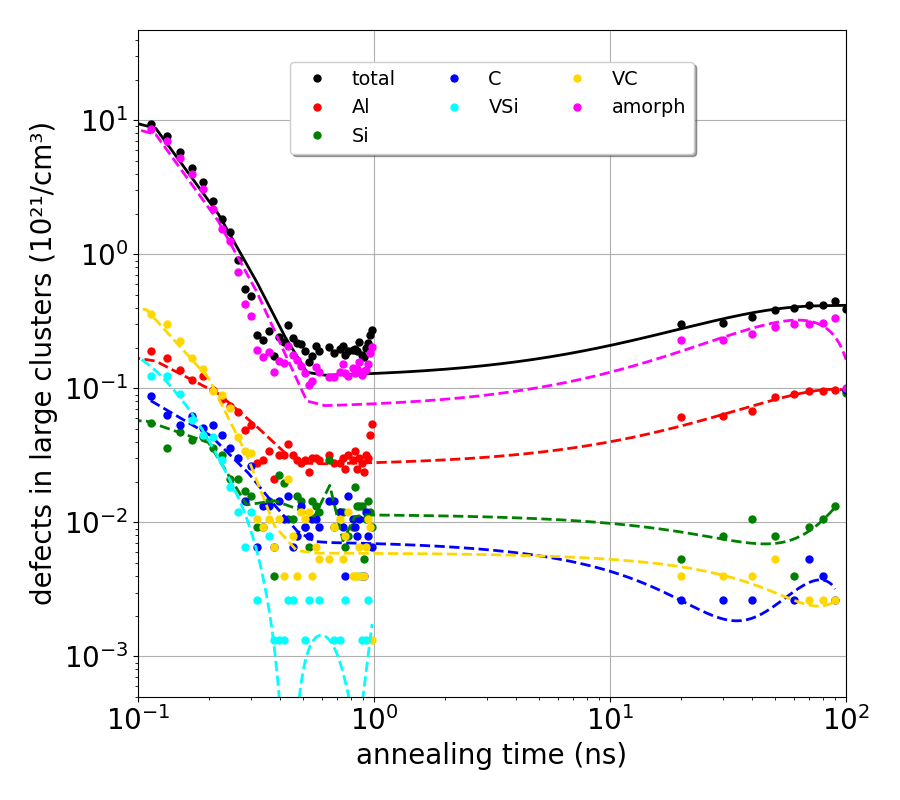}}}\\
   \subfloat[]{\fbox{\includegraphics[width=0.3\textwidth]{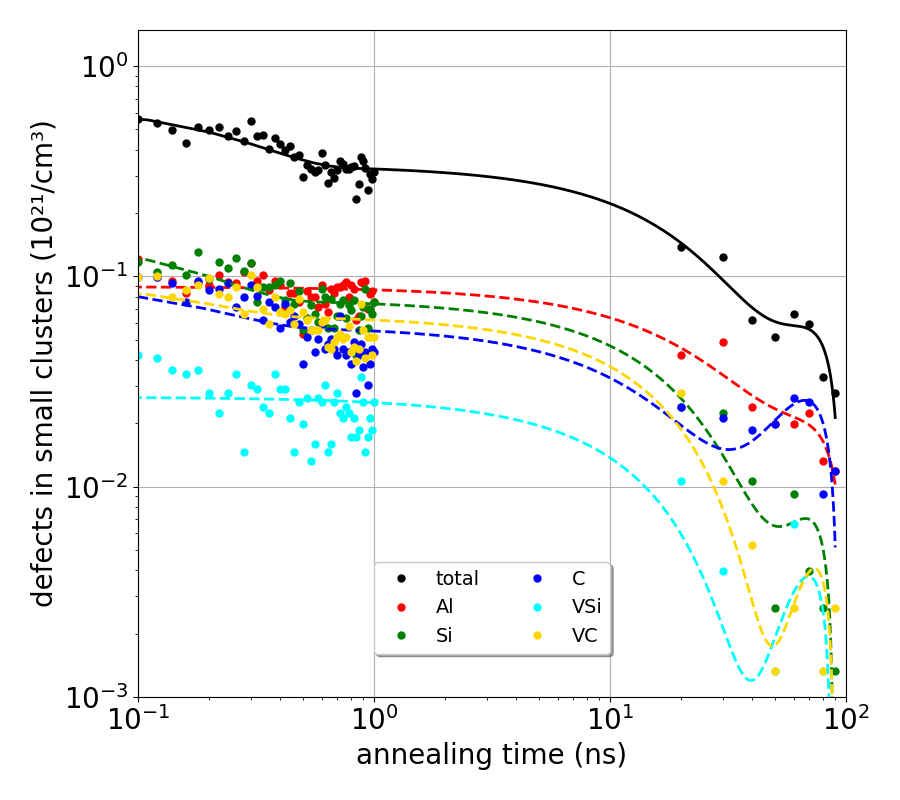}}}\hfill
   \subfloat[]{\fbox{\includegraphics[width=0.3\textwidth]{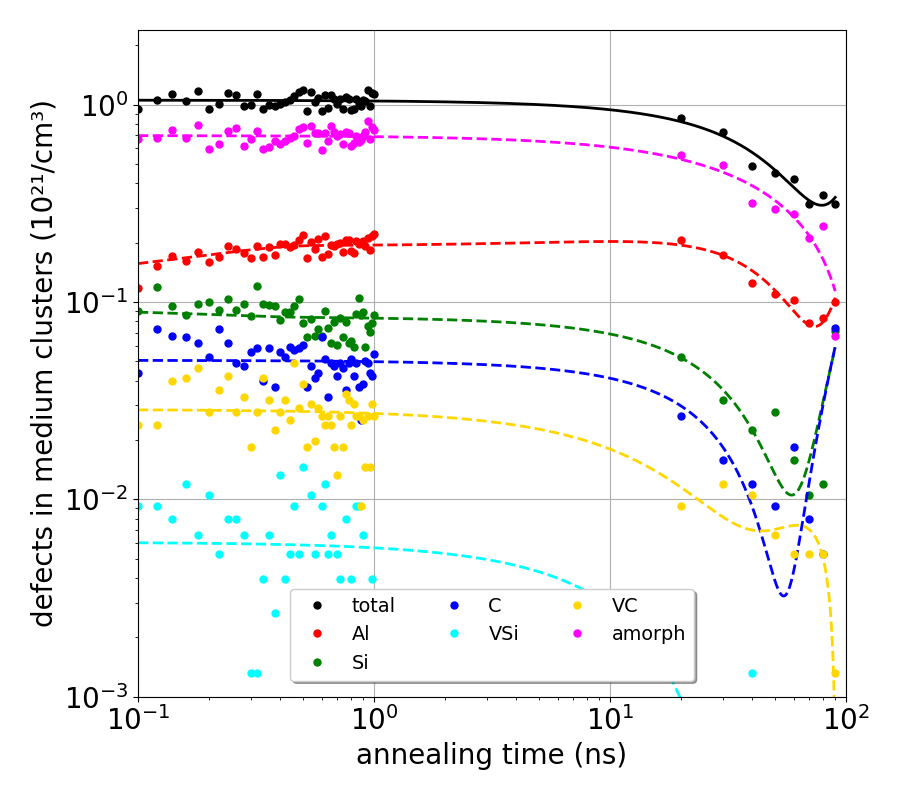}}}\hfill
   \subfloat[]{\fbox{\includegraphics[width=0.3\textwidth]{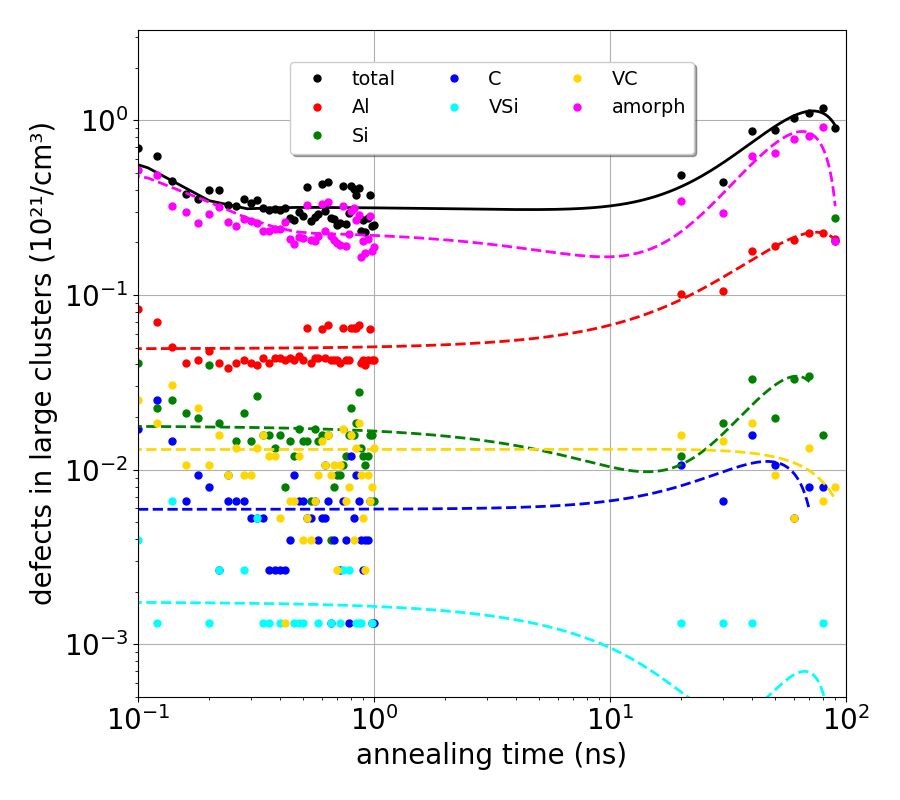}}}
   \caption{Atomistic composition of small (size $<10$) (a,d), medium (10 $\leq$ size <$100$) (b,e), and large clusters (size $\geq$ 100) (c,f) at a dose of \mbox{$5.0\times10^{14}$\,cm$^{-2}$} and annealing temperature of $2350$\,K. The implantation temperature is $500$\,K in the upper panel (a-c) and $900$\,K in the lower panel (d-f). Lines are included as visual guides.}
  \label{fig:cluster_composition}
 \end{sidewaysfigure}

Figure~\ref{fig:cluster_composition} shows the chemical cluster composition in dependence of implantation temperature and cluster size. In general the compositions are very similar as a function of temperature. Al and C interstitials make up the largest proportion of the small clusters. The medium- and large-sized compact clusters have an amorphous core consisting of intrinsic defects that are essentially decorated with Al and residual Si interstitials. C interstitials, on the other hand, are less common in the large clusters. The same applies to the planar clusters that form along the dislocation lines of the stacking faults.
 
\newpage
\subsection{Thermal stability of point defects and complexes}
\begin{sidewaysfigure}[hbtp]
\centering
   \subfloat[]{\fbox{\includegraphics[width=0.3\textwidth]{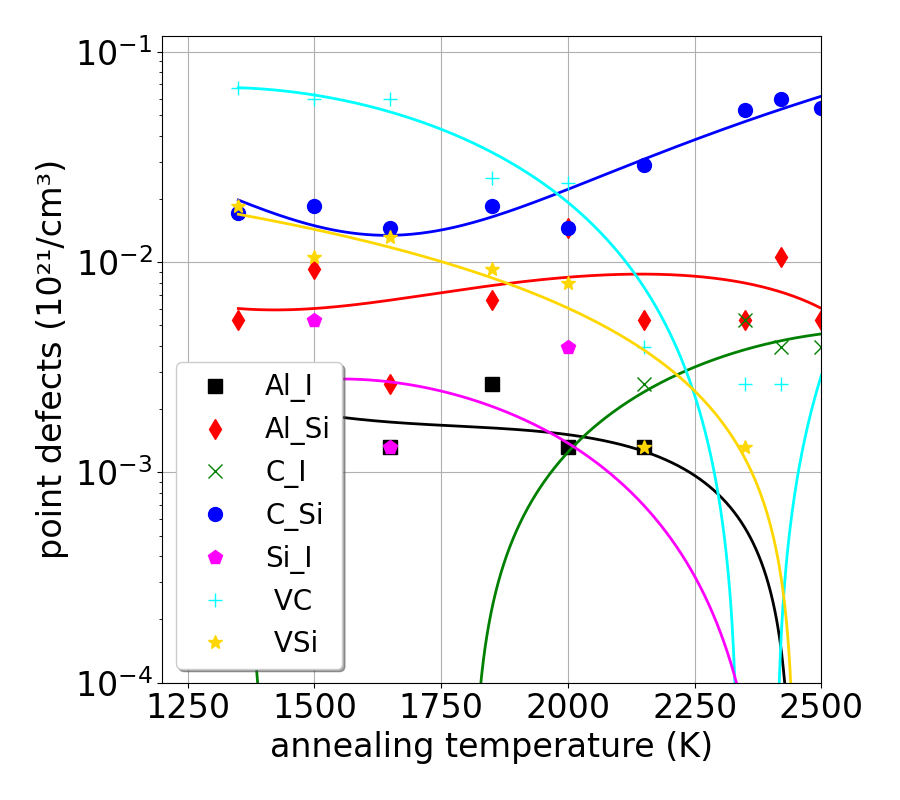}}} \hfill
   \subfloat[]{\fbox{\includegraphics[width=0.3\textwidth]{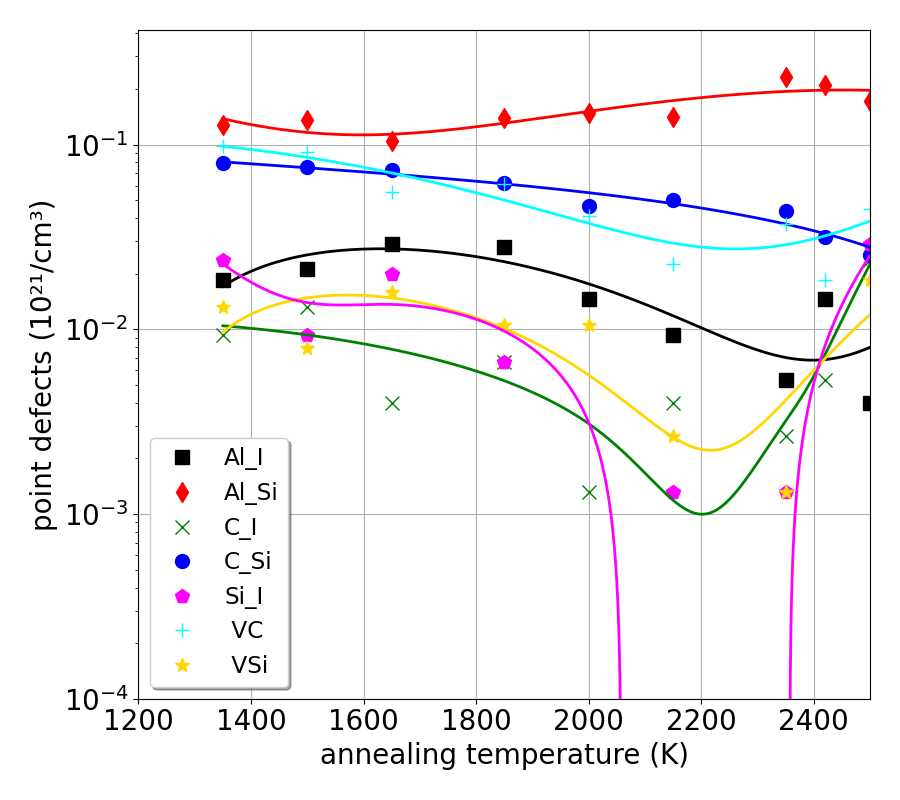}}}\hfill
   \subfloat[]{\fbox{\includegraphics[width=0.3\textwidth]{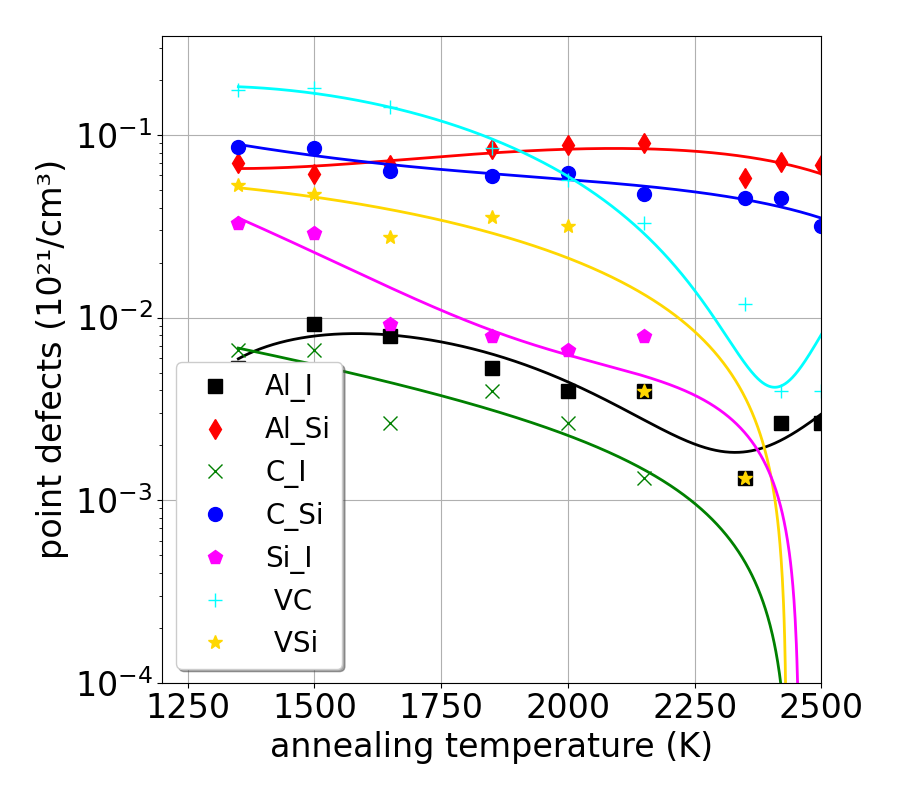}}}\\
   \subfloat[]{\fbox{\includegraphics[width=0.3\textwidth]{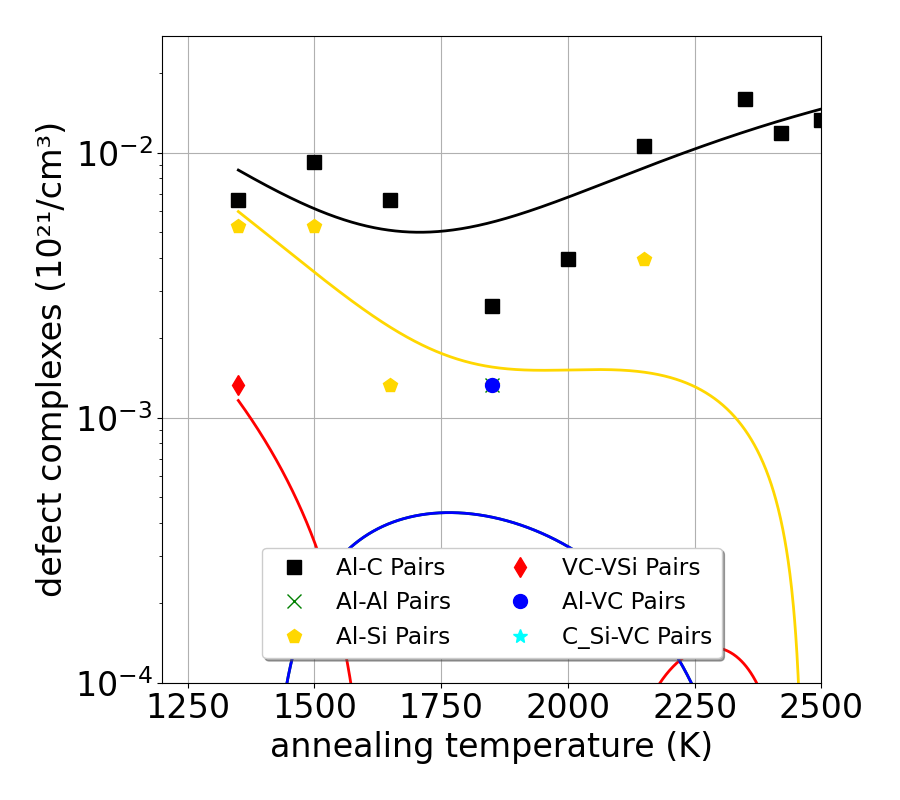}}} \hfill
   \subfloat[]{\fbox{\includegraphics[width=0.3\textwidth]{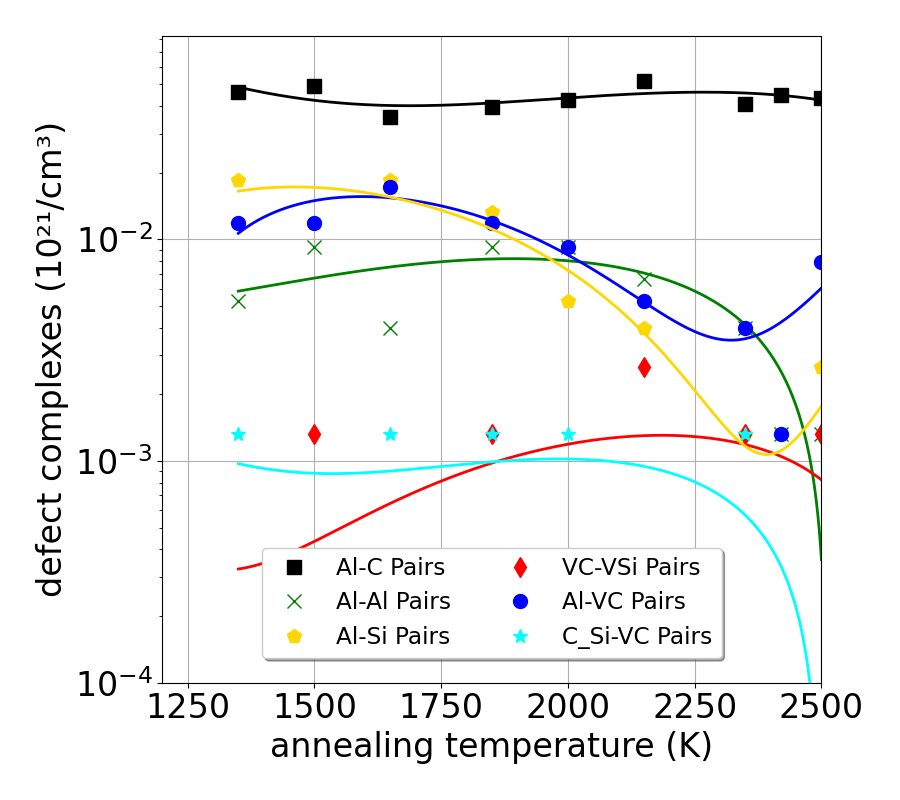}}}\hfill
   \subfloat[]{\fbox{\includegraphics[width=0.3\textwidth]{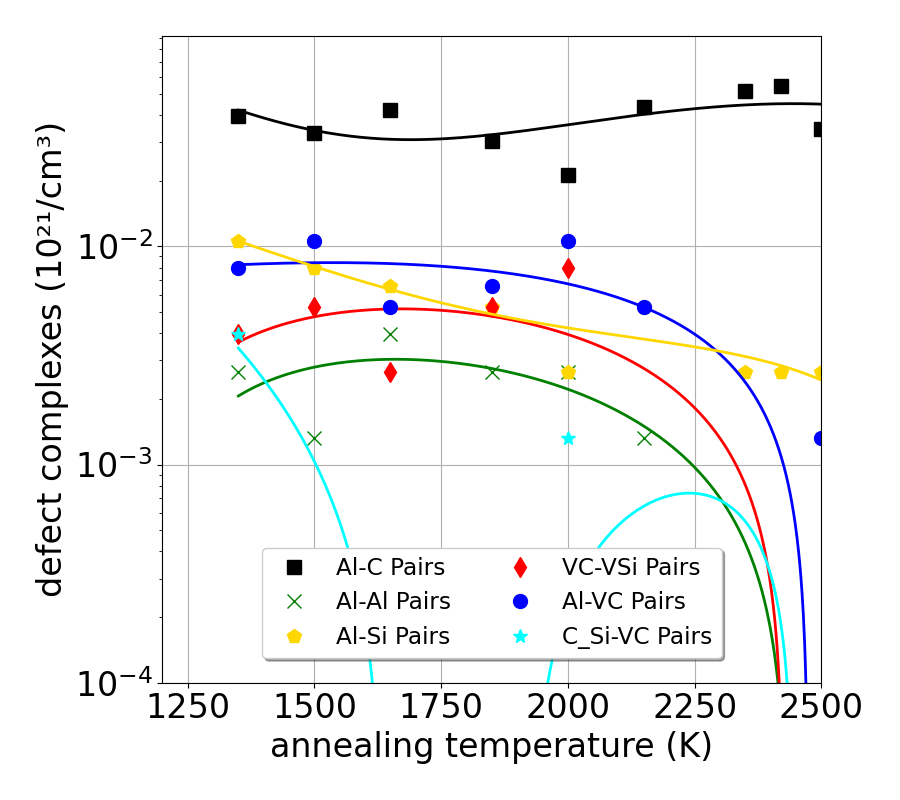}}}
   \caption{Thermal stability of point defects (a, b, c) and defect complexes (d, e, f) for dose $5\times 10^{13}$\,cm$^{-2}$ (a, d), and $7.5\times 10^{14}$\,cm$^{-2}$ for $500$\,K (b, e) and $900$\,K implantations (c, f) after $100$\,ns of annealing. Lines are included as visual guides.}
   \label{fig:selected_defects}
\end{sidewaysfigure}

\end{document}